\documentclass[11pt,letterpaper]{article}
\usepackage[margin=1in]{geometry}
\usepackage[T1]{fontenc}
\usepackage{lmodern}
\usepackage{amsmath,amssymb,amsthm,mathtools}
\usepackage{mathrsfs}
\usepackage{needspace}
\usepackage{tikz}
\usetikzlibrary{arrows.meta}
\usepackage{authblk}

\usepackage{microtype,booktabs,array,enumitem,xcolor}
\usepackage[colorlinks=true,linkcolor=blue!45!black,citecolor=blue!45!black,urlcolor=blue!45!black]{hyperref}
\usepackage[nameinlink,noabbrev]{cleveref}
\setlist{itemsep=3pt,topsep=5pt}
\newtheorem{theorem}{Theorem}[section]
\newtheorem{lemma}[theorem]{Lemma}
\newtheorem{proposition}[theorem]{Proposition}
\newtheorem{corollary}[theorem]{Corollary}
\newtheorem*{specializationlemma}{Lemma}
\theoremstyle{definition}
\newtheorem{definition}[theorem]{Definition}
\newcommand{\G}{\mathcal G}
\newcommand{\Amp}{\operatorname{Amp}_Z}
\newcommand{\Holant}{\operatorname{Holant}}
\newcommand{\ReTract}{\mathsf{ReTract}}
\newcommand{\LMWTract}{\mathsf{Tract}_{\mathrm{LMW}}}
\newcommand{\poly}{\mathrm{poly}}
\newcommand{\FP}{\mathrm{FP}}
\newcommand{\ket}[1]{\lvert #1\rangle}
\newcommand{\bra}[1]{\langle #1\rvert}
\newcommand{\CNOT}{\mathrm{CNOT}}
\newcommand{\CZ}{\mathrm{CZ}}
\newcommand{\SWAP}{\mathrm{SWAP}}
\newcommand{\diag}{\operatorname{diag}}
\newcommand{\Tr}{\operatorname{Tr}}
\newcommand{\Qbar}{\overline{\mathbb Q}}
\newcommand{\Mono}{\mathsf{Mono}}
\newcommand{\ProdPerm}{\mathsf{ProdPerm}}
\newcommand{\DC}{\mathsf{DC}}
\newcommand{\QTract}{\mathsf{QTract}}
\newcommand{\Q}{\mathbb Q}
\newcommand{\R}{\mathbb R}
\newcommand{\C}{\mathbb C}
\newcommand{\Z}{\mathbb Z}
\newcommand{\NF}{\operatorname{NF}}
\newcommand{\SAT}{\#\mathrm{SAT}}

\newcommand{\wt}{\operatorname{wt}}
\newcommand{\Frac}{\operatorname{Frac}}

\definecolor{pinkfoldpink}{HTML}{B74B7D}
\newcommand{\pinkfoldmark}{%
  \raisebox{-0.15em}{%
    \begin{tikzpicture}[x=0.0375em,y=-0.0375em]
      \path[use as bounding box] (0,0) rectangle (24,24);
      \draw[pinkfoldpink,line width=0.06375em,line join=miter,line cap=butt]
        (12,2) -- (20.660,7) -- (20.660,17) -- (12,22) --
        (3.340,17) -- (3.340,7) -- cycle
        (12,12) -- (20.660,7)
        (12,12) -- (3.340,7)
        (12,12) -- (12,22);
    \end{tikzpicture}%
  }%
}

\hypersetup{pdftitle={A Dichotomy for Exact Quantum Circuit Amplitudes},pdfauthor={Chenghua Liu and Boning Meng},pdfsubject={Exact amplitude complexity of finite quantum gate libraries}}

\title{A Dichotomy for Exact Quantum Circuit Amplitudes}
\author[1]{Chenghua Liu}
\author[2]{Boning Meng}
\affil[1]{Institute of Software, Chinese Academy of Sciences, Beijing, China}
\affil[2]{University of Regensburg, Regensburg, Germany}
\affil[ ]{\texttt{liuch.russell@gmail.com}, \texttt{mengboning2013@gmail.com}}
\date{}

\begin{document}
\hypersetup{pageanchor=false}
\begin{titlepage}
\maketitle
\thispagestyle{empty}
\vspace{-2em}

\begin{abstract}
\normalsize
We classify the complexity of exactly computing computational-basis amplitudes
for circuits over any fixed finite library of unitary qubit gates, with each
gate allowed to act on any ordered tuple of distinct wires. Exact evaluation
is polynomial-time computable if all gates permute the computational basis up
to phases, all preserve product states, or all become Clifford under a common
diagonal one-qubit unitary change of basis on every wire. For every other
library, the problem is $\#\mathrm P$-hard under polynomial-time Turing
reductions. Our proof connects
circuit amplitudes to the Boolean Holant classification of tensor contractions
on arbitrary graphs. For an algebraic library containing a nonmonomial gate
and an entangling gate, we show how to simulate the joint use of a symmetric
maximally entangled state and its adjoint at polynomial cost, using only
amplitude queries over the original library. This lets us evaluate
unrestricted contractions of the resulting gate and boundary tensors.
Unitarity and the computational-basis boundaries then restrict the tractable
cases to the three forms above. Algebraic specialization extends the dichotomy
to arbitrary fixed complex gate entries under a symbolic exact-output
convention.
\end{abstract}
\end{titlepage}
\setcounter{page}{1}
\hypersetup{pageanchor=true}

\section{Introduction}
\label{sec:intro}

The difficulty of simulating quantum physics on a classical computer
helped motivate quantum computation itself. In 1982 Feynman asked
whether a classical probabilistic computer could reproduce the
statistics of a quantum system~\cite{Fey82}. A direct simulation of
an $n$-qubit circuit updates $2^n$ complex amplitudes at each step,
but some circuits admit far shorter descriptions. The Gottesman--Knill
theorem gives one landmark example: a Clifford circuit can create a
highly entangled state from a basis input, yet the simulator need only
update the $n$ Pauli generators of its stabilizer~\cite{Got99,AG04}.
Valiant found an efficient simulation for nearest-neighbour matchgate
circuits~\cite{Val02}, whose tractability Terhal and DiVincenzo traced
to the linear mode evolution of noninteracting fermions~\cite{TD02}.
Vidal identified a condition on the evolving state itself: if the
Schmidt ranks across successive cuts of the qubit chain remain
polynomially bounded, the state can be represented and updated
efficiently throughout the computation~\cite{Vid03}.

These distinct mechanisms prompted Aaronson in 2005 to ask whether the
known classical simulation algorithms admit a common generalization,
and what structural features separate classically simulable gate sets
from intractable ones~\cite[Questions~1--2]{Aar05}. Yet the examples
above operate at different levels. Vidal's condition restricts the
entanglement generated along a particular circuit instance, whereas
Valiant's restricts the placement geometry: two circuits over the same
gates may have very different Schmidt ranks, and moving a matchgate to
a distant pair of wires takes it outside the nearest-neighbour
fermionic regime~\cite{JM08}.

To isolate what the gates themselves guarantee, we fix a finite
library $\G$ of unitary qubit gates and consider all circuits over
$\G$. Each $k$-qubit gate may act on any ordered tuple of $k$ distinct wires. One-qubit gates and inverses are available only
if $\G$ supplies them, and there are no intermediate measurements.
Given such a circuit $C$ and computational-basis strings $b$ and $a$,
we ask for the exact transition amplitude $\bra a U_C\ket b$, the
coefficient of $\ket a$ in the output state $U_C\ket b$. Its squared
magnitude is the probability of observing $a$ on input $b$. Even with
both endpoints fixed, exponentially many paths can contribute to this
single coefficient, and their complex weights can reinforce or cancel.
We study the following version of Aaronson's classification question.

\begin{center}
\itshape
What structural condition on $\G$ is necessary and sufficient\\
for exact amplitude evaluation in polynomial time?
\end{center}

We give a complete answer in the form of an $\FP$ versus
$\#\mathrm P$-hard dichotomy for every fixed finite library of unitary
qubit gates. Exact amplitude evaluation is in $\FP$ if (i)~every gate
in $\G$ permutes the computational basis up to phases, (ii)~every gate
preserves product states, or (iii)~a single diagonal one-qubit unitary,
applied on every wire as a change of basis, makes every gate Clifford.
When all three conditions fail, we show how to count the satisfying
assignments of an arbitrary Boolean formula in polynomial time using
an oracle for exact amplitudes over $\G$. This Turing reduction
establishes $\#\mathrm P$-hardness for every library outside the three
classes, ruling out intermediate complexity in this model. Under the
standard assumption $\#\mathrm{SAT}\notin\FP$, exact amplitude
evaluation is therefore in $\FP$ \emph{if and only if} $\G$ satisfies
at least one of the three conditions.

The criterion asks whether one structure is shared by the whole
library. To see why pairwise tractability does not suffice, consider
$H$, $T=\diag(1,e^{i\pi/4})$, and $\CNOT$. Each pair admits efficient
exact amplitude evaluation for a different reason. The gates in
$\{H,T\}$ act on one qubit at a time and hence preserve product states.
With $\{T,\CNOT\}$ a basis input follows a single trajectory up to
phase, and with $\{H,\CNOT\}$ every gate is Clifford. When all three
are available, $H$ branches on a basis input, $\CNOT$ can entangle a
product input, and $T$ remains non-Clifford in every diagonal frame
because it commutes with the change of basis. The full library thus
violates all three conditions, and its exact amplitudes are
$\#\mathrm P$-hard to compute. More generally, three gates always suffice
to witness the failure of the three conditions
(Corollary~\ref{cor:three-gate-witness}).

\subsection{The classification}
\label{sec:classification}

Let $U_C$ be the unitary on $(\C^2)^{\otimes n}$ implemented by an
$n$-wire circuit $C$ over the fixed library $\G$. With output string
$a$ and input string $b$, both in $\{0,1\}^n$ and specifying all
wires including auxiliaries, the exact amplitude problem in the computational
($Z$) basis is
\[
   \Amp(\G)(C,a,b)=\bra a U_C\ket b.
\]
The following three conditions on $\G$ may overlap.
\begin{enumerate}[label=\textup{(\roman*)},leftmargin=2em]
\item \emph{Monomial gates.}
Each $k$-qubit gate $U\in\G$ permutes the computational basis up to
phases. That is, $U\ket x=\omega_x\ket{\pi(x)}$ for a permutation
$\pi$ of $\{0,1\}^k$ and scalars $\omega_x$ of unit modulus. Equivalently, $U$
has exactly one nonzero entry in each row and column.

\item \emph{Product-preserving gates.}
Every $k$-qubit gate maps every product state
$\ket{\psi_1}\otimes\cdots\otimes\ket{\psi_k}$ to a product state.
Equivalently, each $k$-qubit $U\in\G$ has the form
$U=(V_1\otimes\cdots\otimes V_k)P$, where $V_1,\ldots,V_k$ are
one-qubit unitaries and $P$ permutes the $k$ tensor factors~\cite{Johnston11}.

\item \emph{A common diagonal Clifford frame.}
A $k$-qubit unitary is Clifford if conjugation by it preserves the
Pauli group, consisting of tensor products of $I,X,Y,Z$ with phases
in $\{1,i,-1,-i\}$. This definition allows arbitrary global phases. We require
a single $z\in\C$, $|z|=1$, such that every $k$-qubit gate $U\in\G$
can be written as
\begin{equation}
 U=D(z)^{\otimes k} C_U \bigl(D(z)^\dagger\bigr)^{\otimes k},
 \qquad D(z)=\diag(1,z),
 \label{eq:intro-frame}
\end{equation}
where $C_U$ is Clifford. The same $D(z)$ must work on every wire
and for every gate, however many qubits it acts on.
\end{enumerate}

We encode amplitudes symbolically relative to the fixed gate entries.
For algebraic entries we return rational coordinates in a fixed number
field containing them~\cite[Section~6]{ADH97}.
For arbitrary complex entries, let $\alpha=(\alpha_1,\ldots,\alpha_s)$
list their real and imaginary parts. The real and imaginary parts of
each amplitude are polynomials in $\alpha$ with rational coefficients.
To choose a unique representative for each, we use a standard tool of
computer algebra, the Gr\"obner-basis method introduced by
Buchberger~\cite{Buc65,CLO15}. We reduce modulo the full relation ideal
$I_\alpha=\{p\in\Q[x_1,\ldots,x_s]:p(\alpha)=0\}$ in graded
lexicographic order. For example, if $\alpha_j=1/\sqrt{2}$ is a
Hadamard entry, then $x_j^2$ reduces to $1/2$. A finite generating
set for $I_\alpha$ is supplied with the library. The output consists
of the rational coefficients and exponent vectors of the two reduced
polynomials, written in binary. Yamakami and Yao used a related
encoding based on a transcendence basis for quantum Turing machines
with arbitrary complex amplitudes~\cite[Section~4]{YY99}.

Write $\FP$ for the class of functions computable in deterministic
polynomial time and $\#\mathrm{SAT}$ for counting satisfying assignments
of a Boolean formula. We write $f\le_T^{\poly}g$ if $f$ can be computed
in polynomial time with oracle access to $g$, and $f\equiv_T^{\poly}g$
if this holds in both directions.
\begin{theorem}[Exact-amplitude dichotomy]
\label{thm:intro-dichotomy}
If $\G$ satisfies at least one of \textup{(i)}--\textup{(iii)}, then
$\Amp(\G)\in\FP$. Otherwise, $\Amp(\G)$ is $\#\mathrm P$-hard
under polynomial-time Turing reductions.
\end{theorem}
We use this reduction convention throughout.
Under $\#\mathrm{SAT}\notin\FP$, these classes thus exhaust the
libraries with polynomial-time exact amplitude evaluation. Given the
coefficient presentation, one can decide which conditions hold and
construct the corresponding simulation algorithm or hardness reduction.

In \textup{(i)} a simulator tracks one basis string and its phase,
and in \textup{(ii)} it tracks one state vector per wire. In
\textup{(iii)}, replacing each $U$ by $C_U$ on the same wires gives a
Clifford circuit $C'$ whose amplitudes can be computed by a stabilizer
algorithm that retains global phase~\cite{DH22}. The common frame
factors cancel internally.
A general one-qubit change of basis can rotate the endpoints into
nonstabilizer product states. Here $D(z)^{\otimes n}\ket x=z^{|x|}\ket x$,
where $|x|$ is the Hamming weight of $x$. Hence
\[
  \bra a U_C\ket b=z^{|a|-|b|}\bra a U_{C'}\ket b.
\]

For example, let $K=THT^\dagger$. In the frame $D=T$, the non-Clifford
gate $K$ becomes $H$ and $\CZ$ remains unchanged. Since $K$ branches
on $\ket0$ and $\CZ$ entangles $\ket+\ket+$, the library $\{K,\CZ\}$
satisfies only condition \textup{(iii)}. Adding $H$ rules out this
last condition as well, since a common frame $D(z)$ would require
$z^4=1$ for $H$ and $z^4=-1$ for $K$
(Corollary~\ref{cor:phase-compatibility}). Thus exact amplitude
evaluation for $\{H,K,\CZ\}$ is $\#\mathrm P$-hard.

\subsection{Structural consequences of the dichotomy}
\label{sec:consequences}

The example $\{H,T,\CNOT\}$ shows that two gate types need not suffice
to witness hardness. We now show that three always suffice.

\begin{corollary}[Three gates suffice to witness hardness]
\label{cor:three-gate-witness}
If $\G$ fails all three conditions of Theorem~\ref{thm:intro-dichotomy},
then some sublibrary $\G_0\subseteq\G$ with $|\G_0|\le 3$ also fails
all three, so $\Amp(\G_0)$ is $\#\mathrm P$-hard.
\end{corollary}
If a single gate in $\G$ is both nonmonomial and entangling, the bound
drops to two. That gate rules out \textup{(i)} and \textup{(ii)} and,
if it admits a diagonal Clifford frame, fixes the phase up to a fourth
root of unity (Appendix~\ref{app:frame-tests}).

A single gate $U$ already gives $\#\mathrm P$-hardness if two of its
columns have different multisets of squared entry magnitudes. Indeed,
every column of a monomial gate has one nonzero entry of modulus one.
For a one-qubit unitary the two columns have squared magnitudes $p$ and
$1-p$ in opposite orders. Tensor products and wire permutations preserve
equality of the multisets, so all columns of a product-preserving gate
also have the same multiset. Finally, the columns of a Clifford gate
are Pauli translates of one another, and diagonal conjugation preserves
entry magnitudes.

In particular, a gate is hard on its own if it maps one basis input to a
basis state up to phase and another to a superposition of two or more
basis states (Corollary~\ref{cor:deterministic-column}).
For a particle-number-conserving gate, the vacuum supplies the
deterministic column, so it suffices to find one branching column.
With a fixed number $w$ of particles, direct simulation takes place in
a sector of dimension $\binom nw$. For a nonmonomial gate of this kind,
however, padding with idle wires gives hardness at every fixed rational
density $0<\rho<1$, even for the input $1^{\rho n}0^{(1-\rho)n}$ and a
specified output of the same weight, whenever $\rho n\in\Z$
(Corollary~\ref{app:fixed-filling}).

Table~\ref{tab:gate-regimes} illustrates both the column test and the
common-frame condition. For the exchange pulse $U_\theta=e^{i\theta\SWAP}$
and the superconducting $\operatorname{fSim}(\theta,\varphi)$ gate~\cite{Foxen20},
the $\ket{01}$ column branches exactly when $\sin\theta\cos\theta\ne0$.
For the two-ion M{\o}lmer--S{\o}rensen (MS) pulse~\cite{Debnath16}, we
write $M_\phi(\theta)=U_{\mathrm{MS}}(\theta,\phi)$ in the notation
of~\cite[Eq.~(1)]{Yale25}, so that
\[
 M_\phi(\theta)=\exp\!\left(-\frac{i\theta}{2}\sigma_\phi\otimes\sigma_\phi\right),
 \qquad \sigma_\phi=\cos\phi\,X+\sin\phi\,Y.
\]

\begin{table}[htb]
\centering
\small
\caption{Exact computational-basis amplitudes for fixed gate libraries
with unrestricted placements. Evaluation is in $\FP$ under the listed
conditions and is $\#\mathrm P$-hard otherwise.}
\label{tab:gate-regimes}
\begin{tabular*}{\linewidth}{@{\extracolsep{\fill}}lll@{}}
\toprule
Fixed gate library & Tractable parameters & Reason for tractability\\
\midrule
$\{U_\theta\}$ or $\{\operatorname{fSim}(\theta,\varphi)\}$ &
  $\theta\in(\pi/2)\Z$, any $\varphi$ & Monomial\\
$\{M_\phi(\theta)\}$ &
  $\theta\in(\pi/2)\Z$, any $\phi$ & Clifford in frame $D(e^{i\phi})$\\
$\{M_{\phi_1}(\pi/2),M_{\phi_2}(\pi/2)\}$ &
  $\phi_1-\phi_2\in(\pi/2)\Z$ & Common diagonal Clifford frame\\
\bottomrule
\end{tabular*}
\end{table}

For algebraic pulse entries, the single-axis MS classification also
follows from Shepherd's analysis of $X$-programs on graphs at angle
$-\theta/2$~\cite[Section~2.3 and Theorem~1]{She10}. For example, the
placements of $M_0(\pi/4)=e^{-i\pi X\otimes X/8}$ generate a finite
abelian group at each width, yet exact amplitude evaluation is
$\#\mathrm P$-hard. Conjugating by $H$ on every wire gives the diagonal
pulse $e^{-i\pi Z\otimes Z/8}$, whose computational-basis amplitudes
are computable in polynomial time.

The last two rows of Table~\ref{tab:gate-regimes} illustrate the
requirement of a common frame. Each $M_\phi(\pi/2)$ is nonmonomial and
entangling and admits the frame $D(e^{i\phi})$. By the frame
uniqueness used in Corollary~\ref{cor:three-gate-witness}, two such
pulses share a frame precisely when their axis angles differ by a multiple
of $\pi/2$. For a library containing $\CNOT$, any common frame $D(z)$
must satisfy $z^4=1$ and is therefore itself Clifford. Such a library
has polynomial-time exact amplitudes if all its gates are monomial or
all are Clifford in the standard basis, and is $\#\mathrm P$-hard
otherwise (Corollary~\ref{cor:cnot}).

\paragraph*{Exceptional phases in rational families.}
Suppose the entries of a finite library are rational functions of one
shared phase $z$, with algebraic coefficients, and the gates are unitary
at every $|z|=1$ away from their poles. If the library lies outside
the three tractable classes at even one such phase, exact amplitudes
are $\#\mathrm P$-hard at all but finitely many phases in the domain.
The exceptional phases are algebraic and can be computed from the
rational functions (Theorem~\ref{param:finite}). Here the phase is
fixed throughout each circuit. In the full space of libraries of fixed
gate arities, the three tractable classes have Haar measure zero as soon
as one gate has arity at least two (Theorem~\ref{param:geometry}).

\paragraph*{Encoded universality with a fixed exchange pulse.}
Every nonmonomial exchange pulse $U_\theta$ at a polynomial-time
computable angle also suffices for encoded universal quantum computation.
Burkat, Strelchuk, and Studzi\'nski proved this for $U_{\pi/4}$ and asked
which other fixed angles suffice, including
$\pi/6$~\cite[Theorem~5 and Section~VIII]{BSS26}.
Our proof adapts the step of our amplitude reduction that contracts an
entangled pair with its adjoint. Applying it to singlets on overlapping
pairs of wires, we simulate $U_{\pi/4}$ with $U_\theta$ pulses at
inverse-polynomial error and polynomial cost
(Theorem~\ref{thm:exchange-compiler}).

\subsection{Techniques}
\label{sec:proof-overview}

A circuit amplitude can be written as a sum over internal wire bits, with each
summand a product of gate matrix entries. This leads to Boolean Holant theory,
which grew out of Valiant's holographic algorithms and asks which fixed sets
of local tensors allow the corresponding sums to be evaluated efficiently on
arbitrary graphs~\cite{Val08,CLX11}. A succession of dichotomies classified
Boolean Holant problems under different assumptions on symmetry, available
unary tensors, and real or complex
weights~\cite{CLX09,CLX11star,CGW16,Bac17,LW18,CFS20,SC20,Bac21,Bac25,MWXZ25}.
A full dichotomy is now available for arbitrary finite sets of Boolean tensors
with algebraic complex entries~\cite[Theorem~1.1]{LMW26}.

Connections with quantum computation have linked tractable Holant tensors to
Clifford gates and brought entanglement techniques into dichotomy proofs.
In turn, Holant constructions have revealed new properties of multipartite
entangled states~\cite{Bac17,CGW18,CFS20,Bac21,Bac25}. These connections
motivate us to investigate how the full classification extends to quantum
circuit amplitudes. Examining the gadget constructions and changes of basis
developed across this program reveals two fundamental obstacles. First, Holant
contractions need not respect the input-output structure or acyclicity of a
quantum circuit. Second, a common change of basis that simplifies the gate
tensors also transforms the boundary states and need not be unitary. The
resulting tensor classification must therefore be interpreted under the joint
constraints of unitary gates and fixed computational-basis inputs and outputs.

\paragraph{Recovering arbitrary tensor contractions.}
The freedom to connect tensors on arbitrary graphs can change their
computational power. Toffoli circuits have deterministic basis
amplitudes, yet closing the data wires of a reversible Boolean
computation sums over assignments and can encode $\#\mathrm{SAT}$.
To use Holant hardness for circuits, we must recover these more general
contractions through queries over the original gate library.

Entangled pairs suggest how to do this. Applying a gate to one half of
each of several Bell pairs represents the gate as a state, via the
operator-state correspondence~\cite{Jam72,Choi75}. Contracting such
states with the corresponding Bell bras joins their tensor indices.
All vertex states can be prepared before any edges are contracted, so
the resulting computation is acyclic even when the original network
has cycles. This construction requires both the entangled pairs and
their adjoint effects, which the gate library may not supply.

For algebraic gate entries, Theorem~\ref{thm:resources} applies under
two assumptions: some gate branches on a computational-basis input,
and some gate entangles a product state. That product input need not
itself be preparable from the library. From these capabilities, we
obtain a symmetric maximally entangled pair and its adjoint whose joint
use can be simulated through exact amplitude queries over the original
library. Here joint use means evaluating an entire acyclic network
containing any number of copies of these resources, at cost polynomial
in the network size. Exact interpolation recovers these network values
from circuit amplitudes. Symmetry lets us express the pair
as a Bell pair in the same unitary basis on both qubits, so its adjoint
contracts every edge in one common coordinate system. For the tensors
of the gates and their adjoints, together with the computational-basis
boundary tensors in these coordinates, we obtain
\[
 \Amp(\G)\quad\longleftrightarrow\quad
 \text{contracting these tensors on arbitrary graphs}.
\]
Representing a circuit as a tensor network gives the forward direction.
For the reverse direction, the entangled pair and its adjoint let us
evaluate arbitrary contractions using only
amplitude queries over $\G$. Section~\ref{sec:architecture} proves this
equivalence.

\paragraph{A common diagonal Clifford frame.}
We next ask which Holant tractable structures can occur for these gate
and boundary tensors. For a library containing both a branching gate
and an entangling gate, the only surviving structure gives Clifford
gates in a common invertible basis, by the affine--Clifford
correspondence~\cite{CGW18}. This basis need not be unitary, and it also
changes the input and output states. To obtain a stabilizer simulation,
we must relate this common basis to the physical inner product and the
computational-basis boundaries.

Our rigidity argument uses unitarity and the boundary states to show
that the common basis can be replaced by a diagonal unitary
(Theorem~\ref{thm:normal}). A diagonal unitary only multiplies each
computational-basis state by a phase, so the original amplitude differs
from a Clifford-circuit amplitude by a known phase factor. This explains
the common diagonal frame in the third tractability condition.
Monomial and product-preserving libraries already have direct simulators. For every library outside all three classes,
the structural result excludes Holant tractability, and the reduction
above transfers hardness to its circuit amplitudes.

\paragraph{Arbitrary complex gate entries.}
Exact evaluation requires preserving algebraic relations among the gate
entries. For a fixed complex library outside the three tractable
classes, we find an algebraic unitary library that remains outside them
and satisfies every rational polynomial relation among the original
gate coordinates. These preserved relations let us evaluate the
original library's symbolic answers at the replacement coordinates,
recovering amplitudes of the same circuits over the algebraic library.
This transfers hardness to the original library. In the three tractable
classes, the simulators return the required symbolic answers in the
original gate coordinates. Section~\ref{sec:coefficient-extension}
gives the specialization argument and the algorithms for exact symbolic
evaluation.

\Needspace{8\baselineskip}
\subsection{Organization of the paper}
Appendix~\ref{app:prelim} sets the conventions.
Section~\ref{sec:architecture} proves the circuit reduction
(Appendices~\ref{app:calculus}--\ref{app:resources}), and
Section~\ref{sec:normal} classifies the gate libraries, including
arbitrary complex entries (Appendices~\ref{app:holant}--\ref{app:coefficients}).
Appendices~\ref{sec:applications}--\ref{app:applications} develop the circuit
applications of Section~\ref{sec:consequences}. Appendices~\ref{app:parameters}--\ref{app:closure}
treat parameter families and fixed-width gate closures.

\section{From circuit amplitudes to tensor contractions}
\label{sec:architecture}
\label{sec:resources}

Throughout this section the gate entries are algebraic. The library
is fixed, and we use exact coordinates in a fixed number field as in
Appendix~\ref{app:prelim}. We first construct a symmetric maximally
entangled state and its adjoint at polynomial oracle cost. We then
use the state to prepare vertex tensors and the adjoint to contract
their edges, thereby evaluating an arbitrary network by queries over
the original gate library.

\subsection{The tensor set and the resource theorem}

For a $k$-qubit gate $U$, the matrix elements $f_U(y,x)=\bra yU\ket x$
define a Boolean \emph{signature} of arity $2k$, with $k$ output
indices and $k$ input indices. Contracting an edge identifies its two
incident indices and sums over their common bit. Fixing an index to a
basis value is called \emph{pinning}.

To describe the contraction problem in a common one-qubit basis, we
record the transformed gates together with both computational-basis
boundaries. For an invertible $2\times2$ matrix $S$, define
\begin{equation}
\begin{split}
 \mathcal H(\G,S)={}&
 \{f_{(S^{-1})^{\otimes k}US^{\otimes k}}:U\in\G\text{ acts on }k\text{ qubits}\}\\
 &\cup\{S^{-1}e_0,S^{-1}e_1,S^Te_0,S^Te_1\},
 \qquad e_b=\ket b.
\end{split}
\label{eq:holant-language}
\end{equation}
The vectors $S^{-1}e_b$ and $S^T e_b$ encode the input and output bit
$b$, respectively.

\begin{theorem}[Directed entangled resources]
\label{thm:resources}
Suppose an algebraic library $\G$ contains a nonmonomial gate and a
gate that entangles some product input, whether or not $\G$ can prepare
that input. One can effectively find a symmetric unitary algebraic
matrix $A$ such that $\ket A=\sum_{i,j}A_{ij}\ket{ij}$ and its adjoint
effect can be used jointly at polynomial exact Turing cost. There is an
effectively computable unitary algebraic $S$ with $A=SS^T$ such that,
for $\Gamma=\G\cup\G^\dagger$,
\[
 \Amp(\G)\equiv_T^{\poly}\Holant(\mathcal H(\Gamma,S)).
\]
After adjoining all port permutations, the tensor set on the right is
closed under complex conjugation.
\end{theorem}

Here $\Holant(\mathcal F)$ asks for the value of a graph whose vertices
carry tensors from $\mathcal F$, obtained by summing the product of
their entries over all edge labellings by bits. Joint use means that
any acyclic network over $\G$ containing any number of copies of
$\ket A$ and $\bra A$, with all boundary bits fixed, can be evaluated
in polynomial time using amplitude queries over $\G$
(Definition~\ref{calc:definition}).

\subsection{Constructing the entangled state and its adjoint}
\label{sec:construct-pair}

We first seek an entangled two-qubit state with a symmetric invertible
coefficient matrix, then normalize it to obtain $\ket A$. Symmetry
expresses the normalized pair as a Bell pair under the same unitary
change of basis on both qubits, as required for contracting edges.
Throughout, auxiliary tensors are used
in the evaluation sense specified after Theorem~\ref{thm:resources},
with a reduction for their joint use.

Basis projections fit this evaluation model by fixing boundary bits.
To project a wire onto $b$, retire it and include $b$ in the final
output string. If a later operation needs a new input at that logical
position, use a fresh wire initialized to $b$ at the start of the circuit
(Appendix~\ref{calc:admissibility}).

The construction also uses adjoint gates, so we first show how to
simulate them using only gates in $\G$. Fix a gate $U$. Let $F(t)$ be the
amplitude obtained by replacing all $r$ occurrences of $U^\dagger=U^{-1}$
with $U^t$. The desired amplitude is $F(-1)$, while nonnegative
integer powers use only $U$. The spectral decomposition of $U$ gives
\begin{equation}
 F(t)=\sum_{\substack{n_1+\cdots+n_d=r\\n_j\ge0}}
       c_{\boldsymbol n}
       (\lambda_1^{n_1}\cdots\lambda_d^{n_d})^t,
 \label{eq:interpolation}
\end{equation}
where $\lambda_1,\ldots,\lambda_d$ are the distinct eigenvalues of $U$
and $n_j$ counts how many of the $r$ occurrences contribute $\lambda_j$.
Since $d$ is fixed and each $n_j$ is at most $r$, the number of distinct
products $\lambda_1^{n_1}\cdots\lambda_d^{n_d}$ is polynomial in $r$.
After grouping equal products, a Vandermonde system recovers their
coefficients from polynomially many values $F(0),F(1),\ldots$ and hence
determines $F(-1)$. A separate parameter for each of the fixed number
of gate types lets us eliminate all adjoint gates at polynomial cost
by simultaneous interpolation~\cite{CLX11,Xia26}. Appendix~\ref{app:calculus}
gives the bit bounds.

\paragraph{Extracting a one- or two-qubit state.}
A nonmonomial gate $U$ has a column with at least two nonzero entries.
Starting from such a column,
Theorem~\ref{seed:alternative} constructs either a one-qubit
superposition with both coefficients nonzero or an entangled two-qubit
state, using at most $24$ occurrences of $U$ or $U^\dagger$.
The construction uses overlapping placements of circuit fragments over
$\{U,U^\dagger\}$ and computational-basis projections to isolate the
one- or two-qubit output.
In the two-qubit case the coefficient matrix is already symmetric or
the Schmidt coefficients are unequal. In the latter case, contracting
one leg with the corresponding leg of the adjoint gives a one-qubit
map with distinct positive eigenvalues. The same simultaneous-interpolation
principle applies to powers of this map: since its eigenvalue ratio has
infinite order, we can prescribe fixed algebraic weights on its two
eigenspaces in the evaluation reduction (Lemma~\ref{calc:filter}).
This spectral filtering symmetrizes the pair
(Lemma~\ref{calc:symmetrize}) and will also supply the final
normalization. In either two-qubit case we can proceed to that
normalization at the end of this section.

\paragraph{From a one-qubit superposition to an entangled pair.}
If the construction instead yields a one-qubit superposition $s$,
then $E=\{\ket0,\ket1,s\}$ consists of three distinct rays.
An entangling $k$-qubit gate must entangle some product input from
$E^k$ (Lemma~\ref{res:product-test}). To see why, write the output amplitudes as a matrix with one
qubit indexing the rows and the remaining qubits indexing the columns.
The output is a product state exactly when all $2\times2$ minors vanish
for every choice of the row qubit. Each minor is a homogeneous quadratic
in each local input vector, and a nonzero binary quadratic vanishes
on at most two rays. Thus a minor that vanishes on $E^k$ vanishes on
every product input, by applying this observation one input at a time.

The resulting entanglement may involve more than two outputs. To retain
an entangled pair, we must project the other outputs so that the
remaining two-qubit coefficient matrix has nonzero determinant.
Lemma~\ref{res:localization} gives two outputs for which this
determinant, as a polynomial in the effects on the other outputs, is
not identically zero. It is quadratic in each effect, so the same
three-ray argument gives a nonzero value using only adjoints of states
in $E$. If the resulting matrix is symmetric, we can normalize it
directly. If the Schmidt coefficients of the pair are unequal, we can first
symmetrize it by Lemma~\ref{calc:symmetrize}.

\paragraph{When the localized pair is maximally entangled and nonsymmetric.}
Scale its coefficient matrix $B$ so that $BB^\dagger=I$.
The map $BB^\dagger$ is now scalar, so the preceding spectral filtering
cannot distinguish the two Schmidt components. We therefore seek a local
map that separates their weights or makes the pair symmetric directly.
If an available full-rank one-qubit map $T$ is not proportional to a
unitary, then $(TB)(TB)^\dagger=TT^\dagger$ has distinct eigenvalues.
Applying $T$ thus gives unequal Schmidt coefficients, as required for
symmetrization.

It remains to handle the case in which every available full-rank
local map is proportional to a unitary. If two available one-qubit unitaries do not commute,
Proposition~\ref{res:noncommuting} gives either a filter that separates
the Schmidt coefficients or a unitary that makes the pair symmetric
directly. To see how these alternatives arise, contract the second leg
of the pair with the adjoint of a one-qubit state $v$. This leaves
$J_B(v)=B\overline v$ on the first leg. Conjugation by this antiunitary
map takes available unitaries to available unitaries
(Lemma~\ref{res:antiunitary}). Close the group generated by the two
unitaries under this conjugation and include $J_B^2$. Its action on
the Bloch sphere gives a noncyclic rotation group. If the group is
infinite, it contains a rotation of infinite order
(Lemma~\ref{res:infinite-group}). The corresponding unitary has
eigenvalue ratio of infinite order, so interpolation supplies the filter.
If the group is finite, its extension by $J_B$ contains a plane
reflection $QJ_B$ for an available unitary $Q$, and $QB$ is symmetric.

To rule out the commuting case, first obtain four distinct state rays
from the available resources. If $B$ is nonmonomial, pinning one
qubit yields its two orthogonal columns, both distinct from the computational
rays. If $B$ is monomial, $J_Bs$ has two nonzero coordinates and
its ray differs from that of $s$, since a fixed ray of $J_B$ would
force $B$ to be symmetric (Lemma~\ref{res:antiunitary}). Thus
$\ket0,\ket1,s,J_Bs$ give four distinct rays in this case too.
If all full-rank local maps commute, they are diagonal in a common
orthonormal basis. Contract all but one input and one output of a gate with these states
and their adjoints. In the common basis the remaining matrix $T$ is
singular or diagonal, so $(\det T)T_{01}=(\det T)T_{10}=0$.
Each entry of $T$ is linear in each contracted state or effect, making
these expressions homogeneous cubics in each such variable. A nonzero
binary cubic vanishes on at most three distinct rays. Vanishing on
all choices from the four rays therefore makes both expressions vanish
identically. Proposition~\ref{res:four-ray} shows that these identities
force every gate to preserve product states, contradicting the
entangling hypothesis.

Every outcome now gives a symmetric invertible coefficient matrix $M$.
Applying the spectral filter $(MM^\dagger)^{-1/2}$ gives a symmetric
unitary matrix $A$, and a symmetric unitary square root gives $A=SS^T$
(Lemma~\ref{calc:unitary-pair}). To find a construction for the fixed
library, enumerate finite circuit fragments over $\G\cup\G^\dagger$
in order of description length and test their tensors by exact
algebraic arithmetic. The cases above guarantee a finite witness for
one of the constructions, so the search terminates
(Theorem~\ref{res:complete}). Taking the adjoint construction supplies
$\bra A$. With the construction fixed, simultaneous interpolation
eliminates its auxiliary operations in each acyclic network at
polynomial cost, giving joint access to the state and effect over $\G$.

\subsection{Evaluating an arbitrary network with the pair}
\label{sec:network-with-pair}

\begin{figure}[!htbp]
\centering
\begin{tikzpicture}[x=.9cm,y=1cm,font=\small,
 vertex/.style={circle,draw=black,fill=white,minimum size=8mm,inner sep=1pt},
 resource/.style={rectangle,draw=black,fill=white,minimum width=12mm,minimum height=6mm},
 gate/.style={rectangle,draw=black,fill=white,minimum width=7mm,minimum height=6mm},
 wire/.style={draw=black,line width=.65pt,-{Latex[length=1.8mm]}},
 crossing/.style={wire,preaction={draw=white,line width=3pt}},
 every node/.style={text=black}]
\node at (2.3,.75) {(a) Undirected instance};
\node at (11,.75) {(b) Directed auxiliary network};
\node[vertex] (fu) at (3.4,-.6) {$\widetilde f_U$};
\node[vertex] (fv) at (2.1,-3.3) {$\widetilde f_V$};
\node[vertex] (fw) at (4.7,-3.3) {$\widetilde f_W$};
\draw[line width=.65pt] (fu.south west) -- (fv.north east)
 node[pos=.14,left=1.5pt,inner sep=1pt] {$i_U$}
 node[pos=.48,left=1.5pt,inner sep=1pt] {$e_1$}
 node[pos=.86,left=1.5pt,inner sep=1pt] {$i_V$};
\draw[line width=.65pt] (fv.east) -- (fw.west)
 node[pos=.18,above=1.5pt,inner sep=1pt] {$o_V$}
 node[midway,below=1.5pt,inner sep=1pt] {$e_2$}
 node[pos=.82,above=1.5pt,inner sep=1pt] {$o_W$};
\draw[line width=.65pt] (fw.north west) -- (fu.south east)
 node[pos=.14,right=1.5pt,inner sep=1pt] {$i_W$}
 node[pos=.52,right=1.5pt,inner sep=1pt] {$e_3$}
 node[pos=.86,right=1.5pt,inner sep=1pt] {$o_U$};
\node at (5.65,-2.1) {$\Longrightarrow$};
\node[resource] (bin) at (0,-.6) {$S^{-1}e_b$};
\node[resource] (bout) at (0,-3.3) {$S^Te_a$};
\draw[line width=.65pt] (bin.south) -- (bout.north)
 node[midway,left=2pt] {$e_0$};
\node[resource] (bstate) at (6.6,0) {$\ket b$};
\node[resource] (apair) at (8.4,0) {$\ket A$};
\node[gate] (apin) at (8.7,-1.6) {$\bra a$};
\node[resource] (c0) at (7.1,-4.3) {$\bra A_{e_0}$};
\draw[wire] (bstate.south) -- ([xshift=-2mm]c0.north);
\draw[wire] ([xshift=2mm]apair.south) -- (apin.north);
\draw[wire] ([xshift=-2mm]apair.south) .. controls (8,-1.4) and (7.4,-2.8) .. ([xshift=2mm]c0.north);
\node[resource] (au) at (10.1,0) {$\ket A$};
\node[resource] (av) at (12.396,0) {$\ket A$};
\node[resource] (aw) at (14.692,0) {$\ket A$};
\node[gate] (u) at (9.936,-1.6) {$U$};
\node[gate] (v) at (12.232,-1.6) {$V$};
\node[gate] (w) at (14.528,-1.6) {$W$};
\node[resource] (c1) at (10.428,-4.3) {$\bra A_{e_1}$};
\node[resource] (c3) at (12.642,-4.3) {$\bra A_{e_3}$};
\node[resource] (c2) at (14.856,-4.3) {$\bra A_{e_2}$};
\draw[wire] ([xshift=-2mm]au.south) -- (u.north);
\draw[wire] ([xshift=-2mm]av.south) -- (v.north);
\draw[wire] ([xshift=-2mm]aw.south) -- (w.north);
\draw[wire] ([xshift=2mm]au.south) .. controls (10.51,-1.5) and (10.51,-3.1) .. ([xshift=-2mm]c1.north);
\draw[crossing] ([xshift=2mm]av.south) .. controls (12.97,-1.6) and (12.97,-2.7) .. ([xshift=2mm]c1.north);
\draw[crossing] ([xshift=2mm]aw.south) .. controls (15.266,-1.6) and (15.266,-2.7) .. ([xshift=2mm]c3.north);
\draw[crossing] (u.south) .. controls (9.936,-2.6) and (11.822,-3.3) .. ([xshift=-2mm]c3.north);
\draw[crossing] (v.south) .. controls (12.232,-2.6) and (14.036,-3.3) .. ([xshift=-2mm]c2.north);
\draw[crossing] (w.south) .. controls (14.528,-2.5) and (15.184,-3.2) .. ([xshift=2mm]c2.north);
\node at (10.84,-.85) {$i_U$};
\node at (13.10,-.85) {$i_V$};
\node at (15.40,-.85) {$i_W$};
\node[anchor=east,inner sep=0pt] at (9.85,-2.25) {$o_U$};
\node[anchor=east,inner sep=0pt] at (12.15,-2.25) {$o_V$};
\node[anchor=east,inner sep=0pt] at (14.44,-2.25) {$o_W$};
\draw[wire] (15.922,.15) -- (15.922,-4.5) node[midway,right] {order};
\end{tikzpicture}
\caption{Converting gate and boundary tensors into a directed acyclic network. Here $\widetilde f_U=f_{S^{-1}US}$, and similarly
for $V,W$. The vertical edge $e_0$ joins the unary signatures
$S^{-1}e_b$ and $S^Te_a$. They are represented on the right by $\ket b$
and by pinning one output of $\ket A$ with $\bra a$, respectively.
Their remaining outputs are contracted by $\bra A_{e_0}$. The triangle has edges
$e_1=\{i_U,i_V\}$, $e_2=\{o_V,o_W\}$, and $e_3=\{o_U,i_W\}$.
Each input index $i$ is carried by the untouched half of a pair, and each
output index $o$ by a gate output. One effect consumes the two wires of
each original edge. Crossings are not junctions. The auxiliary states
and effects are eliminated jointly by the oracle reduction.}
\label{fig:holant-to-dag}
\end{figure}
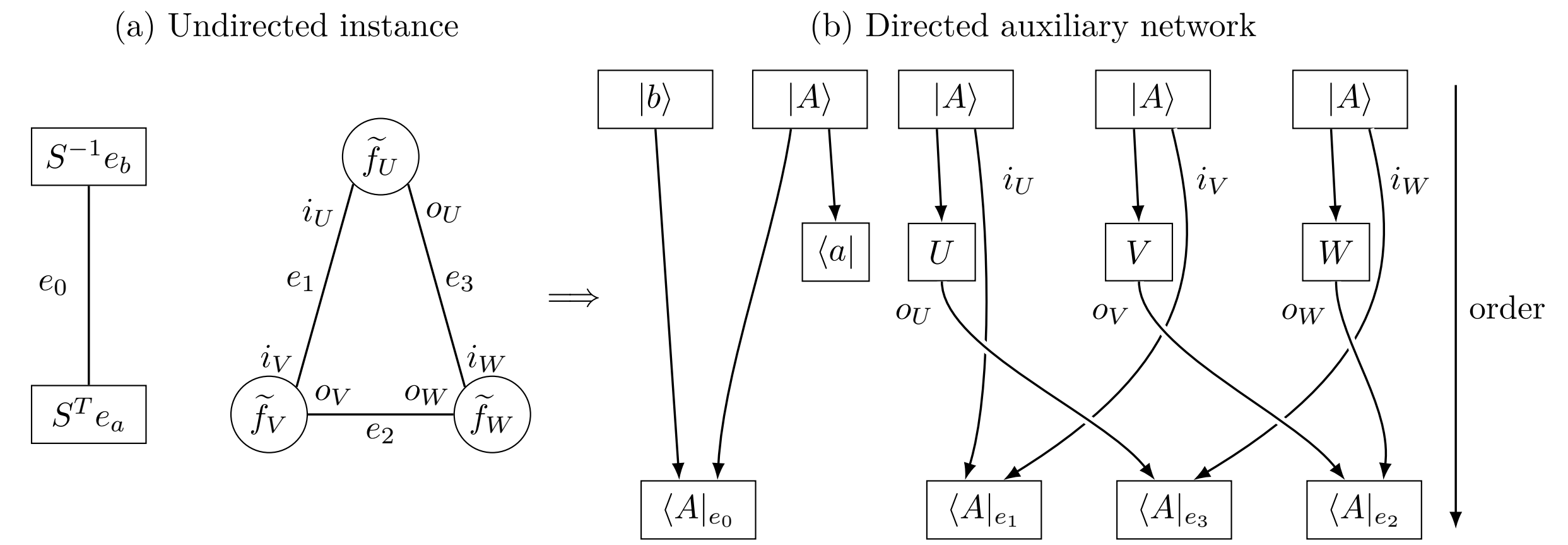

Section~\ref{sec:construct-pair} supplies joint access to $\ket A$ and
$\bra A$ through amplitude queries. We now use them to turn an arbitrary
tensor network into an acyclic evaluation, implementing the construction
outlined in Section~\ref{sec:proof-overview} and illustrated in
Figure~\ref{fig:holant-to-dag}.

First prepare a state for
each vertex of the network. In the unitary
basis $S$ with $A=SS^T$, each $\ket A$ has coordinates
$\ket{00}+\ket{11}$. Thus $k$ copies have coordinates
$\sum_x\ket x\ket x$, with the first qubit of each pair in the first
register. For a vertex labelled by a $k$-qubit gate $U\in\Gamma$,
apply $U$ to that register. If $U_S=(S^{-1})^{\otimes k}US^{\otimes k}$
is its matrix in this basis, the resulting state has coordinates
\[
  \sum_{x,y\in\{0,1\}^k}(U_S)_{y,x}\ket y\ket x .
\]
Its coefficients are exactly the gate tensor in~\eqref{eq:holant-language},
with both $y$ and $x$ now on output wires. For a boundary bit $b$,
preparing $e_b$ gives coordinates $S^{-1}e_b$, while pinning one qubit
of $\ket A$ to $b$ leaves $Ae_b$, with coordinates $S^{-1}Ae_b=S^Te_b$.

Once all vertices are prepared, apply $\bra A$ to the two outputs at
each edge. In the same basis this effect is $\bra{00}+\bra{11}$,
so it sums over the two assignments in which the endpoint bits agree.
Even a self-loop joins two outputs of an already prepared state, so
the computation is acyclic for every graph. By Section~\ref{sec:construct-pair}, we can therefore simulate
all copies of the state and effect together with any adjoint gates
in polynomial time using amplitude queries over $\G$. Conversely, in
a circuit over $\G$ the changes of basis on each internal wire cancel,
so a single contraction over $\mathcal H(\Gamma,S)$ gives its amplitude
(Proposition~\ref{pre:forward}).

It remains to check the conjugate-closure assertion of
Theorem~\ref{thm:resources}. Since $S$ is unitary, $\overline{U_S}$
is the transpose of the matrix for $U^\dagger\in\Gamma$ in the same
basis. Transposing just exchanges the two groups of gate indices.
Conjugation also interchanges the boundary tensors $S^{-1}e_b$ and
$S^Te_b$. Allowing all permutations of indices thus gives the
conjugate-closed set used in Section~\ref{sec:normal}, while relabelling
the incident edges preserves every contraction (Lemma~\ref{hol:port-permutations}).
This completes the proof of Theorem~\ref{thm:resources}
(Proposition~\ref{calc:conjugate-transfer}).

\section{Classification of the gate libraries}
\label{sec:normal}

Libraries consisting entirely of monomial gates or entirely of
product-preserving gates have the direct simulators of
Section~\ref{sec:classification}. Every remaining library contains a
nonmonomial gate and an entangling gate. For algebraic entries,
Theorem~\ref{thm:resources} connects its amplitudes to a Holant problem.
We must determine when the resulting gate and boundary tensors satisfy
the Holant tractability criterion, and show that this forces a common
diagonal Clifford frame. We then extend the dichotomy to arbitrary
fixed complex entries.

\subsection{The Holant criterion for circuit tensors}

After allowing port permutations, the tensor set from
Theorem~\ref{thm:resources} is closed under conjugation and contains a
nonzero one-index boundary tensor. Restricting the full Boolean Holant
dichotomy of Liu, Meng, and Wang~\cite[Theorem~1.1]{LMW26} to such
sets leaves four of its seven tractable cases (Appendix~\ref{hol:four}).
We write $\ReTract(\mathcal F)$ when $\mathcal F$ satisfies at least
one of these four cases. In the first case, every tensor in
$\mathcal F$ factors into
pieces with at most two indices. In each of the other three cases,
one invertible matrix $B$ must work for the whole set. Applying $B^{-1}$
on every index puts each vertex tensor in the chosen family, and the
equality tensor on an edge, now $B^TB$, must belong to that same family
(see~\eqref{hol:transformability}). The choices
are the product-type family (one-index
weights with equality or disequality constraints), the affine family
(stabilizer tensors up to a scalar), and the local-affine family (which
allows local phase corrections to affine tensors).
For the tensor set supplied by Theorem~\ref{thm:resources},
Corollary~\ref{hol:dichotomy} gives $\#\mathrm P$-hardness whenever
$\ReTract$ fails. Recently, Guan, Shao, and Tang~\cite{GST26}
independently classified Boolean Holant problems for the broader class
of all conjugate-closed sets of algebraic signatures.

We now ask when the gate and boundary tensors of a unitary library
can satisfy $\ReTract$ in some basis $S$.

\begin{theorem}[Unitary normal form]
\label{thm:normal}
Suppose an algebraic library $\G$ contains a nonmonomial gate and an
entangling gate. Then
\[
 \exists S\in\mathrm{GL}_2(\Qbar)\;
       \ReTract(\mathcal H(\G,S))
 \quad\Longleftrightarrow\quad
 \G\text{ has a common diagonal Clifford frame}.
\]
\end{theorem}

We first prove the forward implication by excluding three of the four
Holant tractable cases. The next subsection treats the remaining affine
case and proves the converse.
Suppose $\ReTract(\mathcal H(\G,S))$ holds for some invertible algebraic
$S$. A $k$-qubit gate has rank $2^k$ across
its input and output indices, even after invertible changes on the
individual indices. In the tensor-factor case, each factor has
rank at most two. The same bound holds for each connected component of
the product-type constraints, since fixing one bit determines all the
others in that component. Full rank then forces each nonconstant factor
to join exactly one input to one output. Undoing the basis changes
leaves a product of one-qubit gates up to a wire permutation
(Lemma~\ref{norm:tensor-cases}), contrary to the entangling hypothesis.

In the local-affine case, each transformed boundary vector has
only one nonzero entry. The combined transformations on the input
and output indices therefore only rescale and permute basis vectors. A local-affine tensor on $2k$
indices has at most $2^k$ nonzero entries, while an invertible gate
needs one in every row and column. Consequently each row and column
has exactly one, even after undoing the basis changes
(Lemma~\ref{norm:local-affine}). This contradicts the presence of a
nonmonomial gate. It remains to determine which basis the affine
case permits.

\subsection{A common stabilizer basis}

To obtain a common diagonal Clifford frame from the affine case,
we must control the additional basis change $B$ it permits after $S$.
Even when $S$ is unitary, $B$ can be a nonunitary local filter~\cite{DVC00}. Contracting the transformed equality tensor on
the input indices gives a common matrix $C=SB$ (Lemma~\ref{norm:absorb})
such that, for each $k$-qubit gate $U$ and each boundary bit $b$,
\begin{equation}
 Q_U=(C^{-1})^{\otimes k}UC^{\otimes k}\text{ is affine},
 \qquad C^{-1}e_b,\ C^Te_b\text{ are affine}.
 \label{eq:affine-witness}
\end{equation}
By the affine--Clifford correspondence~\cite{CGW18}, the invertible
matrix $Q_U$ is a scalar multiple of a Clifford unitary. Similarity to
$U$ gives $|\det Q_U|=1$, so the scalar has modulus one and $Q_U$
is itself Clifford.

To restrict $C$, consider the output boundaries $C^Te_0,C^Te_1$.
These are the rows $r_0,r_1$ of $C$ written as columns. They are
affine by~\eqref{eq:affine-witness}, and the nonzero one-qubit affine
tensors lie on the six Pauli eigenstate rays. If the rows were
nonorthogonal, their independence would put them on
different Pauli axes. Then $G=C^\dagger C=r_0^\dagger r_0+r_1^\dagger r_1$
would have nonzero Pauli components on both axes.

This same $G$ gives the physical inner product in the $C$ coordinates.
Unitarity of $U$ therefore gives
\begin{equation}
 Q_U^\dagger G^{\otimes k} Q_U=G^{\otimes k}.
 \label{eq:metric}
\end{equation}
Thus $Q_U$, itself unitary, commutes with $G^{\otimes k}$. The next
lemma explains why two nonzero Pauli components of $G$ would prevent
any gate from entangling a product state.

\begin{lemma}[Clifford rigidity for a product metric]
\label{lem:metric}
Let $G$ be a positive definite $2\times2$ matrix whose Bloch vector has
at least two nonzero Pauli components. Every Clifford unitary commuting
with $G^{\otimes k}$ is a product of one-qubit Cliffords and a wire
permutation.
\end{lemma}
\begin{proof}
Write $G=aI+\sum_r g_r\sigma_r$ for $\sigma_r\in\{X,Y,Z\}$.
Taking the logarithm preserves the eigenvectors of $G$, so
$\log G=\alpha I+\beta\sum_r g_r\sigma_r$ with $\beta\ne0$. Hence
\[
 \log(G^{\otimes k})
   =k\alpha I+\beta\sum_{j=1}^k\sum_{r\in\{x,y,z\}}
                         g_r\sigma_r^{(j)},
\]
where $\sigma_r^{(j)}$ acts only on qubit $j$. A Clifford commuting
with $G^{\otimes k}$ preserves this sum. It permutes Pauli operators
up to sign, so distinct terms cannot cancel, and each displayed Pauli
with nonzero coefficient must still act on just one qubit after
conjugation. Two such Paulis on any input qubit anticommute, so their
images must act on the same output qubit. They generate its full Pauli
algebra. Operators on distinct input qubits commute, so their algebras
must go to distinct output qubits. Undoing this wire permutation
leaves a one-qubit Clifford action on each wire, which determines the
unitary up to global phase.
\end{proof}

If the rows of $C$ were nonorthogonal, Lemma~\ref{lem:metric} would
therefore make every $Q_U$ product-preserving. Conjugation by the
same $C$ on every wire preserves that property, contradicting the
presence of an entangling gate in $\G$.

The orthogonal rows thus form a Pauli eigenbasis up to scaling.
Write $C=D_0V$, where $V$ is a one-qubit Clifford and the diagonal
matrix $D_0$ supplies the two row scalings. Absorb $V$ into the
Cliffords by setting $C_U=V^{\otimes k}Q_U(V^\dagger)^{\otimes k}$.
To remove the remaining scalings, write $D_0=LP$ with $L$ diagonal
unitary and $P$ diagonal positive. In these coordinates
\eqref{eq:metric} says that $C_U$ commutes with $(P^2)^{\otimes k}$,
hence with its positive square root $P^{\otimes k}$. The factors
of $P$ cancel, leaving $U=L^{\otimes k}C_U(L^\dagger)^{\otimes k}$
for every gate. Dividing $L$ by its first diagonal entry gives the
common frame $D=\diag(1,z)$. Conversely, taking $S=D$ gives affine
Clifford tensors and computational-basis boundary vectors up to
scalars. For algebraic gates $z$ can be chosen algebraic
(Theorem~\ref{norm:recognition}). This proves Theorem~\ref{thm:normal}.

\begin{theorem}[Algebraic case]
\label{thm:algebraic}
For every fixed finite algebraic unitary qubit library, exact amplitude
evaluation in number-field coordinates is in $\FP$ if one of the three
conditions of Theorem~\ref{thm:intro-dichotomy} holds, and is $\#\mathrm P$-hard
otherwise. The conditions are decidable, with effective algorithms and
reductions. In the common diagonal Clifford frame, the phase can be chosen algebraic.
\end{theorem}
\begin{proof}
On the tractable side, the simulators of Section~\ref{sec:classification}
use polynomial-time arithmetic in a fixed number field
(Appendix~\ref{app:prelim}). Now suppose $\G$ lies outside all three
classes. Then $\Gamma=\G\cup\G^\dagger$ still contains a nonmonomial
gate and an entangling gate, and a common diagonal Clifford frame for
$\Gamma$ would be one for $\G$. Theorem~\ref{thm:normal} thus rules
out $\ReTract(\mathcal H(\Gamma,S))$ for the particular $S$ supplied by
Theorem~\ref{thm:resources}.

Allowing all port permutations preserves the four conditions and the
cost of contraction (Lemma~\ref{hol:port-permutations}). It gives the
conjugate-closed set containing $S^{-1}e_0$ to which
Corollary~\ref{hol:dichotomy} applies. Combining its hardness reduction
with the circuit reduction of Section~\ref{sec:architecture} yields
\[
 \#\mathrm{SAT}\le_T^{\poly}\Holant(\mathcal H(\Gamma,S))
                    \le_T^{\poly}\Amp(\G).
\]
Finally, Theorem~\ref{norm:recognition} gives an effective test for
each of the three tractability conditions and constructs a common
diagonal Clifford frame whenever one exists.
\end{proof}

\subsection{Exact outputs and the proof for arbitrary coefficients}
\label{sec:coefficient-extension}

We now extend the algebraic dichotomy to arbitrary fixed complex entries.
Let $\alpha$ list their real and imaginary parts and $I$ their rational
relation ideal. Under the convention of Section~\ref{sec:classification},
an amplitude query returns two reduced polynomials representing its real
and imaginary parts. For algebraic entries this encoding is
polynomial-time equivalent to number-field coordinates
(Lemma~\ref{fg:algebraic-encoding}).

Suppose $\G$ lies outside the three tractable classes. We seek a real
algebraic tuple $\beta$ such that $\G(\beta)$ remains unitary and outside
those classes, allowing us to apply Theorem~\ref{thm:algebraic}.
Every relation in $I$ must vanish at $\beta$ so that reduction modulo
$I$ preserves values there. It suffices to require the supplied generators
of $I$ to vanish at $\beta$. Unitarity and the three tractability conditions are expressible
by polynomial equations and quantifiers over real variables, with
rational coefficients, as are their negations (Lemma~\ref{fg:specialize}).
Our requirements therefore form a rational first-order condition
satisfied by $\alpha$. Real closed field transfer~\cite[Theorem~4.4]{Pla23}
gives a real algebraic witness $\beta$. Enumerating such tuples and
testing the condition by quantifier elimination finds one in a
terminating preprocessing step for the fixed library. The resulting
$\beta$ serves every input circuit.

To evaluate a circuit over $\G(\beta)$, query the same circuit and
boundary strings over $\G$. If the answer is $(p,q)$, return
$p(\beta)+i q(\beta)$. Each amplitude is a polynomial in the gate
entries, and the preserved relations ensure that reducing it before
substitution does not change this value. The returned polynomials have
polynomial degree and bit length, so evaluation at the fixed $\beta$
takes polynomial time (Lemma~\ref{fg:hard-reduction}). Thus
\[
 \#\mathrm{SAT}\le_T^{\poly}\Amp(\G(\beta))\le_T^{\poly}\Amp(\G).
\]

On the tractable side, the simulators return the prescribed normal forms
in the original gate coordinates in polynomial time: Lemma~\ref{fg:mono-prod}
handles monomial and product-preserving libraries, and
Lemma~\ref{fg:dc-simulation} handles a common diagonal Clifford frame,
expressing the required phase factor through the original gate entries.
Appendix~\ref{app:coefficients} gives the bit bounds and effective
procedures to recognize the three classes and preprocess their simulators
from $I$. This completes Theorem~\ref{thm:intro-dichotomy} and the
recognition claim.

\clearpage
\Needspace{13\baselineskip}
\section*{AI methodology statement}
During our research, OpenAI's GPT-6 Astra helped us explore mathematical
ideas and develop proofs. In particular, it helped us derive the reduction
from arbitrary contractions of circuit tensors to exact amplitude queries
over the original gate library (Theorem~\ref{thm:resources}).
We also used the model to extend our results to arbitrary complex
coefficients, incorporate this extension into the paper, and develop the
proof of the exchange compilation theorem (Theorem~\ref{thm:exchange-compiler}).

We used OpenAI's GPT-6 Astra to improve the writing and presentation.
The authors independently checked the claims, proofs, and computational
results on exact amplitude evaluation and take full responsibility for
the correctness and originality of the paper.
The \href{https://github.com/liuchliuch/exact-quantum-circuit-amplitudes-lean}{\pinkfoldmark\nobreak\hspace{0.2em}Lean formalization}
is available on GitHub.

\section*{Acknowledgments}
We thank Zhengfeng Ji for fruitful discussions and for his encouragement and support.

\clearpage
\appendix

\section{Circuit, tensor, and arithmetic conventions}
\label{app:prelim}

We write $\Qbar$ for the field of algebraic complex numbers. For a matrix
$A$, the symbols $A^T$, $\overline A$, and $A^\dagger$ denote transpose,
entrywise conjugation, and conjugate transpose, respectively. Tensor
contraction is bilinear unless an adjoint is explicitly specified.
The column vectors $e_0=(1,0)^T$ and $e_1=(0,1)^T$ are the computational
basis vectors. For a bit string $x$, $\wt(x)=|x|$ denotes its Hamming
weight. A signature is a tensor with Boolean indices; each index position
is called a port. Its arity counts these ports, so a $k$-qubit gate has
a coefficient tensor of arity $2k$. The input/output
partition is additional information, retained in every directed
construction. An unrestricted Holant instance uses the coefficients and
port ordering but allows arbitrary connections between the ports.

\subsection{Exact arithmetic}

A circuit over a fixed finite library $\G=\{U_1,\ldots,U_s\}$ is an explicit
sequence of occurrences, each carrying an ordered tuple of distinct qubit
positions. Its input includes both length-$n$ boundary strings. Identity
wires are permitted. Relabelling the wires and the boundary strings does
not require an additional physical gate.

For an algebraic gate library, let $K$ be the number field generated
by its entries and their conjugates. Exact outputs are represented by rational coordinates
in a fixed basis of $K/\mathbb Q$. Arithmetic in a fixed number field has
polynomial bit complexity. To bound the output length, choose one integer
denominator clearing the coordinates of all gate entries and all structure
constants for multiplication in the chosen basis. An amplitude of a circuit
of length $m$ is a sum of products of $m$ gate entries. Its denominator
divides a fixed integer raised to $O(m)$, each summand has coordinate
magnitudes at most $c^{O(m)}$ for a fixed $c$, and the logarithm of the
number of summands is polynomial in the input size. Its rational
coordinates therefore have polynomial bit length. This output bound
applies to every library, including those on the hard side of the dichotomy.

Spectral decompositions and resource constructions can require an extension
of $K$. Only finitely many algebraic constants are introduced for each
fixed library, so one fixed extension contains all of them. An answer known
to lie in $K$ is converted to its prescribed coordinates by fixed rational
linear algebra. All reductions below use exact arithmetic and
polynomial-size oracle queries. Their constants may depend on $\G$.

\subsection{Product states and Clifford gates}

A nonzero pure state is a product state if it is a tensor product of
one-qubit vectors. A gate is entangling if it takes some product input
to a nonproduct state. This definition quantifies over all product
inputs. For example, CNOT entangles $\ket+\ket0$, while its
computational-basis columns are product states.

\begin{lemma}[Product-preserving unitaries]
\label{pre:product}
A $k$-qubit unitary preserves all product states if and only if it has the
form $(U_1\otimes\cdots\otimes U_k)P$, where the $U_j$ are one-qubit
unitaries and $P$ permutes the tensor factors.
\end{lemma}
\begin{proof}
The linear preserver theorem~\cite[Theorem~6.1]{Johnston11} gives a tensor
product of invertible local matrices followed by a permutation of equally
dimensional factors. If these matrices are $B_j$, unitarity gives
$\bigotimes_j(B_j^\dagger B_j)=I$. Taking the partial trace over every
factor but $j$ shows that $B_j^\dagger B_j=c_jI$, with $c_j>0$ and
$\prod_jc_j=1$. Replacing $B_j$ by $B_j/\sqrt{c_j}$ proves the required
form. Its converse follows by applying the local factors and the
permutation to a product input.
\end{proof}

The Pauli group on $k$ qubits consists of tensor products of $I,X,Y,Z$
multiplied by $\pm1$ or $\pm i$. A Clifford unitary normalizes this group
under conjugation. We include arbitrary overall phases in the term
``Clifford unitary'', and retain those phases in amplitude computations.
For each fixed $k$ the projective $k$-qubit Clifford group is finite.

For later use, $\Mono(\G)$, $\ProdPerm(\G)$, and $\DC(\G)$ denote
conditions (i), (ii), and (iii) of Theorem~\ref{thm:intro-dichotomy}, respectively,
and $\QTract(\G)=\Mono(\G)\lor\ProdPerm(\G)\lor\DC(\G)$.
Each predicate is a condition on the whole library. Exact Clifford amplitudes can be evaluated by a
quadratic-form representation that retains the overall phase~\cite{DH22}.
Together with the two direct algorithms and
the boundary-phase identity in Section~\ref{sec:classification}, this proves the tractable implication
of Theorem~\ref{thm:algebraic} with the number-field output convention.

\subsection{Binary resources}

For a $2\times2$ matrix $A$, write
$\ket A=\sum_{i,j=0}^1A_{ij}\ket{ij}$. A nonzero state of this form
is entangled exactly when $A$ is invertible, and is maximally entangled
up to normalization exactly when $AA^\dagger$ is a positive scalar
multiple of the identity. It is symmetric when $A=A^T$, in the fixed
computational coordinates. Its adjoint effect has coefficients
$\overline{A}_{ij}$.

A one-qubit state $s_0\ket0+s_1\ket1$ is coherent when $s_0s_1\ne0$.
An entangled two-qubit state is nonmaximal when its two Schmidt
coefficients, equivalently the singular values of its coefficient
matrix, are unequal. A one-qubit map is scalar-unitary when its
$2\times2$ matrix is a nonzero scalar multiple of a unitary.

\begin{proposition}[Change of tensor coordinates]
\label{pre:forward}
For every $S\in\mathrm{GL}_2(\Qbar)$, one evaluation of
$\Holant(\mathcal H(\G,S))$ computes a given amplitude over $\G$.
\end{proposition}
\begin{proof}
For each gate occurrence use $(S^{-1})^{\otimes k}US^{\otimes k}$.
Use $S^{-1}e_{b_j}$ at input $j$ and the unary tensor $S^Te_{a_j}$ at
output $j$. All internal $S$ and $S^{-1}$ factors cancel. The contraction
is
\[
 \bigl(\bra a S^{\otimes n}\bigr)
 (S^{-1})^{\otimes n}U_C S^{\otimes n}
 \bigl((S^{-1})^{\otimes n}\ket b\bigr)
 =\bra a U_C\ket b.
\]
The construction has linear size and uses precisely the signatures in
equation~\eqref{eq:holant-language}.
\end{proof}

\section{Exact oracle resources and tensor contraction}
\label{app:calculus}

Appendices~\ref{app:calculus}--\ref{app:resources} prove
Theorem~\ref{thm:resources} for a fixed finite algebraic unitary gate
library $\G$, using the conventions of Appendix~\ref{app:prelim}.
This appendix gives interpolation rules for evaluating auxiliary
tensors and proves the contraction reduction once a symmetric unitary
pair and its adjoint are jointly available. Appendix~\ref{app:seed}
constructs an initial one- or two-qubit state from a nonmonomial gate.
Appendix~\ref{app:resources} completes that state to the symmetric pair
and combines the constructions with the reduction proved here.

We first define which directed tensor compositions can be evaluated
through $\Amp(\G)$, then establish the interpolation rules that preserve
this property. These rules also normalize a symmetric pair so that its
adjoint contracts edges in one common unitary basis. Preparing all vertex
states before contracting their edges gives an acyclic evaluation of an
arbitrary tensor network.

\subsection{Directed gadgets and admissibility}
\label{calc:admissibility}

A \emph{basis-postselected gadget} is a circuit $F$ with selected input
and output ports fixed to bits in the computational basis. The remaining
ports are open. Its tensor is the corresponding submatrix of the circuit
operator: with the boundary order fixed, its entries are
$T_{y,x}=\bra{y,\eta_{\rm out}}U_F\ket{x,\eta_{\rm in}}$, where
$x,y$ label the open inputs and outputs, and
$\eta_{\rm in},\eta_{\rm out}$ are the prescribed boundary strings.
A state gadget has no open inputs. An effect gadget has no open
outputs. Gadgets are composed by connecting output ports to input ports and
summing over their common indices. Such compositions must be acyclic.

To realize a scalar composition (one with no open ports) as an amplitude
with global computational-basis boundaries, initialize every private ancilla at the beginning and
execute the gates in a topological order. Once an output is pinned, retire its wire: no later gate acts on that
wire, and its prescribed value is included in the final computational-basis
string. If another gadget needs a new input at the same logical position,
use a fresh initialized wire. In particular, the open input and output of a
one-qubit map need not occupy the same physical wire. A scalar contraction of
such gadgets is therefore a legal amplitude instance. Arbitrary ordered gate
placements provide routing by wire labels, without requiring a SWAP gate.

The adjoint of a gadget over $\G\cup\G^\dagger$ is again a gadget over that
library: reverse the gate sequence, take the adjoint of every gate, and
exchange initialized and pinned boundary wires.

Some resources will be introduced by interpolation rather than by a single
gadget. Every such tensor has a fixed ordered partition of its ports into
inputs and outputs. These partitions are respected in every composition.

\begin{definition}[Oracle admissibility]
\label{calc:definition}
A fixed finite family $\mathcal R$ of algebraic tensors with specified input
and output ports is \emph{oracle-admissible over $\G$} if evaluation of every
acyclic scalar network over $\G\cup\mathcal R$, with computational-basis
boundaries and with connections only from output ports to input ports,
has a polynomial-time Turing reduction to $\Amp(\G)$.
The number of occurrences of resources in $\mathcal R$ is part of the
instance size.
\end{definition}

Fixed gadgets over $\G$ are oracle-admissible by substitution.
Interpolation supplies further oracle-admissible tensors by recovering
the value of a whole network from circuit queries. A fixed finite
composition of these reductions preserves admissibility, as does forming
fixed acyclic gadgets from a jointly admissible family. In particular,
using a tensor and its adjoint in the same network requires the reduction
for their joint family.

We may represent a fixed tensor up to a known nonzero algebraic scalar.
If a gadget realizes $cR$ rather than $R$, substitute the gadget at every
occurrence and divide the final answer by the product of the scalar factors.
There are polynomially many factors, and their exact product has polynomial
coordinate bit length. All arithmetic is performed in a fixed number field
containing the finitely many algebraic constants used in the reduction.
Enlarging this field changes only constants in the running time.

\Needspace{11\baselineskip}
\subsection{Simultaneous interpolation}
\label{calc:interpolation}

The resource constructions use adjoint gates and one-qubit filters.
Inverse interpolation lets us use the adjoints through queries over the
original gate library. Two-eigenvalue replacement then lets us weight
the two Schmidt components separately, as needed for symmetrization
and normalization. Both reductions handle all occurrences jointly
through acyclic queries with polynomial bit complexity. For spectral
interpolation in counting problems, see Xia~\cite[Section~2.7]{Xia26}.

\begin{lemma}[Inverse interpolation]
\label{calc:inverse}
Let $A$ be a fixed invertible diagonalizable algebraic $D\times D$ matrix,
where $D=2^q$, regarded as a tensor with $q$ inputs and $q$ outputs.
Evaluation of acyclic scalar networks over a fixed finite set of algebraic
tensors together with $A^{-1}$ reduces in polynomial time to evaluation
over that set, provided the set contains $A$. All tensors have specified
input and output ports.
The conclusion holds simultaneously for any fixed finite collection of
such matrices. In particular,
\[
  \Amp(\G\cup\G^\dagger)\equiv_T^{\poly}\Amp(\G).
\]
\end{lemma}

\begin{proof}
Consider an acyclic scalar network with $r\ge1$ designated occurrences of
$A^{-1}$. Replace all of them by $A^t$ and let $F(t)$ be the resulting
contraction. Let $\lambda_1,\ldots,\lambda_d$ be the distinct eigenvalues
of $A$, with spectral idempotents $P_1,\ldots,P_d$. These constants are
algebraic and every $\lambda_j$ is nonzero. For every integer $t$,
\[
  A^t=\sum_{j=1}^d\lambda_j^tP_j.
\]
Multilinearity of contraction gives
\begin{equation}
 F(t)=\sum_{\substack{n_1+\cdots+n_d=r\\n_j\ge0}}
 c_{\boldsymbol n}
 \bigl(\lambda_1^{n_1}\cdots\lambda_d^{n_d}\bigr)^t.
 \label{calc:frequency-expansion}
\end{equation}
Here $n_j$ counts the occurrences assigned the spectral projector $P_j$.
The coefficients $c_{\boldsymbol n}$ collect the corresponding contractions
with the unchanged tensors. There are at most $\binom{r+d-1}{d-1}$ displayed frequencies. Compute these
frequencies exactly and merge equal ones, obtaining distinct nonzero values
$\mu_1,\ldots,\mu_N$ and an identity
\[
  F(t)=\sum_{j=1}^N b_j\mu_j^t.
\]
The Vandermonde matrix $(\mu_j^t)_{0\le t<N,\,1\le j\le N}$ is
nonsingular. Thus $F(0),\ldots,F(N-1)$ determine the coefficients $b_j$,
and the desired contraction is
\[
  F(-1)=\sum_{j=1}^N b_j\mu_j^{-1}.
\]
Merging equal frequencies accounts for the multiplicative relations
among the eigenvalues.

Each query replaces a designated occurrence by a chain of $t$ copies of
$A$, or by identity wires when $t=0$. These replacements preserve
acyclicity. If $A$ is itself represented by a fixed gadget, every copy uses
fresh private ancillas, and its pinned outputs are retired. If the gadget
represents $cA$ for a known $c\ne0$, the query acquires the known scalar
$c^{rt}$, which is divided out. Since $d$ is fixed,
\[
  N\le\binom{r+d-1}{d-1}
\]
is polynomial in the instance size, as are the lengths of all query
networks.

The interpolation also has polynomial bit complexity. Work in one fixed
number field containing the eigenvalues and all tensor entries. Frequencies
and Vandermonde entries are products of fixed algebraic constants with
exponents at most $rN$, so their rational coordinates have polynomial bit
length. Equality tests used to merge frequencies are exact. In a fixed
integral basis, a power of a fixed denominator clears every Vandermonde
entry. This common denominator has polynomial bit length. Determinant
bounds then give polynomial coordinate lengths for the determinant and
minors. The sum-of-products estimate in Appendix~\ref{app:prelim}
also gives polynomial coordinate bit length for the query answers over
any fixed finite algebraic tensor set. Exact linear algebra over the
fixed field therefore recovers the $b_j$ in polynomial time. The inverses
$\mu_j^{-1}$ and the final sum satisfy the same bounds: inversion in a
fixed number field is a linear system of fixed dimension over $\mathbb Q$.

For finitely many matrix types, use a separate replacement parameter for
each type and perform the interpolations on their product grid. The number
of types is fixed, so the product of their polynomial query counts remains
polynomial. Every gate $U\in\G$ is diagonalizable and satisfies
$U^{-1}=U^\dagger$. Applying the construction to the finitely many gate
types gives the reduction from $\Amp(\G\cup\G^\dagger)$ to $\Amp(\G)$.
The reverse reduction simply regards a circuit over $\G$ as one over
$\G\cup\G^\dagger$.
\end{proof}

Lemma~\ref{calc:inverse} also applies inside an oracle-admissible extension:
first eliminate the formal inverses and then invoke the reduction witnessing
admissibility. For the filters used below, we also need to prescribe
weights on two spectral subspaces. The next lemma recovers networks
containing such a filter from powers of the original map.

\begin{lemma}[Two-eigenvalue replacement]
\label{calc:filter}
Let $M$ be a fixed diagonalizable algebraic matrix with exactly two
distinct nonzero eigenvalues $\lambda_1,\lambda_2$, whose ratio is not a
root of unity. Let $P_1,P_2$ be its spectral projectors. For any fixed
algebraic numbers $c_1,c_2$, the matrix
\[
  F=c_1P_1+c_2P_2
\]
can be adjoined at polynomial-time Turing cost for evaluation of acyclic
scalar networks over a fixed finite set of algebraic tensors containing
$M$, all with specified input and output ports. In particular, the
conclusion applies to every positive
definite $2\times2$ matrix with distinct eigenvalues.
\end{lemma}

\begin{proof}
For $r$ occurrences of $F$, replace all of them by $M^t$ and denote the
resulting contraction by $H(t)$. Multilinearity gives
\[
  H(t)=\sum_{j=0}^r h_j\bigl(\lambda_1^j\lambda_2^{r-j}\bigr)^t.
\]
The $r+1$ frequencies are distinct: equality between two would make a
nonzero power of $\lambda_1/\lambda_2$ equal to one. A nonsingular
Vandermonde system therefore recovers the coefficients $h_j$ from
$H(0),\ldots,H(r)$. The required value is then
\[
  \sum_{j=0}^r h_jc_1^jc_2^{r-j}.
\]
The powers used in the queries have polynomial size. The frequencies,
Vandermonde entries, and solutions have polynomial coordinate bit length
in one fixed number field, by the product and determinant bounds in the
proof of Lemma~\ref{calc:inverse}. Both query construction and exact
postprocessing consequently take polynomial time. Positive unequal
eigenvalues have a positive ratio different from one, which cannot be a
root of unity.
\end{proof}

If $M$ is normal, its spectral projectors are Hermitian, so
Lemma~\ref{calc:filter} also adjoins
$F^\dagger=\overline{c_1}P_1+\overline{c_2}P_2$.
Use separate replacement parameters for occurrences of $F$ and $F^\dagger$.
This proves their joint admissibility, including when they occur within
state and effect gadgets. More generally, a fixed finite collection of
such replacements is handled by the product of the interpolation grids.

\subsection{Symmetric entangled states}
\label{calc:symmetric-pairs}

We use the coefficient matrix notation $\ket A$ from
Appendix~\ref{app:prelim}. The state is a tensor with two output ports,
called a \emph{cup}. Its adjoint is a tensor with two input ports,
called a \emph{cap}, with coefficients $\overline{A}_{ij}$.
For an entangled pair with unequal Schmidt coefficients, the first lemma
makes its coefficient matrix symmetric. Once the matrix is symmetric,
the second lemma makes it unitary. Both steps preserve joint
oracle-admissibility of the state and its adjoint.

\begin{lemma}[Symmetrization]
\label{calc:symmetrize}
Suppose that $\ket A$ and its adjoint effect are jointly oracle-admissible
over $\G$, where $A$ is invertible and $AA^\dagger$ is nonscalar.
One can compute an invertible symmetric algebraic matrix
$A'$ such that $\ket{A'}$ and its adjoint effect are jointly
oracle-admissible over $\G$.
\end{lemma}

\begin{proof}
Connect the second cup port to the second cap port, leaving the first cup
port as output and the first cap port as input. This directed composition
has matrix
\[
  M_{ij}=\sum_{\ell=0}^1 A_{i\ell}\overline{A}_{j\ell},
  \qquad M=AA^\dagger.
\]
The matrix $M$ is positive definite with two distinct eigenvalues, so
Lemma~\ref{calc:filter} allows us to weight its two eigenspaces
independently. We choose the weights so that applying the resulting
filter to the first qubit makes the coefficient matrix symmetric
and invertible. Take an algebraic singular-value decomposition
\[
  A=U\diag(s_1,s_2)V^\dagger,\qquad s_1,s_2>0.
\]
We will replace $s_1,s_2$ by nonzero algebraic weights $t_1,t_2$
so that the two off-diagonal entries agree. For the columns $u_j,v_j$
of $U,V$, set
\[
  d_j=(u_j)_0\overline{(v_j)_1}
       -(u_j)_1\overline{(v_j)_0}.
\]
The number $d_j$ is the difference between the off-diagonal entries of
$u_jv_j^\dagger$. Thus our target is $t_1d_1+t_2d_2=0$ with
$t_1t_2\ne0$. We have $d_1=0$ if and only if $d_2=0$. Indeed, $d_1=0$ means that
$u_1$ and $\overline{v_1}$ are proportional. Their Hermitian orthogonal
complements coincide and are spanned by $u_2$ and $\overline{v_2}$,
respectively, so $d_2=0$. Interchanging the indices proves the converse.

If both $d_j$ vanish, $A$ is already symmetric. Otherwise both are
nonzero. Set $t_1=d_2$, $t_2=-d_1$, and define
\[
  A'=U\diag(t_1,t_2)V^\dagger,
  \qquad
  F=\frac{t_1}{s_1}P_1+\frac{t_2}{s_2}P_2,
  \qquad P_j=u_ju_j^\dagger.
\]
Then $A'=FA$, and the difference between its off-diagonal entries is
$t_1d_1+t_2d_2=0$. Since $t_1t_2\ne0$, the matrix $A'$ is symmetric
and invertible. Lemma~\ref{calc:filter} jointly adjoins $F$ and
$F^\dagger$. Applying $F$ to the first cup output gives $\ket{A'}$.
Precomposing the corresponding cap input with $F^\dagger$ gives its
adjoint. Both are fixed acyclic compositions of a jointly admissible
family, proving the claim for arbitrarily many occurrences.

The construction is effective: diagonalize $M$, normalize its eigenvectors
with algebraic square roots, and recover the right singular vectors by
$v_j=A^\dagger u_j/s_j$. These operations compute the singular-value
decomposition and hence all the filter coefficients above.
\end{proof}

We next make the symmetric coefficient matrix unitary. Its adjoint effect
will then have the inverse coefficient matrix, as required to join
arbitrary tensor ports. The normalization uses a positive filter,
introduced through the evaluation reductions above.

\begin{lemma}[Unitary normalization of a symmetric pair]
\label{calc:unitary-pair}
Suppose that $A$ is an invertible symmetric algebraic $2\times2$ matrix
and that $\ket A$ and its adjoint effect are jointly oracle-admissible
over $\G$. One can effectively construct a symmetric unitary algebraic
matrix $B$ such that $\ket B$ and its adjoint effect are jointly
oracle-admissible over $\G$, also jointly with the original pair.
Moreover, one can compute a unitary algebraic matrix $S$ with $B=SS^T$.
\end{lemma}

\begin{proof}
Joining the second state port to the second adjoint-effect port gives the
one-qubit map $M=AA^\dagger$, as in the proof of
Lemma~\ref{calc:symmetrize}. It is positive definite. To obtain
$BB^\dagger=I$, put
\[
 F=M^{-1/2},\qquad B=FA.
\]
If $M=\lambda I$, the filter is the known scalar
$\lambda^{-1/2}I$. Otherwise its two eigenvalues
$\lambda_1,\lambda_2$ are distinct positive algebraic numbers, and
$\lambda_1/\lambda_2$ is not a root of unity. With $P_1,P_2$ its
spectral projectors, Lemma~\ref{calc:filter} adjoins
\[
 F=\lambda_1^{-1/2}P_1+\lambda_2^{-1/2}P_2.
\]
The positive square roots and projectors are effectively computable
algebraic constants. Since $F=F^\dagger$, applying $F$ to the first
state output gives $\ket B$, and precomposing the first effect input
with $F$ gives its adjoint.

We now verify that this filter preserves symmetry. Since $A=A^T$, we have
$A^\dagger=\overline A$ and
\[
 MA=AM^T.
\]
On the one- or two-point spectrum of $M$, the function
$x\mapsto x^{-1/2}$ agrees with a polynomial $p(x)$ with real
algebraic coefficients. Hence $F=p(M)$ and $FA=AF^T$. Therefore
\[
 B^T=A^TF^T=AF^T=FA=B,
 \qquad BB^\dagger=FMF=I.
\]

To verify joint admissibility, consider a network of size $N$ containing
a total of $r$ occurrences of the new state and effect. Expanding these
occurrences introduces exactly $r$ copies of $F$. Replacing all of them
by $M^t$, for $0\leq t\leq r$, gives $r+1$ queries, each containing
$O(N+r^2)$ occurrences of the original state and effect. The powers are
serial acyclic compositions. After interpolation, invoke the joint
admissibility of the original pair. The bit bounds of
Lemma~\ref{calc:filter} apply in one fixed number field. Existing occurrences
of the original state and effect are unchanged, so the reduction also
allows their joint use with the new pair. The scalar case uses only the
scalar accounting in Section~\ref{calc:admissibility}.

We can compute the required unitary factorization directly by spectral
calculus. Let $\mu_1,\ldots,\mu_s$ be the distinct eigenvalues of $B$,
and let
\[
 E_j=\prod_{\ell\ne j}\frac{B-\mu_\ell I}{\mu_j-\mu_\ell}
\]
be its spectral projectors, with the empty product equal to $I$.
As $B$ is normal, these projectors are Hermitian and mutually orthogonal.
Choose algebraic square roots $\nu_j^2=\mu_j$ and put
$S=\sum_j\nu_jE_j$. Unitarity of $B$ gives $|\mu_j|=1$, hence
$|\nu_j|=1$, so $S^\dagger S=I$ and $S^2=B$.
Every $E_j$ is a polynomial in the symmetric matrix $B$, whence
$S^T=S$ and $SS^T=B$. Eigenvalue computation and algebraic root
selection make the factorization effective. The matrix $S$ specifies
the coordinates used in the tensor-contraction reduction below.
\end{proof}

\subsection{Converting tensor contractions to circuit queries}
\label{calc:holant-transfer}

The symmetric unitary pair prepares a state for each tensor vertex and
supplies the effect that contracts each edge. We first use this pair
to prove evaluation equivalence between circuit amplitudes and arbitrary
contractions of the gate and boundary tensors.

\begin{proposition}[Symmetric-resource transfer]
\label{calc:transfer}
Suppose that a state with symmetric unitary algebraic coefficient matrix $A$
and its adjoint effect are jointly oracle-admissible over $\G$.
For every unitary algebraic factorization $A=SS^T$,
\[
  \Amp(\G)\equiv_T^{\poly}\Holant(\mathcal H(\G,S)),
\]
where $\mathcal H(\G,S)$ is defined in~\eqref{eq:holant-language}.
Such a factorization is effectively computable.
\end{proposition}

\begin{proof}
Lemma~\ref{calc:unitary-pair} computes a unitary algebraic factorization
$A=SS^T$. Fix any such $S$. Let $(V,E)$ be an instance of
$\Holant(\mathcal H(\G,S))$, with tensor $f_v$ at vertex $v$ and an
ordering of its $r_v$ incident ports. Each edge joins two distinct port
occurrences, possibly at the same vertex. For each label in the fixed
set~\eqref{eq:holant-language}, choose a representing gate or boundary
tensor once and for all. We construct a directed scalar network
$\mathcal D$ with the same value.

\smallskip\noindent\emph{Construction of the directed network.}
For a gate vertex $v$ represented by a $k$-qubit gate $U\in\G$, prepare
$k$ copies of $\ket A$ on fresh pairs of wires $(p_j,q_j)$, and apply
$U$ to $(p_1,\ldots,p_k)$ in that order. Keep all $2k$ outputs open.
The outputs of $U$ carry the original output indices of $f_v$, and the
untouched wires $(q_1,\ldots,q_k)$ carry its original input indices.
Thus every port of $f_v$, including each original input port, is now
represented by a distinct output wire. This block uses the original
$U$ specified by the label.

For a vertex labelled by $S^{-1}e_b$, prepare $e_b$. For a vertex
labelled by $S^Te_b$, prepare one copy of $\ket A$ and pin its second
output to $b$, leaving its first output open. The latter state has
coefficient vector $Ae_b$. Use disjoint wires for all vertex blocks.

For each edge $e$ joining port $i$ of $v$ to port $j$ of $w$, connect
the two corresponding output wires to a fresh copy of the effect
$\bra A$. This effect has two inputs and no outputs. Each port occurs
on exactly one edge, so every retained output is consumed exactly once.
In particular, a self-loop consumes two distinct outputs of one block,
and parallel edges consume different pairs of outputs. This completes
$\mathcal D$. Figure~\ref{fig:holant-to-dag} illustrates the construction.

\smallskip\noindent\emph{Acyclicity.}
Assign level $0$ to the pair and basis-state preparations, level $1$
to the gates $U$, and level $2$ to all edge effects and basis-pinning
effects. Every directed wire goes from level $0$ to level $1$ or $2$,
or from level $1$ to level $2$. A directed cycle would strictly increase
this level and return to its starting level, which is impossible.
Equivalently, prepare all pairs and basis states, apply the mutually
disjoint vertex gates, and then apply all effects. This order works for
every input graph, including self-loops and parallel edges. Each
original gate acts on an ordered tuple of distinct wires, as required
by the circuit model.

\smallskip\noindent\emph{Equality of values.}
We verify the coefficients of the constructed network in the same
basis $S$ at every port. For a gate block, list the acted-on outputs
first and the untouched outputs second. Its coefficient matrix is
$R_U=UA^{\otimes k}$, so
\[
 (S^{-1})^{\otimes k}R_U(S^{-T})^{\otimes k}
  =(S^{-1})^{\otimes k}U(SS^T)^{\otimes k}(S^{-T})^{\otimes k}
  =(S^{-1})^{\otimes k}US^{\otimes k}.
\]
This is precisely its prescribed gate tensor. The two kinds of unary
blocks give $S^{-1}e_b$ and $S^{-1}Ae_b=S^Te_b$, respectively.
Thus, if $R_v$ is the physical coefficient tensor of any vertex block,
then $R_v=S^{\otimes r_v}f_v$.

Symmetry and unitarity give $\overline A=A^\dagger=A^{-1}$, so the
coefficient matrix of each edge effect is $A^{-1}$. After substituting
$R_v=S^{\otimes r_v}f_v$ at every vertex, the factor associated with
an edge whose endpoint coordinates are $x,y$ is
\[
 \sum_{a,b\in\{0,1\}}S_{ax}(A^{-1})_{ab}S_{by}
   =(S^TA^{-1}S)_{xy}=\delta_{xy}.
\]
Each edge therefore identifies its two endpoint bits. Expanding the
finite sums and grouping them by edge assignments yields
\[
 \operatorname{val}(\mathcal D)
   =\sum_{\sigma:E\to\{0,1\}}
      \prod_{v\in V}f_v\bigl(\sigma|_{\mathrm{ports}(v)}\bigr),
\]
which is the original Holant value. Here the tuple at $v$ follows its
specified port order, and a self-loop contributes the same edge bit
at its two ports. The equality is exact, including all scalar factors.

\smallskip\noindent\emph{Reduction to circuit queries.}
The maximum gate size is fixed, so $\mathcal D$ has $O(|V|+|E|)$
gate, state, and effect occurrences. Its directed incidence structure
is obtained by scanning the vertices and edges. By the joint
oracle-admissibility of $\ket A$ and $\bra A$, its value can be
computed using polynomially many amplitude queries over $\G$, with
polynomial circuit size and exact arithmetic bit complexity.
Section~\ref{calc:admissibility} places all private ancillas and pinned
outputs at the global computational-basis boundaries of those queries.
If a resource gadget represents a tensor up to a known nonzero scalar,
the product of these factors is tracked and divided out as in that
section. This proves
$\Holant(\mathcal H(\G,S))\le_T^{\poly}\Amp(\G)$.

For the reverse reduction, Proposition~\ref{pre:forward} sends each
circuit over $\G$ to one Holant instance over $\mathcal H(\G,S)$.
The factors $S$ and $S^{-1}$ cancel on internal wires, and the transformed
boundary tensors preserve the input and output bits. The construction
has linear size and preserves the amplitude, completing the equivalence.
\end{proof}

The tensor set in Proposition~\ref{calc:transfer} already contains
nonzero one-index boundary tensors. To apply
Corollary~\ref{hol:dichotomy}, we also need closure under conjugation.
We obtain it by adjoining adjoint gates and allowing permutations of
tensor ports, while preserving the reduction to queries over $\G$.

\begin{proposition}[Conjugate-closed transfer]
\label{calc:conjugate-transfer}
Suppose a state with symmetric unitary algebraic coefficient matrix $A$
and its adjoint effect are jointly oracle-admissible over $\G$.
Let $S$ be any unitary algebraic matrix with $A=SS^T$, and set
$\Gamma=\G\cup\G^\dagger$. Then
\[
 \Amp(\G)\equiv_T^{\poly}\Holant(\mathcal H(\Gamma,S)).
\]
The finite set $\operatorname{Perm}(\mathcal H(\Gamma,S))$ obtained by
adjoining all permutations of tensor ports is closed under entrywise
complex conjugation, and its Holant problem is equivalent to the one
displayed above. Every final amplitude query is an acyclic circuit over
$\G$ with computational-basis boundaries.
\end{proposition}

\begin{proof}
To apply Proposition~\ref{calc:transfer} to $\Gamma$, we first check
joint admissibility in networks that also contain adjoint gates.
Hold the state and effect tensors fixed and use Lemma~\ref{calc:inverse}
to remove every gate in $\Gamma\setminus\G$. Joint admissibility over
$\G$ then reduces the remaining network to $\Amp(\G)$. Since each
query over $\G$ is also a query over $\Gamma$, the pair is jointly
oracle-admissible over $\Gamma$. Proposition~\ref{calc:transfer}
applied to $\Gamma$, followed by inverse interpolation, gives
\[
 \Holant(\mathcal H(\Gamma,S))
       \leq_T^{\poly}\Amp(\Gamma)
       \equiv_T^{\poly}\Amp(\G).
\]
The forward reduction in that proposition already sends a circuit over
$\G\subseteq\Gamma$ to one instance of the displayed Holant problem.
The acyclic preparation-then-contraction schedule and polynomial bounds
are exactly those proved in Proposition~\ref{calc:transfer}.

To verify conjugate closure, for each $k$-qubit $U\in\Gamma$ write
\[
 V_U=(S^\dagger)^{\otimes k}US^{\otimes k}.
\]
Since $S$ is unitary and $\Gamma$ is closed under adjoints,
$V_{U^\dagger}=V_U^\dagger$, and therefore
\[
 \overline{f_{V_U}}(y,x)=f_{V_{U^\dagger}}(x,y).
\]
The right side exchanges the two $k$-port blocks of a tensor already in
$\mathcal H(\Gamma,S)$. The boundary tensors also occur in conjugate
pairs:
\[
 \overline{S^\dagger e_b}=S^Te_b,
 \qquad \overline{S^Te_b}=S^\dagger e_b
 \quad (b=0,1).
\]
Conjugation commutes with any permutation of ports. The permutation
closure is therefore conjugate-closed, and it is finite
because the library and the number of qubits per gate are fixed.
Given an instance over the enlarged set, replace each permuted tensor
by the original one and relabel its incident ports accordingly. This
preserves every summand, including at self-loops, and gives the Holant
equivalence by relabelling the signature indices.
\end{proof}

\section{A finite resource construction from basis branching}
\label{app:seed}

From a gate that branches on a computational-basis input, we extract
either a coherent one-qubit state or an entangled two-qubit state whose
coefficient matrix is symmetric or whose Schmidt coefficients are unequal.
Overlapping circuit fragments reduce the two-qubit construction to four
pairs of amplitudes. The entangled outcomes lead to a symmetric unitary
pair by Lemmas~\ref{calc:symmetrize} and~\ref{calc:unitary-pair}.
The coherent outcome is the input to Appendix~\ref{app:resources},
where an entangling gate is also used.

All fixed boundary ports of these gadgets have computational-basis conditions. We may use
$U^\dagger$ while constructing them because Lemma~\ref{calc:inverse}
eliminates all such occurrences at polynomial oracle cost. Known nonzero
algebraic scalar factors are recorded and restored in the final value.

\begin{theorem}[Finite resource alternative]
\label{seed:alternative}
Given a nonmonomial algebraic $k$-qubit unitary $U$, a finite algorithm
constructs a gadget over $\{U,U^\dagger\}$ producing at least one of the
following:
\begin{enumerate}[label=\textup{(\roman*)},leftmargin=2em]
\item a coherent one-qubit state $s_0\ket0+s_1\ket1$ with $s_0s_1\ne0$;
\item a two-qubit state with an invertible symmetric coefficient matrix;
\item an entangled two-qubit state with unequal Schmidt coefficients.
\end{enumerate}
The gadget uses at most $24$ occurrences of $U$ or $U^\dagger$ and
$O(k)$ wires. Every choice is determined by an exact algebraic test.
\end{theorem}

Each of two circuit fragments below has two available forms. The four
choices give pairs of selected output amplitudes to which the following
lemma applies. Its conclusion supplies either unequal magnitudes or a phase
ratio that can be corrected to make the two-qubit matrix symmetric.

\begin{lemma}[Four amplitude pairs]
\label{seed:four-pairs}
Let $u,r\in\mathbb C$, and put $v=2u-1$ and $s=2r-1$.
Among the four ordered pairs
\begin{equation}
  (u,r),\qquad(v,2r),\qquad(2u,s),\qquad(v,s),
  \label{seed:pairs}
\end{equation}
there is a pair with two nonzero entries that either have unequal absolute
values or have ratio $1$, $-1$, or a primitive third root of unity.
\end{lemma}

\begin{proof}
Assume that every pair in~\eqref{seed:pairs} with two nonzero entries has
entries of equal absolute value. First suppose that $u,v,r,s$ are all
nonzero. The first three pairs give
\begin{equation}
  |u|=|r|,\qquad |2u-1|=2|u|,\qquad |2r-1|=2|r|.
  \label{seed:equal-moduli}
\end{equation}
Squaring the last two equalities gives
$\operatorname{Re}u=\operatorname{Re}r=1/4$. Together with $|u|=|r|$,
this implies $r=u$ or $r=\overline u$. In the first case the first pair
has ratio $1$. In the second,
\[
  v=2u-1=-2\overline u=-2r,
\]
so the second pair has ratio $-1$.

Suppose next that $u=0$, so $v=-1$. If $r=0$, the fourth pair is
$(-1,-1)$ and has ratio $1$. If $s=0$, then $r=1/2$, and the second
pair has ratio $-1$. Otherwise the second and fourth pairs imply
\begin{equation}
  |r|=\frac12,\qquad |2r-1|=1,
  \qquad r=\frac{1\pm i\sqrt3}{4}.
  \label{seed:third-root-values}
\end{equation}
The second ratio $-1/(2r)$ is then a primitive third root of unity.

If $v=0$, then $u=1/2$. When $r=0$, the third pair has ratio $-1$.
When $s=0$, the first pair has ratio $1$. In the remaining case, the
first and third pairs again give~\eqref{seed:third-root-values}, and
the third ratio $1/(2r-1)$ is a primitive third root of unity.

Finally, exchanging $(u,v)$ with $(r,s)$ reverses the ordered pairs
and exchanges the middle two pairs. For a pair with two nonzero entries,
reversal reciprocates its ratio, preserving the allowed ratios and
inequality of absolute values. The preceding cases therefore also cover
a zero among $r,s$. Every case supplies a pair with two nonzero entries.
\end{proof}

\begin{proof}[Proof of Theorem~\ref{seed:alternative}]
We first construct a one-qubit map that either yields a coherent state
or gives $Z$ up to a known nonzero scalar. In the latter case, $Z$ and
a basis projection will supply the two forms of each circuit fragment.
Since $U$ is nonmonomial, some column has nonzero entries with both bit
values at an output position $j$. Let $P_j$ be the projector onto value
one of that output qubit, and put
\begin{equation}
  H=U^\dagger P_jU.
  \label{seed:projector}
\end{equation}
This is a nondiagonal projector. Indeed, the chosen column gives a
diagonal entry of $H$ strictly between zero and one, whereas every
diagonal projector has only zeros and ones on its diagonal. Choose
distinct basis strings $x,y$ with $h=H_{yx}\ne0$.

Write $\ket{\psi_x}=U\ket x$ and $\ket{\psi_y}=U\ket y$.
Preparing $\ket{\psi_x}$ and applying the effect $\bra{\psi_y}$ on all
matching ports except $j$ gives the one-qubit map
\begin{equation}
  L=\operatorname{Tr}_{\ne j}
       \bigl(\ket{\psi_x}\bra{\psi_y}\bigr),
  \qquad L_{11}=h,\qquad L_{00}=-h.
  \label{seed:reduced-map}
\end{equation}
The identity $L_{11}=h$ follows from
$\bra{\psi_y}P_j\ket{\psi_x}=H_{yx}$, and the other identity follows
from $\operatorname{tr}L=\langle\psi_y\mid\psi_x\rangle=0$.
To realize this as a directed gadget, prepare $\ket{\psi_x}$ on fresh
wires and retain its $j$th output as the open output. Feed the open
input and the other $k-1$ prepared outputs into $U^\dagger$, with the
open input at position $j$, and pin the resulting outputs to $y$.
Thus $L$ uses one $U$ and one $U^\dagger$ and involves no closed wire.
If an off-diagonal entry of $L$ is nonzero, one of its columns has two
nonzero entries because $L_{00}$ and $L_{11}$ are both nonzero.
Pinning the input to that column gives a coherent one-qubit state, proving
the first alternative.

We may therefore assume $L=-hZ$. The gadget gives $Z$ up to a known
nonzero scalar. The rank-one map $P_0=\ket0\bra0$ is also available
using boundary pins alone: retire its input wire with final pin zero
and use a fresh wire initialized to zero as its output, following
Section~\ref{calc:admissibility}. These two one-qubit maps give the
circuit fragments
\begin{equation}
  K(z)=U^\dagger\bigl(I+(z-1)P_j\bigr)U
      =I+(z-1)H,\qquad z\in\{0,-1\},
  \label{seed:K}
\end{equation}
with directed implementations up to known nonzero scalars. For $z=0$,
the middle map is $P_0$ on qubit $j$. For $z=-1$, it is $Z$ on that
qubit. A fresh output wire replacing an active wire is used as the
same logical wire in subsequent gates. Hence $K(0)$ uses two
occurrences of $U$ or $U^\dagger$, and $K(-1)$ uses four, including
the two occurrences in the gadget for $Z$.

To separate the contributions of the two fragments to two selected
outputs, we place them on tuples differing in one wire. Choose a bit
position $\ell$ changed by the transition $x\to y$, so
$x_\ell\ne y_\ell$. Let
$S=(q_1,\ldots,q_k)$ be a tuple of logical wires, and let $T$ be the
same tuple with $q_\ell$ replaced by a fresh wire $f$.
Initialize $S$ to $x$ and $f$ to $x_\ell$, and denote this full basis
input by $I_0$. Choose $z,w$ independently from $\{0,-1\}$ and
apply $K(z)$ on $S$, followed by $K(w)$ on $T$.
Consider the output strings
\begin{equation}
\begin{aligned}
  a_{\mathrm{out}} &: \quad S=y,\qquad f=x_\ell,\\
  c_{\mathrm{out}} &: \quad T=y,\qquad q_\ell=x_\ell.
\end{aligned}
\label{seed:outputs}
\end{equation}
They agree on all shared wires. Their bits on $q_\ell,f$ are opposite
and exchanged. To track the term involving both copies of $H$, define
\begin{equation}
  \alpha=\bra{a_{\mathrm{out}}}H_TH_S\ket{I_0},
  \qquad
  \beta=\bra{c_{\mathrm{out}}}H_TH_S\ket{I_0},
  \label{seed:cross-terms}
\end{equation}
where a subscript specifies the ordered tuple on which a map acts.
Expanding~\eqref{seed:K}, the two output amplitudes are
\begin{equation}
\begin{aligned}
 F_a(z,w)&=(z-1)\bigl[h+(w-1)\alpha\bigr],\\
 F_c(z,w)&=(w-1)\bigl[h+(z-1)\beta\bigr].
\end{aligned}
\label{seed:amplitudes}
\end{equation}
The identity term contributes to neither output. The $H_S$ term
contributes $h$ to $a_{\mathrm{out}}$ and zero to $c_{\mathrm{out}}$,
because it cannot change $f$. The $H_T$ term contributes $h$ to
$c_{\mathrm{out}}$ and zero to $a_{\mathrm{out}}$, because it cannot
change $q_\ell$. The joint contribution is given by the cross terms
in~\eqref{seed:cross-terms}.

Pin each shared output wire to its bit in $y$, leaving $q_\ell,f$ as
the two outputs. All internal ancillas retain the same prescribed
boundary pins for both amplitudes. To match the four pairs in
Lemma~\ref{seed:four-pairs}, express the cross terms relative to the
known nonzero value $h$ by setting
\begin{equation}
  u=1-\alpha/h,\qquad r=1-\beta/h,
  \qquad v=2u-1,\qquad s=2r-1.
  \label{seed:parameters}
\end{equation}
Substitution in~\eqref{seed:amplitudes} gives the four pairs
\begin{equation}
\begin{array}{c|c}
  (z,w)&\bigl(F_a(z,w),F_c(z,w)\bigr)\\ \hline
  (0,0)&-h(u,r)\\
  (0,-1)&-h(v,2r)\\
  (-1,0)&-h(2u,s)\\
  (-1,-1)&-2h(v,s).
\end{array}
\label{seed:four-evaluations}
\end{equation}
The common nonzero factor in each row preserves the properties tested
in Lemma~\ref{seed:four-pairs}. That lemma therefore selects a pair
whose two entries are nonzero and satisfy one of its conclusions.
Order the two output ports so that these amplitudes occur at $01$ and
$10$, and denote the respective coefficients by $a,b$.
If the state also has a
nonzero coefficient at $00$ or $11$, pinning one output gives a
coherent one-qubit state. Otherwise its coefficient matrix is
\begin{equation}
  B=\begin{pmatrix}0&a\\ b&0\end{pmatrix},\qquad ab\ne0.
  \label{seed:off-diagonal-state}
\end{equation}
Exchanging the output order only reciprocates $a/b$, preserving all
conclusions of Lemma~\ref{seed:four-pairs}.

If $|a|\ne|b|$, the singular values of $B$ are unequal, giving the
third alternative. If $a/b=1$, the matrix is symmetric already.
If $a/b=-1$, applying the available $Z$ to one output makes it
symmetric. It remains to consider $a/b=\omega$, where $\omega$ is
a primitive third root of unity. We cancel this relative phase using
a diagonal map whose diagonal-entry ratio is $\omega^2$.

Take a copy of the state with coefficient matrix $B$ and its adjoint
effect, obtained by reversing the gadget as in
Section~\ref{calc:admissibility}. Connect the second state port to the
first effect port. The remaining state port is the open output, and
the remaining effect port is the open input. This port ordering gives
\begin{equation}
  M=B\overline B
   =|b|^2\begin{pmatrix}\omega&0\\0&\omega^{-1}\end{pmatrix}.
  \label{seed:third-root-map}
\end{equation}
Apply this map to the first output of another copy of the state $B$.
The new coefficient matrix is $MB$, and its nonzero entries satisfy
\begin{equation}
  \frac{(MB)_{01}}{(MB)_{10}}
    =\omega^2\frac ab=\omega^3=1.
  \label{seed:symmetric-third-root}
\end{equation}
Thus $MB$ is symmetric and invertible. The construction consists of
two state preparations followed by one effect, so it is acyclic.

All steps inspect finite lists of entries, candidate indices, and
four fixed circuits. Nonvanishing, equality, and comparison of squared
absolute values are decidable for algebraic numbers. Each candidate
two-qubit state uses at most eight occurrences of $U$ or $U^\dagger$.
A $Z$ correction adds two. Equations~\eqref{seed:third-root-map}--
\eqref{seed:symmetric-third-root} use three copies of a candidate
gadget or its adjoint, giving the bound of $24$.
Only a constant number of $k$-wire gadgets and their private ancillas
are composed, so the number of wires is $O(k)$. Every retired wire
remains untouched until its final basis pin. The output is therefore
a gadget in the required circuit model.
\end{proof}

The symmetric outcome of Theorem~\ref{seed:alternative} can be
normalized by Lemma~\ref{calc:unitary-pair}. The unequal-Schmidt
outcome first becomes symmetric by Lemma~\ref{calc:symmetrize}.
For the coherent-state outcome, Theorem~\ref{res:coherent} supplies
a symmetric pair when the library also contains an entangling gate.
These subsequent constructions use the evaluation reductions of
Appendix~\ref{app:calculus}. The finite gadgets constructed here use
$U$ and $U^\dagger$, and Lemma~\ref{calc:inverse} removes the adjoint
gates when these resources are used in reductions to $\Amp(\G)$.

\section{Symmetric entangled resources}\label{app:resources}

We treat the coherent one-qubit outcome of Theorem~\ref{seed:alternative}.
Together with the two basis states, it gives three distinct state rays
and their adjoint effects. We use these states to obtain a nonproduct
output of an entangling gate, then use the effects to retain an entangled
pair on two outputs. A pair with unequal Schmidt coefficients can be
symmetrized by Lemma~\ref{calc:symmetrize}. The remaining case has equal
Schmidt coefficients and a nonsymmetric coefficient matrix $A$. Here
$AA^\dagger$ is scalar, so spectral filtering of this matrix cannot
separate the two Schmidt components. We seek additional local maps that
either separate them or make the coefficient matrix symmetric directly.

Throughout this appendix, availability refers to
fixed gadgets over the adjoint completion $\Gamma=\G\cup\G^\dagger$, with
computational-basis conditions on their other boundary ports. Known nonzero
algebraic scalar factors are allowed. The adjoint of an available state is
therefore an available effect. Every composition below is directed and
acyclic, with initialized and retired wires handled as in
Section~\ref{calc:admissibility}. Lemma~\ref{calc:inverse} makes any fixed finite
collection of available gadgets jointly oracle-admissible over $\G$.

\begin{theorem}[Coherent-state resource construction]\label{res:coherent}
Suppose an algebraic unitary qubit library has an available coherent
one-qubit state and contains a gate that is not product-permutation. Then one can
effectively construct an invertible symmetric algebraic matrix $A$ such that
$\ket A$ and its adjoint effect are jointly oracle-admissible over $\G$.
A finite certificate for the construction can be found from the gate
matrices and the circuit preparing the coherent state.
\end{theorem}

Theorem~\ref{res:coherent} supplies the invertible symmetric matrix for
the coherent-state branch. In Section~\ref{res:completeness-section}, we
combine this branch with the other outcomes of
Theorem~\ref{seed:alternative} and apply Lemma~\ref{calc:unitary-pair}
to obtain the symmetric unitary resource. For the nonsymmetric maximally
entangled case, a full-rank local map that is not scalar-unitary gives
unequal Schmidt coefficients. Here scalar-unitary means a nonzero scalar
multiple of a unitary, as in Appendix~\ref{app:prelim}. If all full-rank
local maps have this form, two noncommuting ones supply a symmetrization
through the group argument below. If they all commute, they force every
gate to preserve product states, contradicting the hypothesis.

\subsection{Localizing entanglement}

A one-qubit ray is a nonzero vector in $\mathbb C^2$ modulo nonzero scalar
multiplication. We use the same convention for covectors. The following
lemma applies to either kind of ray.

\begin{lemma}[Interpolation on products of rays]\label{res:ray-interpolation}
Let $p$ be a polynomial separately homogeneous of degree $d$ in each of
$m$ pairs of variables. If $p$ vanishes on a product of sets containing
$d+1$ distinct rays in each pair, then $p$ is identically zero.
\end{lemma}
\begin{proof}
A nonzero homogeneous polynomial of degree $d$ in two variables has at most
$d$ distinct projective zeros: each zero supplies a distinct linear factor,
and the product of $d+1$ such factors has degree larger than $d$.
For several pairs, fix the first $m-1$ pairs at points of their grids and
apply this fact to the last pair. Every coefficient in the last pair
vanishes on the grid for the other pairs. Induction on $m$ makes all those
coefficient polynomials zero.
\end{proof}

\begin{lemma}[Three-ray localization]\label{res:localization}
Let $E$ contain three distinct one-qubit covector rays. Every nonproduct pure
state on $n\geq2$ qubits has a contraction, using members of $E$ on all but
two output ports, whose two-qubit coefficient matrix is invertible.
\end{lemma}
\begin{proof}
We first find two output ports whose contraction has a determinant
that is not identically zero as a polynomial in the remaining effects.
At least one one-qubit flattening, the matrix with that qubit indexing
rows and all other qubits indexing columns, has rank two. Indeed, if
every such flattening had rank one, we could factor off one qubit and
repeat on the remaining tensor, obtaining a product state. Relabel a
rank-two flattening as the first qubit and write the state as
$\ket0 A+\ket1 B$, with linearly independent coefficient tensors $A,B$.
Associate to them the following multiaffine polynomials, in which each
variable has degree at most one:
\[
 a(z_2,\ldots,z_n)=
 \sum_{x\in\{0,1\}^{n-1}} A_x\prod_{\ell=2}^n z_\ell^{x_\ell},
 \qquad
 b(z_2,\ldots,z_n)=
 \sum_{x\in\{0,1\}^{n-1}} B_x\prod_{\ell=2}^n z_\ell^{x_\ell}.
\]
For each $j\in\{2,\ldots,n\}$, write $a=a_0+z_ja_1$ and
$b=b_0+z_jb_1$. Then
\begin{equation}\label{res:local-determinant}
 a\partial_jb-b\partial_ja=a_0b_1-a_1b_0.
\end{equation}
The expression on the right is the determinant of the coefficient matrix
obtained by retaining qubits $1,j$ and contracting every other qubit with
the covector $(1,z_\ell)$.

This determinant is a nonzero polynomial for some $j$. Otherwise every
partial derivative of the rational function $b/a$ would vanish, forcing
$b/a$ to be constant and contradicting independence. The assertion about
rational functions follows algebraically in characteristic zero. In one
variable, write a reduced fraction $p/q$ over the field of the other
variables. The identity $p'q-pq'=0$ and $\gcd(p,q)=1$ imply
$p\mid p'$ and $q\mid q'$. A nonconstant polynomial cannot divide its
derivative, so both degrees are zero. Apply this argument successively
to all variables.

For this choice of $j$, replace each contracting covector $(1,z_\ell)$
by a homogeneous pair. The determinant is then homogeneous of degree
two in each contracted pair. It remains a nonzero polynomial, since its affine specialization
\eqref{res:local-determinant} is nonzero. By
Lemma~\ref{res:ray-interpolation}, it cannot vanish on the whole product
of the three-ray sets $E$. A nonzero value gives the required contraction.
\end{proof}

\begin{lemma}[Three-ray product-input test]\label{res:product-test}
Let $E$ contain three distinct one-qubit state rays. If a unitary $U$ maps
every product of states from $E$ to a product state, then $U$ is
product-permutation. Consequently, an available coherent one-qubit state
and a gate that is not product-permutation supply an available invertible
two-qubit state.
\end{lemma}
\begin{proof}
For each output qubit, every $2\times2$ minor of its flattening is
homogeneous of degree two in each input coordinate pair. By the
hypothesis, these minors vanish on the product grid $E^k$, where $U$ acts
on $k$ qubits. Lemma~\ref{res:ray-interpolation} makes every minor
identically zero. Thus all one-qubit flattenings of the output of an
arbitrary product input have rank at most one. The output is nonzero by
unitarity, so factoring off one output qubit at a time shows that it is a
product state. Lemma~\ref{pre:product} now makes $U$ product-permutation.

For the consequence, let $s$ be the coherent state and take
$E=\{\ket0,\ket1,s\}$. These are three distinct rays. By the first
assertion, an entangling gate has a nonproduct output on some product
input from $E^k$. Apply Lemma~\ref{res:localization} to that output,
using the three available adjoint effects. The resulting two-qubit state is
available and invertible. This is a finite search: there are $3^k$ input
choices and at most $\binom{k}{2}3^{k-2}$ contractions for each output.
All determinant tests use exact algebraic arithmetic.
\end{proof}

\subsection{Maximal entanglement and local unitary maps}

Suppose an available invertible two-qubit state is maximally entangled.
After dividing its coefficient matrix by a known nonzero algebraic scalar,
write it as $A$ with $AA^\dagger=I$. Contracting its second output
with the adjoint of a state $v$ leaves the one-qubit state
\[
 J_A(v)=A\overline v.
\]
This conjugate-linear, norm-preserving map is antiunitary. To symmetrize
the pair by a local unitary, we seek an available $Q$ for which $QA$ is
symmetric. The next
lemma shows how $J_A$ generates available local maps and relates symmetry
to a plane reflection on the Bloch sphere. For this geometric description,
use the Bloch vector defined by
$vv^\dagger/\lVert v\rVert^2=(I+xX+yY+zZ)/2$. A unitary induces a
rotation of this vector. Coordinatewise complex conjugation, denoted by
$\mathsf K$, induces the reflection $\kappa(x,y,z)=(x,-y,z)$.
Every rotation has a unitary lift: conjugation
by $\exp(-i\theta(n_xX+n_yY+n_zZ)/2)$ rotates through angle $\theta$
about the unit vector $n$. The kernel of this action consists of scalar
unitaries, since a matrix acting trivially must commute with all three
Pauli matrices. An antiunitary consequently induces an orientation-reversing
orthogonal transformation.

\begin{lemma}[Antiunitary symmetry]\label{res:antiunitary}
For an available unitary two-qubit coefficient matrix $A$, the one-qubit unitary
$M=A\overline A$ is available. If a one-qubit unitary $R$ is available, then
so is
\begin{equation}\label{res:tau}
 \tau(R)=A\overline R A^{-1},
 \qquad \tau(M)=M,
 \qquad \tau^2(R)=MRM^{-1}.
\end{equation}
If $J_A$ induces a plane reflection on the Bloch sphere, then $A=A^T$.
The same conclusion holds if $J_A$ fixes a state ray.
\end{lemma}
\begin{proof}
Prepare the state with coefficient matrix $A$, and connect its second
output to the first input of its adjoint effect. The first state port is
the remaining output and the second effect port is the remaining input.
The resulting map has entries
\[
 M_{ij}=\sum_\ell A_{i\ell}\overline{A_{\ell j}},
 \qquad M=A\overline A.
\]
Inserting a one-qubit map $R$ on the connecting wire gives
\[
 \sum_{\ell,a} A_{i\ell}R_{a\ell}\overline{A_{aj}}
       =(AR^T\overline A)_{ij}.
\]
Use $R^\dagger$ in this construction and compose on the right with
$M^{-1}$. The latter is available from the adjoint circuit for the
unitary $M$. The resulting map is
\[
 A\overline R\,\overline A(A\overline A)^{-1}
       =A\overline R A^{-1}=\tau(R).
\]
The other identities follow directly:
\[
 \tau(M)=A(\overline A A)A^{-1}=A\overline A,
 \qquad
 \tau^2(R)=A\overline A\,R\,\overline A^{-1}A^{-1}=MRM^{-1}.
\]
These constructions have a state before an effect and no directed cycle.

Every plane reflection is $O\kappa O^{-1}$ for some rotation $O$.
Choose a unitary lift $V$ of $O$. The antiunitary $V\mathsf K V^\dagger$
has coefficient matrix $VV^T$, since
$V\overline{V^\dagger v}=VV^T\overline v$. Two antiunitaries with the
same Bloch action differ by a scalar phase, by the kernel observation
above. Hence $J_A$ having this action implies $A=\omega VV^T$ for a
phase $\omega$, and $A$ is symmetric.

Finally, if $J_A$ fixes a state ray, its Bloch action fixes a unit vector
$u$. The plane $u^\perp$ is invariant, and the restricted orthogonal
map has determinant $-1$. A two-dimensional orthogonal map of
determinant $-1$ is a reflection, with eigenvalues $1,-1$. The full
three-dimensional map therefore has eigenvalues $1,1,-1$ and is a plane
reflection. The preceding argument applies.
\end{proof}

Available one-qubit unitaries are closed under composition and adjoints,
so their Bloch rotations form a group. We use this group to find a
transformation of the two-qubit state. An infinite subgroup over a fixed
number field contains a rotation that yields a spectral filter. In the
finite noncyclic case, adjoining the Bloch action of $J_A$ to a suitable
rotation subgroup gives a finite orthogonal group containing a plane
reflection. The corresponding local unitary makes the coefficient matrix
symmetric. The passage from projective matrices to rotations and
the use of number fields to bound finite orders also appear in Xia's
binary-group framework~\cite[Theorem~4.6 and Corollary~4.7]{Xia26}.
The required rotation-group arguments follow.

\begin{lemma}[Infinite algebraic rotation groups]\label{res:infinite-group}
An infinite subgroup of $\mathrm{SO}(3)$ whose matrix entries lie in a
fixed number field contains an element of infinite order.
\end{lemma}
\begin{proof}
Let the field be $K$, and suppose that every group element has finite
order. Its eigenvalues are roots of unity of degree at most
$3[K:\mathbb Q]$ over $\mathbb Q$. There are only finitely many such
roots, so only finitely many traces of group elements. The following
explicit order bound also supplies the root-of-unity test in
Proposition~\ref{res:noncommuting}. A primitive $m$th root has degree
$\varphi(m)$, and
\[
 \frac{\varphi(m)^2}{m}
   =\prod_{p^a\parallel m}p^{a-2}(p-1)^2\geq\frac12.
\]
Each factor for an odd prime is at least one, and the only possible
factor below one is $1/2$ from $2^1$. Thus a bound on the degree bounds
$m$ as well.

Let $W$ be the complex linear span of the group matrices, and choose a
basis $g_1,\ldots,g_d$ of $W$ from the group. The map
\[
 g\longmapsto\bigl(\Tr(gg_1),\ldots,\Tr(gg_d)\bigr)
\]
has finite image, because each product $gg_j$ is another group element.
It is injective. Indeed, if $h=g-g'$ pairs to zero with every $g_j$
under $(X,Y)\mapsto\Tr(XY)$, it pairs to zero with all of $W$.
The space $W$ is closed under adjoints: for each group matrix,
$g^\dagger=g^{-1}$ is in the group. Thus $h^\dagger\in W$, giving
$\Tr(hh^\dagger)=0$. This is the sum of the squared absolute values
of the entries of $h$, so $h=0$. A finite image and injectivity would
make the group finite, a contradiction.
\end{proof}

\begin{theorem}[Finite rotation-group classification
{\cite{OConnor03}}]\label{res:finite-classification}
Every finite subgroup of $\mathrm{SO}(3)$ is conjugate to a cyclic group
$C_m$, a dihedral rotation group $D_m$ of order $2m$, or the rotational
symmetry group of a tetrahedron, cube, or icosahedron, isomorphic to
$A_4,S_4,A_5$, respectively. Here the elements outside the principal
cyclic subgroup of $D_m$ are half-turns in three-space.
\end{theorem}

\begin{lemma}[A reflection in the improper coset]\label{res:finite-reflection}
Let $H\leq\mathrm O(3)$ be finite and contain an orientation-reversing
element. If $K=H\cap\mathrm{SO}(3)$ is noncyclic, then $H$ contains a
plane reflection.
\end{lemma}
\begin{proof}
Write $J=-I_3$. If $J\in H$, the noncyclic possibilities for $K$ in
Theorem~\ref{res:finite-classification} are $D_m$ with $m\geq2$,
$A_4,S_4,A_5$. Each has an element of order two: one may take a
dihedral generator, a double transposition in $A_4$ or $A_5$, and a
transposition in $S_4$. In $\mathrm{SO}(3)$ such an element $r$ is a
half-turn, with eigenvalues $1,-1,-1$. Thus $Jr\in H$ has eigenvalues
$-1,1,1$ and is a plane reflection.

Suppose $J\notin H$. The set
\[
 K'=K\cup J(H\setminus K)
\]
is a finite subgroup of $\mathrm{SO}(3)$. The two pieces are disjoint:
if $Jh\in K$ for $h\in H\setminus K$, then $J=(Jh)h^{-1}\in H$.
Closure follows because $J$ is central, the product of two improper
elements lies in $K$, and multiplication by a proper element preserves
the improper coset. Hence $K$ has index two in $K'$. The group $K'$ is
noncyclic because it contains the noncyclic subgroup $K$. We examine
all its remaining possibilities.

Neither $A_4$ nor $A_5$ has an index-two subgroup. Both are generated
by 3-cycles, and a homomorphism to $C_2$ must kill every 3-cycle.
Thus no such homomorphism can be surjective. If $K'$ is $S_4$, any
homomorphism $S_4\to C_2$ takes the same value on all transpositions,
since they are conjugate. Transpositions generate $S_4$, so the only
nontrivial homomorphism is the sign map. Consequently $K=A_4$ and
$K'\setminus K$ contains a transposition, an order-two rotation.

It remains to consider $K'=D_m=\langle r,s\mid r^m=s^2=1,
srs=r^{-1}\rangle$. Its elements are $r^j$ and $r^js$, and every
$r^js$ is a half-turn. An index-two subgroup is the kernel of a
surjective homomorphism $\chi:D_m\to C_2$, written additively.
If $\chi(r)=0$, then $\chi(s)=1$ and the kernel is the cyclic
subgroup $\langle r\rangle$, contrary to the hypothesis. Therefore
$\chi(r)=1$, which forces $m$ even. Writing $\chi(s)=\varepsilon$
with $\varepsilon\in\{0,1\}$ gives
\[
 K=\langle r^2,r^\varepsilon s\rangle.
\]
The complementary coset contains $r^{\varepsilon+1}s$, which is a
half-turn. This also checks both possible noncyclic index-two kernels
in the dihedral case.

Thus in every possible case a half-turn $r_0$ belongs to
$K'\setminus K$. By the definition of $K'$, the element $Jr_0$ is
in $H$, and its eigenvalues show that it is the required reflection.
\end{proof}

\begin{proposition}[Noncommuting local unitaries]\label{res:noncommuting}
Suppose an invertible maximally entangled two-qubit state and one-qubit unitaries
$R,S$ with $RS\ne SR$ are available. Then a two-qubit state with an
invertible symmetric coefficient matrix and its adjoint effect are jointly
oracle-admissible over $\G$.
The construction has an effectively findable finite certificate.
\end{proposition}
\begin{proof}
Normalize the two-qubit coefficient matrix to a unitary $A$, and use
$M,\tau$ from Lemma~\ref{res:antiunitary}. To include $J_A$ in the
subsequent group analysis, we need a unitary subgroup preserved by
conjugation by $J_A$ and containing $J_A^2=M$. We obtain both properties
by adjoining $M$ and the $\tau$-images of $R,S$:
\begin{equation}\label{res:local-group}
 \Lambda=\langle R,S,M,\tau(R),\tau(S)\rangle.
\end{equation}
Its elements and inverses are available one-qubit unitaries. The map $\tau$
preserves the group. It sends $M$ to itself and the first two generators
to the last two. It sends those last two to $MRM^{-1},MSM^{-1}$.
Thus $\tau(\Lambda)\subseteq\Lambda$. Since $\tau^2$ is conjugation
by $M\in\Lambda$, this inclusion is an equality.

Let $K\leq\mathrm{SO}(3)$ be the Bloch image of $\Lambda$. All its
entries lie in one number field, generated by the entries and conjugates
of the finitely many generators, together with $i$. For example, its
matrix entries are $\frac12\Tr(\sigma_aQ\sigma_bQ^\dagger)$ for
Pauli matrices $\sigma_a,\sigma_b$ and $Q\in\Lambda$. The group $K$
is noncyclic. If it were generated by the image of $Q_0\in\Lambda$,
every $Q\in\Lambda$ would be a scalar multiple of a power of $Q_0$.
Then $R$ and $S$ would commute, a contradiction.

If $K$ is infinite, Lemma~\ref{res:infinite-group} gives an element
of infinite order. Choose a word $Q\in\Lambda$ lifting it. Its two
eigenvalues $\lambda_0,\lambda_1$ have ratio that is not a root of
unity: a power of a unitary has trivial Bloch action precisely when it
is scalar. To separate the Schmidt coefficients, assign distinct
positive weights to its two eigenspaces. With $P_0,P_1$ denoting the
spectral projectors, choose weights $1,2$ and set
\[
 F=P_0+2P_1.
\]
The projectors are algebraic and Hermitian. Lemma~\ref{calc:filter}
adjoins $F$ at polynomial exact Turing cost, and $F^\dagger=F$.
The state with coefficient $FA$, together with its adjoint effect, is
therefore jointly admissible. Its Gram matrix is
$(FA)(FA)^\dagger=F^2=P_0+4P_1$, so its singular values are unequal.
Lemma~\ref{calc:symmetrize} supplies the symmetric resource.

If $K$ is finite, let $j_A$ denote the Bloch action of $J_A$.
Conjugation by $j_A$ on $K$ is induced by $\tau$, and
$j_A^2$ is the action of $M$. Thus
\[
 H=K\cup K j_A
\]
is a finite subgroup of $\mathrm O(3)$ whose rotation subgroup is
$K$. Lemma~\ref{res:finite-reflection} gives a plane reflection in
its improper coset. It has the form $\rho(Q)j_A$ for a word
$Q\in\Lambda$, where $\rho$ is the Bloch action. The corresponding
antiunitary is $QJ_A$, with coefficient matrix $QA$.
Lemma~\ref{res:antiunitary} therefore makes $QA$ symmetric. Applying
$Q$ to the first output of $\ket A$ realizes this invertible symmetric
state, and reversing the gadget realizes its adjoint effect.

Both cases can be found by one word search. Enumerate words $Q$ in the generators of
\eqref{res:local-group} and their inverses. For each word, test whether
$QA$ is symmetric and whether the eigenvalue ratio of $Q$ has infinite
order.
The finite case guarantees success of the first test for some word. The
infinite case guarantees success of the second. Both are decidable
algebraic tests. In particular, for an algebraic number $\zeta$ of
degree $d$, membership among roots of unity can be checked by testing
$\zeta^m=1$ for $1\leq m\leq2d^2$, using the degree bound in
Lemma~\ref{res:infinite-group}. A repeated eigenvalue has ratio one
and never passes the second test.

All words found in this search are fixed for the library. The filters,
state gadgets, and adjoint gadgets form a fixed finite family. Their
joint admissibility follows from Lemmas~\ref{calc:filter} and
\ref{calc:symmetrize}, followed by the simultaneous elimination of
the original gate adjoints in Lemma~\ref{calc:inverse}.
\end{proof}

The matrix condition $RS\ne SR$ ensures that the Bloch group is
noncyclic. For example, two anticommuting Pauli matrices generate
commuting half-turns about distinct axes, giving a noncyclic rotation
group to which the proposition applies.

\subsection{Four-ray rigidity}

We now show that, given four available state rays, the remaining
commuting case forces every available unitary to be product-permutation.
Section~\ref{res:construction-section} will construct these four rays
from the nonsymmetric maximally entangled pair. When all full-rank local
maps are scalar-unitary and commute, the four rays impose polynomial identities
on contractions of each gate. To recover the gate structure from these
identities, we use the following factorization criterion: a contraction
determinant is not identically zero exactly when its retained input and output
variables belong to the same irreducible factor of the coefficient
polynomial.

\begin{lemma}[Mixed derivatives and irreducible factors]\label{res:ufd}
Let $f$ be a nonzero multiaffine polynomial over $\mathbb C$.
Each variable occurs in at most one nonconstant irreducible factor.
For distinct variables $x,y$ occurring in $f$,
\begin{equation}\label{res:factor-test}
 ff_{xy}-f_xf_y\ne0
 \quad\Longleftrightarrow\quad
 x,y\text{ occur in the same irreducible factor of }f.
\end{equation}
\end{lemma}
\begin{proof}
The degree in any given variable adds under multiplication, whereas
its degree in $f$ is at most one. Thus it occurs in only one factor,
and no factor containing it can be repeated. In the rational-function
field, the mixed logarithmic derivative is
\[
 \partial_y\left(\frac{f_x}{f}\right)
       =\frac{ff_{xy}-f_xf_y}{f^2}.
\]
If $x,y$ occur in different factors, this expression is zero.
If they occur in the same irreducible factor $g$, it equals
$(gg_{xy}-g_xg_y)/g^2$. The numerator cannot vanish: otherwise
$g\mid g_xg_y$. The polynomial ring is a unique factorization domain,
so the irreducible polynomial $g$ is prime and must divide $g_x$ or
$g_y$. Both derivatives are nonzero, since their variables occur in
$g$, and each has smaller degree than $g$ in its differentiated
variable. Either divisibility is impossible.
\end{proof}

\begin{proposition}[Four-ray rigidity]\label{res:four-ray}
Suppose four distinct one-qubit state rays and their adjoint effects are
available. If every full-rank available one-qubit map is scalar-unitary and
these maps commute pairwise, then every available unitary is
product-permutation.
\end{proposition}
\begin{proof}
Choose a common orthonormal eigenbasis $V$ for the full-rank one-qubit maps.
This exists even if the family is infinite. A nonscalar member is a
normal $2\times2$ matrix with two distinct eigenvalues, and every
commuting map preserves its one-dimensional eigenspaces. If all members
are scalar, any orthonormal basis works. For an available $k$-qubit
unitary $U$, use coordinates
\[
 \widetilde U=(V^\dagger)^{\otimes k}UV^{\otimes k}.
\]
These coordinates diagonalize every full-rank available one-qubit map.
We use them to study the coefficients of $U$ and its contractions.

Retain one input $j$ and one output $i$ of $U$, and contract all other
ports with variable product states and effects. Express the remaining
one-qubit map as a $2\times2$ matrix $T$ in the basis $V$. Its entries are
separately homogeneous of degree one in each boundary coordinate pair.
On the product of the four-ray grids, this map is available: prepare
the other inputs, apply $U$, and contract the other outputs with the
chosen effects. These operations form an acyclic gadget. Its matrix $T$
is either singular or diagonal in the common basis. Consequently
\begin{equation}\label{res:minor-identities}
 (\det T)T_{01}=0,
 \qquad (\det T)T_{10}=0
\end{equation}
on the entire grid. The input grid has coordinates $V^\dagger s$ and
the effect grid has coordinates $s^\dagger V$. Each consists of four
distinct rays. The expressions in \eqref{res:minor-identities} have
degree three in each boundary pair. Lemma~\ref{res:ray-interpolation}
makes both expressions polynomial identities. State and effect
coordinates are independent polynomial variables. Conjugation enters
through the fixed values defining the adjoint-effect grid.

If the polynomial $\det T$ is nonzero, the polynomial ring has no
zero divisors, so both off-diagonal entries of $T$ vanish identically.
Their coefficients are the entries of $\widetilde U$ with unequal
retained input and output bits. Hence in this case
\begin{equation}\label{res:bit-matching}
 \widetilde U_{a,b}\ne0\quad\Longrightarrow\quad a_i=b_j.
\end{equation}
This conclusion also applies when $k=1$, where there are no contracted
pairs and \eqref{res:minor-identities} are constant identities.

We next identify the input-output pairs with nonzero determinant
polynomial and show that they form a matching of the wires. Encode
$\widetilde U$ by the multiaffine coefficient polynomial
\[
 f(x_1,\ldots,x_k,y_1,\ldots,y_k)
   =\sum_{a,b\in\{0,1\}^k}\widetilde U_{a,b}
      \prod_{i=1}^k y_i^{a_i}\prod_{j=1}^k x_j^{b_j}.
\]
Its factors on disjoint variable sets describe tensor factors of
$\widetilde U$. For a fixed pair $x_j,y_i$, write
$f=\alpha+\beta x_j+\gamma y_i+\delta x_jy_i$, with the
coefficients depending on the other variables. Then
\[
 ff_{x_jy_i}-f_{x_j}f_{y_i}=\alpha\delta-\beta\gamma.
\]
The right side is precisely the determinant polynomial of the
corresponding contraction, after specializing the other homogeneous
pairs to $(1,z)$. A nonzero homogeneous polynomial cannot become
identically zero under this dehomogenization, because it has a fixed
degree in each pair. Therefore Lemma~\ref{res:ufd} identifies exactly
the input-output pairs for which the determinant polynomial is
nonzero, and for each such pair \eqref{res:bit-matching} holds.

Every input and output variable occurs in $f$: a missing input
variable would make all columns with its bit equal to one zero, and
a missing output variable would make the corresponding rows zero.
Both contradict unitarity. Unitarity also requires each nonconstant
irreducible factor to contain both input and output variables.
To prove this, suppose a factor involves only a nonempty set $S$ of
input variables. Lemma~\ref{res:ufd} then makes the tensor factor
across those ports and all remaining ports. After grouping
columns according to $S$, its matrix would have the form of one row
vector tensored with a matrix having $2^{k-|S|}$ columns. Its rank
would be at most $2^{k-|S|}<2^k$, a contradiction. The same argument
with rows excludes factors containing only output variables.

The bit-matching constraint now limits each factor to one input and
one output. Suppose instead that a factor contains two inputs $x_j,x_m$
and an output $y_i$. The mixed derivative test applies to both pairs
$(x_j,y_i)$ and $(x_m,y_i)$, so every nonzero coefficient obeys
$b_j=a_i=b_m$. All columns with $b_j\ne b_m$ would be zero.
Similarly, a factor with two outputs and one input would force zero
rows. Every nonconstant factor therefore contains exactly one input
variable and one output variable. Their variables are disjoint and
together exhaust all input and output variables.

Each resulting input-output factor gives a diagonal one-qubit matrix.
Indeed, \eqref{res:bit-matching} forbids the separate
$x_j$ and $y_i$ terms, leaving $c+dx_jy_i$. To see that this
restriction on the whole coefficient tensor also restricts the factor,
fix any nonzero coefficient in the product of the other factors.
Their variables are disjoint, so multiplying that coefficient introduces
no cancellation. Both $c$ and $d$ must be nonzero, since the resulting
one-qubit factor must have rank two for $\widetilde U$ to be invertible.
Thus $\widetilde U$ is a tensor product of invertible diagonal one-qubit
matrices, followed by a wire permutation.

Unitarity normalizes these local matrices. In the tensor product of
their Gram matrices, varying only one local basis index shows that
the two diagonal entries of each local Gram matrix agree. Hence each
local factor is a positive scalar times a unitary, and the product
of the positive scalars is one. Absorb any overall phase into one
local unitary. Returning to the original coordinates preserves the
product-permutation form, proving the proposition.
\end{proof}

\subsection{Construction and effective search}
\label{res:construction-section}

\begin{proof}[Proof of Theorem~\ref{res:coherent}]
Lemma~\ref{res:product-test} supplies an available two-qubit state with
invertible coefficient matrix $A$. If $A=A^T$, the state and its adjoint are
already the required available resources. If $AA^\dagger$ is
nonscalar, apply Lemma~\ref{calc:symmetrize}. It remains to treat
the case $AA^\dagger=cI$ with $c>0$ and $A^T\ne A$. Normalize by
$\sqrt c$ so that $A$ is unitary.

To apply the rigidity result in the commuting case, we first construct
four distinct available one-qubit rays. If $A$ is nonmonomial, all its
entries are nonzero: in a $2\times2$ unitary,
one zero entry forces the opposite entry to vanish as well, giving
a monomial matrix. Pinning the second output of $\ket A$ to the two
computational-basis effects therefore gives two coherent orthogonal
columns. Their rays are distinct from one another and from both
computational rays. If $A$ is monomial, let $s$ be the coherent
state in the hypothesis. Contracting the second output against
$s^\dagger$ gives $A\overline s=J_As$, which is coherent because a
monomial matrix only permutes and rescales coordinates. Its ray
cannot equal that of $s$: a fixed ray of $J_A$ would make $A$
symmetric by Lemma~\ref{res:antiunitary}. Thus
$\ket0,\ket1,s,J_As$ give four distinct rays in this case as well.
All their adjoint effects are available over $\Gamma$.

If there is a full-rank available one-qubit map $T$ that is not
scalar-unitary, then $TA$ is the invertible coefficient matrix of an
available two-qubit state, and
\[
 (TA)(TA)^\dagger=TT^\dagger
\]
is nonscalar. Lemma~\ref{calc:symmetrize} again gives the result,
jointly with its adjoint. Otherwise every full-rank available one-qubit
map is scalar-unitary. If two such maps do not commute, normalize
them to unitary matrices $R,S$. The normalization uses known
algebraic scalars and preserves noncommutation, so
Proposition~\ref{res:noncommuting} applies. If all these maps
commute, Proposition~\ref{res:four-ray} makes every available
unitary product-permutation, contradicting the assumed gate in
$\G$. These cases exhaust the possibilities.

The preceding cases give a terminating search for a construction.
First perform the finite input and contraction search in
Lemma~\ref{res:product-test}. Test its two-qubit output for symmetry
and for unequal Schmidt coefficients. If neither holds, enumerate
one-qubit gadgets over $\Gamma$ by complete encoded description length,
including their wire counts, gate placements, open ports, and basis
conditions. Compute each resulting $2\times2$ matrix exactly.
Test every full-rank matrix for scalar-unitarity, retaining normalized
unitaries and testing all pairs for noncommutation. The preceding
contradiction guarantees that either a full-rank map that is not
scalar-unitary or a noncommuting pair is eventually found. In the latter case,
use the terminating word search in
Proposition~\ref{res:noncommuting}.

All tests are exact tests on algebraic numbers. The gadgets, spectral
projectors, normalizations, and filters in the resulting certificate
are fixed once the library is fixed. Their joint use, including
adjoints, has polynomial exact Turing cost by the simultaneous
replacement lemmas and Lemma~\ref{calc:symmetrize}. Finally
Lemma~\ref{calc:inverse} eliminates the temporary original gate
adjoints. This proves the required joint oracle-admissibility over
$\G$ and the effectivity assertion.
\end{proof}

\subsection{Resource completeness}
\label{res:completeness-section}

\begin{theorem}[Symmetric-resource completeness]\label{res:complete}
Let $\G$ be a finite algebraic unitary qubit library. Suppose that
some gate is nonmonomial and some gate entangles a product input.
Then a symmetric unitary algebraic matrix $A$ can be computed
such that $\ket A$ and its adjoint effect are jointly
oracle-admissible over $\G$. A unitary algebraic factorization $A=SS^T$
can also be computed, and for every such unitary factorization, with
$\Gamma=\G\cup\G^\dagger$,
\begin{equation}\label{res:complete-equivalence}
 \Amp(\G)\equiv_T^{\mathrm{poly}}
       \Holant\bigl(\mathcal H(\Gamma,S)\bigr).
\end{equation}
The port-permutation closure of $\mathcal H(\Gamma,S)$ is
conjugate-closed and has the same Holant complexity.
Every final oracle query is an acyclic circuit over the original
library with computational-basis inputs and outputs.
\end{theorem}
\begin{proof}
Apply Theorem~\ref{seed:alternative} to a nonmonomial gate. Its
symmetric invertible state outcome gives an initial symmetric pair,
together with the adjoint gadget. Its invertible state with unequal
Schmidt coefficients gives a symmetric pair through
Lemma~\ref{calc:symmetrize}. In the remaining outcome it supplies
an available coherent one-qubit state. The entangling gate is not
product-permutation by Lemma~\ref{pre:product}, so
Theorem~\ref{res:coherent} supplies a symmetric pair in that case.
All pairs are jointly oracle-admissible with their adjoints over $\G$.

Apply Lemma~\ref{calc:unitary-pair} to this initial pair. It produces
the symmetric unitary matrix $A$ in the statement, preserves joint
admissibility with the adjoint, and computes a unitary algebraic $S$
with $A=SS^T$. Proposition~\ref{calc:conjugate-transfer} gives
\eqref{res:complete-equivalence} and the conjugate-closed permutation
extension.

All constructions use algebraic numbers effectively. The seed
construction is finite. The positive-certificate searches in
Theorem~\ref{res:coherent} terminate by the alternatives proved
above. Once a certificate is fixed, all gadget sizes and algebraic
constants depend only on $\G$.

To bound the cost of the composed reduction, remove its formal
resources in reverse order of construction. First remove the positive
filter that makes the symmetric pair unitary. Next remove any spectral
replacement that symmetrizes a nonmaximal two-qubit state. Then remove any filter
obtained from an infinite-order local unitary. Finally remove the
original gate adjoints. Each stage has only finitely many fixed
resource types and handles all their occurrences simultaneously.
The number of stages is constant for the fixed library. Their
composition therefore has polynomial query count, circuit size,
and exact arithmetic bit complexity. All gadget compositions are
acyclic, and every final query uses only $\G$ and global
computational-basis boundaries. This proves all assertions of
Theorem~\ref{thm:resources}.
\end{proof}

The branchwise constructions above also give a single certificate
search for Theorem~\ref{res:complete}. Enumerate finite tuples of gadget
descriptions over $\Gamma$, including their wire counts, ordered ports,
and basis boundaries. Test each tuple for one of three certificates:
\begin{enumerate}
\item an invertible symmetric two-qubit state;
\item an invertible two-qubit state with unequal Schmidt coefficients;
\item a maximally entangled two-qubit state together with a full-rank
  scalar-unitary one-qubit map whose eigenvalue ratio is not a root of unity.
\end{enumerate}
The first certificate supplies the symmetric pair, and the second uses
Lemma~\ref{calc:symmetrize}. The third uses the spectral filter from
the infinite-group case of Proposition~\ref{res:noncommuting} to obtain
the second certificate as an oracle-admissible pair. The proof of
Theorem~\ref{res:complete} guarantees that one of these certificates
exists. Its conditions are decidable by exact algebraic tests, including
the root-of-unity test in Proposition~\ref{res:noncommuting}.
Lemma~\ref{calc:unitary-pair} then normalizes the pair. This gives an
effective reduction certificate for every library satisfying the
hypotheses of Theorem~\ref{res:complete}.

\section{Holant signatures and algebraic recognition}
\label{app:holant}

The circuit reduction of Theorem~\ref{thm:resources} gives, after
including port permutations, a conjugate-closed signature set containing
nonzero unary signatures. This appendix derives the Holant criterion
for such sets from the full dichotomy of Liu, Meng, and Wang~\cite{LMW26}.
Section~\ref{hol:four} derives the four structural conditions used in
the circuit classification. Section~\ref{hol:encoding} expresses them
by finite polynomial formulas, which decide membership for algebraic
data, produce algebraic witnesses, and support the parameter-family analysis
in Appendix~\ref{app:parameters}.
Appendix~\ref{app:normal} determines which of these conditions can hold
for the gate and boundary tensors of a unitary library. Together with
the circuit reduction, this yields the algebraic dichotomy proved in
Section~\ref{sec:normal}. Appendix~\ref{app:coefficients}
then extends it to arbitrary fixed complex entries by algebraic
specialization on the hard side and symbolic simulation on the tractable side.

\subsection{Signatures and the four tractable families}
\label{hol:classes}

A signature of arity \(r\) is a function
\(f:\mathbb F_2^r\longrightarrow\mathbb C\), identified with its
coefficient tensor in \((\mathbb C^2)^{\otimes r}\). Its support is
\(\operatorname{supp}(f)=\{x:f(x)\ne0\}\). A nullary signature is a
scalar. For a fixed finite signature set \(\mathcal F\), an instance
of \(\Holant(\mathcal F)\) is a finite graph whose vertices carry
signatures of the corresponding degrees, with an ordering of each
vertex's ports. The value is
\[
 \sum_{\sigma:E\to\{0,1\}}
       \prod_{v\in V}f_v\bigl(\sigma|_{E(v)}\bigr).
\]
Parallel edges and self-loops are allowed. Each edge identifies two
indices and sums their common value. This contraction is bilinear.
Complex conjugation is a separate operation. For the computational
problems in this appendix, the fixed signatures have algebraic entries.
We define the structural predicates on arbitrary complex entries so
that they can also be used in parameter spaces.

For an invertible \(2\times2\) matrix \(B\), write
\(Bf=B^{\otimes r}f\). The tensor closure
\(\langle\mathcal C\rangle\) of a signature class consists of tensor
products of its members, up to a port permutation and an overall
scalar. It includes zero signatures and the nullary empty product.
Let \(\mathscr{T}\) be the class of unary and binary signatures. The
condition \(\mathcal F\subseteq\langle\mathscr{T}\rangle\) means
that all tensors split into components with at most two ports.

The other three classes are as follows.
\begin{enumerate}[label=\textup{(\alph*)},leftmargin=2em]
\item The \emph{affine class} \(\mathscr{A}\) consists of signatures
  of the form
  \[
   f(x)=\lambda\mathbf 1[x\in V]i^{Q(x)},\qquad
   Q(x)=c+\sum_{j=1}^r a_jx_j
       +2\sum_{1\le j<\ell\le r}b_{j\ell}x_jx_\ell\pmod4,
  \]
  where \(V\subseteq\mathbb F_2^r\) is an affine subspace,
  \(\lambda\in\mathbb C\), \(c,a_j\in\mathbb Z_4\), and
  \(b_{j\ell}\in\mathbb F_2\). The nonzero tensors in this class,
  up to normalization and overall phase, are stabilizer states.
\item The \emph{product class} \(\mathscr{P}\) consists of ordinary
  products of arbitrary unary signatures and binary equality or
  disequality constraints on the variables, multiplied by an overall
  scalar. Factors can share variables. Thus this class describes a
  constraint factorization, rather than necessarily a product state.
\item Put \(T=\operatorname{diag}(1,e^{\pi i/4})\). The
  \emph{local-affine class} \(\mathscr{L}\) consists of signatures
  satisfying
  \[
   \left(\bigotimes_{j=1}^rT^{\alpha_j}\right)f\in\mathscr{A}
       \qquad\text{for every }\alpha\in\operatorname{supp}(f).
  \]
  Each support word \(\alpha\) specifies a local phase correction.
  Every corrected signature specified in this way must belong to
  \(\mathscr{A}\).
\end{enumerate}
All three classes include zero and all nullary scalars. The
stabilizer interpretation of $\mathscr{A}$ and its relation to Clifford
unitaries are recalled in Section~\ref{norm:affine-section}.
Section~\ref{hol:encoding} uses the definitions above to test membership
and transformability effectively.

For \(\mathcal C\in\{\mathscr{P},\mathscr{A},\mathscr{L}\}\), call a
signature set \(\mathcal F\) \(\mathcal C\)-\emph{transformable} if
there is one \(B\in\mathrm{GL}_2(\mathbb C)\) such that
\begin{equation}
 B^TB\in\mathcal C,\qquad
 (B^{-1})^{\otimes r}f\in\mathcal C
       \quad\text{for every arity-\(r\) }f\in\mathcal F.
 \label{hol:transformability}
\end{equation}
A binary tensor is identified here with its \(2\times2\) coefficient
matrix. The tensor \(B^TB\) represents the equality edge in the
transformed coordinates: if two physical legs are \(Bu,Bv\), their
contraction is \(u^TB^TBv\). Thus the same $B$ transforms every vertex
signature, and $B^TB$ supplies the edge signature in those coordinates.
This agrees with
\cite[Lemma~2.10]{LMW26} for the ordinary equality edge \(E=I_2\).

Define \(\ReTract(\mathcal F)\) to be the disjunction
\begin{enumerate}[label=\textup{(\arabic*)},leftmargin=2.2em]
\item \(\mathcal F\subseteq\langle\mathscr{T}\rangle\);
\item \(\mathcal F\) is \(\mathscr{P}\)-transformable;
\item \(\mathcal F\) is \(\mathscr{A}\)-transformable;
\item \(\mathcal F\) is \(\mathscr{L}\)-transformable.
\end{enumerate}
Every condition concerns the whole signature set. Zero and nullary
signatures can be deleted without changing \(\ReTract\). In an evaluation
instance, a zero signature makes the answer zero, while nullary
signatures contribute a known product of scalar factors.

\subsection{Conjugation and the Holant dichotomy}
\label{hol:four}

For a signature \(f\), let \(\overline f(x)=\overline{f(x)}\).
A signature set is \emph{conjugate-closed} when it contains
\(\overline f\) for every member \(f\). Write
\(\operatorname{Perm}(\mathcal F)\) for the finite set obtained by
including every permutation of the ports of every member of a fixed
finite \(\mathcal F\).

\begin{lemma}[Port permutations]
\label{hol:port-permutations}
For every fixed finite signature set \(\mathcal F\),
\[
 \Holant(\operatorname{Perm}(\mathcal F))
       \equiv_T^{\poly}\Holant(\mathcal F),\qquad
 \ReTract(\operatorname{Perm}(\mathcal F))
       \Longleftrightarrow\ReTract(\mathcal F).
\]
The evaluation equivalence uses one query with only a relabelling of
ports. In particular, if the conjugate of each member of \(\mathcal F\)
is a port permutation of a member, then
\(\operatorname{Perm}(\mathcal F)\) is conjugate-closed.
\end{lemma}
\begin{proof}
A vertex labelled by a permuted signature is replaced by the original
signature and the corresponding ordering of its incident ports. This
preserves every summand in the contraction, including at self-loops.
The reverse reduction follows from containment. Each of
\(\mathscr{A},\mathscr{P},\mathscr{L}\) and
\(\langle\mathscr{T}\rangle\) is invariant under port permutations.
Moreover, a common tensor-power change of coordinates commutes with
such a permutation, so the same \(B\) in~\eqref{hol:transformability}
works for all permuted signatures. This proves the predicate
equivalence. Finally, entrywise conjugation commutes with port
permutation, proving the last assertion.
\end{proof}

\paragraph{Specializing the full dichotomy.}
Let \(\LMWTract\) denote the seven-case criterion of
Liu, Meng, and Wang~\cite[Section~2.13.3]{LMW26}. Its cases
(3), (5), (6), and (7) correspond, respectively, to conditions
(1), (3), (2), and (4) of \(\ReTract\).
In particular, \(\ReTract\) implies \(\LMWTract\) for every finite algebraic
signature set. We prove the converse under the two hypotheses supplied
by the circuit reduction.

\begin{specializationlemma}[Conjugate-closed signature sets with a unary]
Let \(\mathcal F\) be a finite conjugate-closed set of algebraic Boolean
signatures containing a nonzero unary. Then
\(\LMWTract(\mathcal F)\) holds if and only if \(\ReTract(\mathcal F)\) holds.
\end{specializationlemma}
\begin{proof}
Delete zero and nullary signatures, as in both criteria. We handle the
three remaining cases of \cite[Section~2.13.3]{LMW26}: the nonzero unary
excludes case (1), case (2) gives a common affine or product-class
transform, and case (4) forces factors of arity at most two.
To use conjugate closure in these cases, choose coordinates in which
conjugation also complements every bit. For a signature \(f\) of
arity \(r\), write
\[
 K=\frac{1}{\sqrt{2}}\begin{pmatrix}1&1\\ i&-i\end{pmatrix},
 \qquad X=\begin{pmatrix}0&1\\1&0\end{pmatrix},
 \qquad \widehat f=(K^{-1})^{\otimes r}f .
\]
Since \(\overline K=KX\),
\[
 \widehat{\overline f}
   =X^{\otimes r}\overline{\widehat f}.
\]
Conjugate closure of \(\mathcal F\) therefore makes the transformed
set contain the partner
\(g^\star(x)=\overline{g(\mathbf 1\oplus x)}\) of every
\(g=\widehat f\). Taking partners interchanges weights \(w\) and
\(r-w\). Also \(K^TK=X\), the binary disequality signature.

\emph{Common one-sided support (1).}
This case places all supported words on one common side of the
middle weight, allowing equality. The transformed nonzero unary has a
supported bit \(b\), and its partner has the supported bit \(1-b\).
Their weights lie strictly on opposite sides of \(1/2\), ruling out
this case.

\emph{Single-weight signatures (2).}
Case (2) tests signatures after completing them to even arity and
support at exactly half that arity. For a signature \(g\) of arity
\(r\) supported at weight \(w\), put \(m=\max\{w,r-w\}\).
Form its completion \(h\) by appending \(2w-r\) bits pinned to zero
if \(2w\ge r\), or \(r-2w\) bits pinned to one otherwise.
The new arity is \(2m\) and the support weight is \(m\).
These tensor completions, and the slices recovering \(g\) below,
serve to test membership in the signature classes.
Completing \(g^\star\) gives \(h^\star\), since complementing all
bits also complements each appended pin. Case (2) imposes two conditions
on this set of completions~\cite[Definition~2.8 and Eq.~(10)]{LMW26}.

First, for any three support words of a completion, their xor lies in
the support or lies strictly on a prescribed side of the middle weight.
The side is common to all completions. If a triple for \(h\) had xor
\(\delta\) outside its support, the complemented triple for \(h^\star\)
would have xor \(\mathbf 1\oplus\delta\) outside its support. These
two xors cannot lie strictly on the same side. Hence the support of
each \(h\) is closed under ternary xor. Choose a support word \(a\).
The translated support contains zero and is closed under binary xor:
for support words \(x,y\),
\((x\oplus a)\oplus(y\oplus a)=(x\oplus y\oplus a)\oplus a\)
lies in that translated support. Thus the translated support is linear,
and the support of \(h\) is affine.

The second condition controls restrictions to opposite bits on paired
ports. To recover the full signature from such a restriction, we need
one pairing satisfied by every word in its support. An affine support
at the middle weight admits such a pairing. To construct it,
parametrize the support by \(t\in\mathbb F_2^d\), writing
its coordinates as \(x_j(t)=a_j\cdot t+b_j\). The middle-weight
identity gives
\[
 0=\sum_{j=1}^{2m}(-1)^{x_j(t)}
  =\sum_{a\in\mathbb F_2^d}
      \left(\sum_{j:\,a_j=a}(-1)^{b_j}\right)(-1)^{a\cdot t}
 \qquad\text{for every }t .
\]
By orthogonality of the characters of \(\mathbb F_2^d\), each inner
sum vanishes. For each \(a\), pair ports with \(a_j=a\) and \(b_j=0\)
bijectively with those with \(a_j=a\) and \(b_j=1\). This also pairs
the constant coordinates.

More precisely, case (2) supplies one class
\(\mathcal C\in\{\mathscr{A},\mathscr{P}\}\) for the entire signature set.
An oriented pairing \(\pi=((p_j,q_j))_{j=1}^m\) partitions all
\(2m\) ports into pairs and orders the two ports within each pair.
For every such \(\pi\), the restriction \(h^\pi\) obtained by setting
\(x_{p_j}=z_j,\ x_{q_j}=1-z_j\) belongs to \(\mathcal C\).
For each \(h\), choose the pairing just constructed, with either
orientation for each pair. This pairing may depend on \(h\), while
\(\mathcal C\) remains common to all completions. We recover
\[
 h(x)=h^\pi(x_{p_1},\ldots,x_{p_m})
          \prod_{j=1}^{m} X(x_{p_j},x_{q_j}).
\]
If \(x\) satisfies every paired disequality, the restriction on the
right gives \(h(x)\). If any pair violates disequality, the right side
is zero, and so is \(h(x)\), because every support word satisfies all
these constraints. Thus the identity holds for every \(x\).
For \(\mathscr{A}\) the product imposes affine equations while
preserving the allowed quadratic phase. For \(\mathscr{P}\) it adds
binary disequality factors. Thus \(h\in\mathcal C\) in either case.
Fixing the appended bits to their prescribed values preserves both
classes and recovers \(g\). The same class contains every \(\widehat f\)
and the edge \(K^TK=X\). Taking \(B=K\) in
\eqref{hol:transformability} gives \(\ReTract(\mathcal F)\).

\emph{Matching signatures (4).}
A tensor-prime factor is a nonzero tensor that cannot be split into
two factors on nonempty disjoint sets of ports. In one orientation of
this case, every such factor of every \(\widehat f\) is supported on
words of weight at most one
\cite[Remark~2.9]{LMW26}. Taking partners preserves tensor products and
indecomposability, since it acts by conjugation and a bit flip on every
port and is an involution. If \(q\) is a nonzero prime factor of arity
\(s\), then \(q^\star\) is a prime factor of the partner signature.
For any supported word of \(q\) of weight \(w\), we therefore have
\(w\le1\) and \(s-w\le1\), so \(s\le2\). The other orientation gives
the same bounds after flipping all bits. Applying \(K\) back on each
port preserves the factorization, and hence
\(\mathcal F\subseteq\langle\mathscr{T}\rangle\).
\end{proof}

\begin{corollary}[Holant dichotomy for the circuit reduction]
\label{hol:dichotomy}
Let \(\mathcal F\) be a fixed finite set of algebraic complex-valued
Boolean signatures. If \(\ReTract(\mathcal F)\) holds, then
\(\Holant(\mathcal F)\in\FP\). If \(\mathcal F\) is conjugate-closed,
contains a nonzero unary, and fails \(\ReTract\), then
\(\Holant(\mathcal F)\) is \(\#\mathrm P\)-hard.
\end{corollary}
\begin{proof}
The first implication follows from \(\ReTract\Rightarrow\LMWTract\) and
the tractable direction of \cite[Theorem~1.1]{LMW26}. Under the
hypotheses of the second implication, the preceding lemma gives
\(\neg\LMWTract(\mathcal F)\). The hardness direction of the same
theorem gives \(\#\mathrm{SAT}\leq_T^{\poly}\Holant(\mathcal F)\).
\end{proof}

Proposition~\ref{calc:conjugate-transfer} supplies the conjugate-closed
signature set used in the hardness implication. To decide whether its
structural conditions hold, we next express $\ReTract$ by finite
polynomial systems. This recognition step also applies to general
signature sets.

\subsection{Finite algebraic encoding}
\label{hol:encoding}

An existential polynomial formula is a finite existential formula whose
atoms are polynomial equations or non-equations over \(\Qbar\), with
quantified variables ranging over \(\Qbar\). Its Boolean combinations
encode finite choices of supports, factorizations, and common transforms.

\begin{theorem}[Algebraic recognition]
\label{hol:recognition}
For every fixed finite list of arities, \(\ReTract\) has an effectively
constructible existential polynomial formula in the signature entries.
Consequently, for a finite algebraic matrix library \(\G\), the condition
\[
 \exists S\in\mathrm{GL}_2(\Qbar)\;
       \ReTract\bigl(\mathcal H(\G,S)\bigr)
\]
is decidable, where \(\mathcal H\) is defined by
\eqref{eq:holant-language}. If the condition holds, an algebraic witness
\(S\) can be computed. No polynomial bound in the description of the
library is asserted.
\end{theorem}

\begin{proof}
Fix the arity list. We first encode membership in the individual
signature classes.

At arity \(r\), the affine class has finitely many possible affine
supports \(V\subseteq\mathbb F_2^r\) and phase polynomials \(Q\) from
its definition. For each choice, introduce a scalar variable \(\lambda\)
and impose the \(2^r\) linear equations
\[
 f(x)=\lambda\mathbf 1[x\in V]i^{Q(x)}
       \qquad(x\in\mathbb F_2^r).
\]
The finite disjunction of these systems describes \(\mathscr{A}\),
including zero.

For \(\mathscr{P}\) at positive arity \(r\), enumerate relation patterns
that assign equality, disequality, or no constraint to each unordered
pair of distinct variables. Introduce unary parameters \(u_{j,0},u_{j,1}\)
and impose, for every \(x\in\mathbb F_2^r\),
\begin{equation}
 f(x)=
 \prod_{j=1}^r u_{j,x_j}
 \prod_{\{a,b\}\in E_=}\mathbf 1[x_a=x_b]
 \prod_{\{a,b\}\in E_{\ne}}\mathbf 1[x_a\ne x_b].
 \label{hol:product-encoding}
\end{equation}
These are polynomial equations. The enumeration is complete: multiple
unary factors on one variable can be multiplied together, and the
overall scalar can be absorbed into any unary factor. Repeated
constraints are idempotent. A repeated-variable equality is redundant.
A repeated-variable disequality, simultaneous equality and disequality,
or any inconsistent relation pattern forces the zero signature. Zero
is already represented by setting a unary factor to zero. Nullary
signatures are unrestricted scalars.

Membership in \(\mathscr{L}\) is the finite conjunction over
\(\alpha\in\mathbb F_2^r\) of
\[
 f(\alpha)=0
 \quad\text{or}\quad
 \left(\bigotimes_{j=1}^rT^{\alpha_j}\right)f\in\mathscr{A}.
\]
Use the preceding affine encoding after the indicated fixed linear
substitution, with separate auxiliary variables for each occurrence.
Finite conjunctions and disjunctions of existential polynomial formulas
remain existential polynomial formulas.

To encode \(\langle\mathscr{T}\rangle\), enumerate partitions of the port
set into blocks of size at most two. Introduce the entries of a factor
on each block and equate \(f\) to their tensor product. Include an
overall scalar as an additional variable. The number and arities of
these factors are bounded by the fixed arity of \(f\), so this is a
finite disjunction of polynomial systems.

For \(\mathcal C\)-transformability, introduce matrices \(B,R_B\) with
\(BR_B=R_BB=I_2\). Impose the polynomial encodings of
\(B^TB\in\mathcal C\) and \(R_B^{\otimes r}f\in\mathcal C\) for every
arity-\(r\) member of the signature set. The same \(B,R_B\) are used
throughout. All transformed entries are polynomial expressions in
these variables and the original signature entries. This encodes
cases (2)--(4), including their equality-edge conditions.
Zero and nullary signatures impose no further restriction. Together
with the tensor-closure encoding, these formulas describe \(\ReTract\).

We next encode the existence of a common circuit basis. Introduce
independent matrix variables \(S,R_S\) with \(SR_S=R_SS=I_2\).
The entries of \(\mathcal H(\G,S)\) are polynomial expressions in the
supplied matrices, \(S\), and \(R_S\): the gate tensors have matrices
\(R_S^{\otimes k}US^{\otimes k}\), and the boundary tensors are
\(R_Se_b,S^Te_b\). Substitute these entries into the formula for
\(\ReTract\), retaining its separate transformability variables
\(B,R_B\). This gives an existential polynomial formula for the
common-basis condition.

To decide this formula for fixed algebraic input data, expand its
Boolean part into disjunctive normal form. In each branch replace
every non-equation \(p\ne0\) by
\(pt-1=0\), with a fresh variable \(t\). The coefficients lie in a
computable number field \(F\). Each branch is now a finite polynomial
system over \(F\). Compute a Gr\"obner basis of its ideal and test
whether the ideal contains \(1\). Extension to an algebraic closure
preserves whether the ideal is proper. By the weak Nullstellensatz,
the branch has a solution over \(\Qbar\) if and only if the ideal is
proper. Testing all branches decides the existential condition.

Once a branch is known to be satisfiable, extract an algebraic witness.
One may use an effective algebraic solving algorithm to obtain a solution.
Alternatively, enumerate tuples of
algebraic numbers of the required finite length and test the equations
exactly. The established existence of a solution guarantees termination.
The \(S\)-coordinates give the required common basis, and
\(SR_S=R_SS=I_2\) guarantees its invertibility. The same argument shows
that allowing complex, possibly transcendental witnesses for algebraic
input data would not enlarge the positive instances.
\end{proof}

For the parameter families in Appendix~\ref{app:parameters}, keep the
matrix entries as free variables in the recognition formula and eliminate
the auxiliary witnesses. This describes the set of matrix tuples
satisfying the structural conditions in a common basis. We also include
the all-monomial case, which
has a direct amplitude simulator, to match the full unitary classification
in Theorem~\ref{norm:reduction}. The following proposition gives this
polynomial description. Its interpretation for algebraic unitary tuples
uses the normal-form theorem proved in Appendix~\ref{app:normal}.

\begin{proposition}[Constructible extension]
\label{hol:constructible}
Fix the number of gates in the library and the dimensions of their matrices. Extend
\[
 \Psi(\G)=
 \text{\(\G\) is all-monomial}
 \quad\text{or}\quad
 \exists S\in\mathrm{GL}_2(\mathbb C)\;
       \ReTract\bigl(\mathcal H(\G,S)\bigr)
\]
to arbitrary complex matrix tuples, without imposing unitarity.
Its positive locus is an effectively constructible set over \(\Qbar\):
it has an effectively computable Boolean description in finitely many
polynomial equalities in the matrix entries. On algebraic unitary tuples,
\(\Psi\) is exactly the disjunction of the three structural conditions
in Theorem~\ref{thm:intro-dichotomy}.
\end{proposition}

\begin{proof}
For a fixed matrix size, monomiality is a finite disjunction over
permutations: the entries off the permutation support are zero and
the entries on that support are nonzero. The other clause has the
existential polynomial encoding from Theorem~\ref{hol:recognition}.
That encoding applies to arbitrary matrix entries. Elimination over
algebraically closed fields gives a Boolean combination of polynomial equations
over the same algebraic coefficient field. Equivalently, replace
non-equations by inverse variables, project the resulting finite
polynomial systems, and compute their constructible images. These
standard elimination procedures are effective, so the Boolean
description can be computed from the prescribed matrix dimensions.

For algebraic matrix entries, the existence of a complex witness is
equivalent to the existence of an algebraic witness, by the last
paragraph of Theorem~\ref{hol:recognition}. On unitary tuples the
equivalence with the three quantum structures follows from
Theorem~\ref{norm:reduction}.
\end{proof}

\section{Unitary normal forms}
\label{app:normal}

\providecommand{\Mono}{\mathsf{Mono}}
\providecommand{\ProdPerm}{\mathsf{ProdPerm}}
\providecommand{\DC}{\mathsf{DC}}
\providecommand{\AffWit}{\mathsf{AffWit}}

We now classify unitary gate libraries whose tensors satisfy one of
the four conditions of $\ReTract$. The input/output partition of each
gate tensor lets us use matrix invertibility and unitarity to restrict
these tensor structures. For an invertible algebraic \(S\), recall the
signature set
\[
 \mathcal H(\G,S)=
 \{f_{(S^{-1})^{\otimes k}US^{\otimes k}}:
                   U\in\G\text{ acts on }k\text{ qubits}\}
 \;\cup\;\{S^{-1}e_b,S^Te_b:b=0,1\}.
\]
The predicate $\ReTract$ is defined in Section~\ref{hol:classes}.
We analyze it for every invertible algebraic $S$ and every finite
algebraic unitary library $\G$. Theorem~\ref{hol:recognition} ensures
that any further transformability witness can also be chosen algebraic.

Write \(\Mono(\G)\) when every gate is monomial, and
\(\ProdPerm(\G)\) when every gate is a tensor product of one-qubit
unitaries followed by a wire permutation. By
Lemma~\ref{pre:product}, the latter condition is equivalent to every gate
preserving all product states. Write \(\DC(\G)\) when there is one
algebraic phase \(z\), \(|z|=1\), such that, for \(D=\operatorname{diag}(1,z)\),
every \(k\)-qubit gate satisfies
\[
 U=D^{\otimes k}C_U(D^\dagger)^{\otimes k},
 \qquad C_U\text{ a Clifford unitary}.
\]
As in the main theorem, Clifford unitaries include arbitrary global
phases. In this display all entries of \(C_U\) are algebraic.
For algebraic gates, allowing an arbitrary complex phase \(z\) gives the
same condition: the entry-ratio equations in
Theorem~\ref{norm:recognition} either force \(z\) to be algebraic or
permit the choice \(z=1\).

The next theorem applies to all algebraic unitary libraries, with the
monomial case retained explicitly. Under the hypotheses of
Theorem~\ref{thm:normal}, the nonmonomial and entangling gates exclude
\(\Mono\) and \(\ProdPerm\), leaving the common diagonal Clifford frame.

\begin{theorem}[Unitary normal forms]
\label{norm:reduction}
For every finite algebraic unitary qubit library \(\G\),
\[
 \begin{split}
 &\Mono(\G)\ \lor\
   \exists S\in\mathrm{GL}_2(\Qbar)\;
                  \ReTract\bigl(\mathcal H(\G,S)\bigr)\\
 &\hspace{3em}\Longleftrightarrow\
   \Mono(\G)\ \lor\ \ProdPerm(\G)\ \lor\ \DC(\G).
 \end{split}
\]
\end{theorem}

An invertible input/output flattening severely restricts a factorization
of the coefficient tensor: factors of rank at most two can only connect
one input to one output. The local-affine case instead has so few
support entries that invertibility forces a deterministic basis map.
The affine case permits entanglement. There the boundary tensors
restrict a common change of coordinates, and unitarity restricts the
inner product in those coordinates. We use both restrictions to obtain
either product preservation or a common diagonal Clifford frame.

\subsection{Tensor factors and product evolution}
\label{norm:tensor-section}

\begin{lemma}[Balanced factors]
\label{norm:balanced}
Suppose that the coefficient tensor of an invertible \(k\)-qubit linear
map factors over a partition of its \(2k\) ports. A factor containing
\(p\) output ports and \(q\) input ports has \(p=q\), and its matrix across
this cut is invertible. If every non-nullary factor has matrix rank at
most two, every such factor has one input and one output. If the full
map is unitary, it is then product-permutation.
\end{lemma}

\begin{proof}
Permute the output ports and input ports separately. The full matrix
becomes a tensor product of the factor matrices. Let their ranks be
\(r_j\), and their respective output and input port counts be \(p_j,q_j\).
Then
\[
 2^k=\prod_j r_j,\qquad
 r_j\le 2^{p_j},\quad r_j\le 2^{q_j},\qquad
 \sum_jp_j=\sum_jq_j=k.
\]
Equality in both product bounds forces
\(r_j=2^{p_j}=2^{q_j}\) for every \(j\). Hence \(p_j=q_j\), and each
factor matrix is square and invertible. If its rank is at most two,
then \(p_j=q_j\le1\). A non-nullary factor consequently has
\(p_j=q_j=1\).

In this case the map is \((\bigotimes_j L_j)P\) for a wire permutation
\(P\). Unitarity gives \(\bigotimes_jL_j^\dagger L_j=I\).
Fix eigenvectors for all factors except one and compare the resulting
eigenvalues. Every \(L_j^\dagger L_j\) must be a scalar matrix
\(c_jI\), with \(c_j>0\) and \(\prod_jc_j=1\).
Replacing \(L_j\) by \(c_j^{-1/2}L_j\) makes all factors unitary
without changing their tensor product.
\end{proof}

\begin{lemma}
\label{norm:tensor-cases}
Suppose that \(S\) is invertible. If
\(\mathcal H(\G,S)\subseteq\langle\mathscr{T}\rangle\), or if
\(\mathcal H(\G,S)\) is \(\mathscr{P}\)-transformable, then
\(\ProdPerm(\G)\) holds.
\end{lemma}

\begin{proof}
In the tensor-closure case, each transformed gate tensor is a product
of factors of arity at most two. Every factor has rank at most two
across its input/output cut. Lemma~\ref{norm:balanced} therefore
leaves only binary factors, each connecting one input to one output.

In the product-transformable case, let \(B\) be the common
transformation. After applying \(B^{-1}\) to every port, each gate
tensor belongs to \(\mathscr{P}\). Split its equality/disequality
constraint graph into connected components and assign every unary
factor to its component. A component has at most two support words:
the value at one vertex determines all other values in that component.
Its matrix consequently has rank at most two across any input/output
cut. Lemma~\ref{norm:balanced} then forces each component to connect
exactly one input to one output.

In both cases, undoing the invertible transformations on the individual
ports preserves the decomposition into binary input/output factors.
Thus the original \(U\) has such a decomposition. The last part of
Lemma~\ref{norm:balanced} uses its unitarity to normalize the local
factors, proving \(\ProdPerm(\G)\).
\end{proof}

\subsection{Sparse support and deterministic basis evolution}
\label{norm:support-section}

For the local-affine case, we first bound the support of a signature.
Applied to the boundary tensors, this bound restricts the change of
coordinates. Applied to a gate tensor, it forces one nonzero entry per
row and column. For a binary linear subspace \(C\subseteq\mathbb F_2^r\),
write
\[
 C^\perp=\{v\in\mathbb F_2^r:v\cdot c=0
                   \text{ for every }c\in C\},
\]
where the dot product is computed in \(\mathbb F_2\).

\begin{lemma}[Support of a local-affine signature]
\label{norm:local-support}
If \(0\ne f\in\mathscr{L}\) has arity \(r\), its support is an affine
space \(t+C\), where \(C\subseteq C^\perp\) for the binary dot product.
Consequently \(\dim C\le r/2\). In particular, a nonzero unary member
of \(\mathscr{L}\) is a computational-basis ray.
\end{lemma}

\begin{proof}
Put \(\omega=e^{\pi i/4}\). For every
\(\alpha\in\operatorname{supp}(f)\), the signature
\[
 f_\alpha(x)=\omega^{\sum_j\alpha_jx_j}f(x)
\]
belongs to \(\mathscr{A}\). These transformations preserve support, so
that support has the form \(t+C\). Ratios of nonzero values of an
affine signature are fourth roots of unity. Compare two corrections
\(f_\alpha,f_\beta\) to cancel the common ratio
\(f(x\oplus d)/f(x)\) and isolate their phase difference. For
\(\alpha,\beta,x\in t+C\) and \(d\in C\),
\[
 \frac{f_\alpha(x\oplus d)/f_\alpha(x)}
      {f_\beta(x\oplus d)/f_\beta(x)}
 =
 \omega^{(\alpha-\beta)\cdot((x\oplus d)-x)}
 \in\{1,-1,i,-i\}.
\]
The exponent, with the differences interpreted over the integers,
is even. Its parity is \((\alpha\oplus\beta)\cdot d\).
Every \(c\in C\) has the form \(\alpha\oplus\beta\), so
\(c\cdot d=0\) for every \(c,d\in C\). This is
\(C\subseteq C^\perp\). Nondegeneracy of the binary dot product gives
\(2\dim C\le r\). For \(r=1\), the support has one point.
\end{proof}

\begin{lemma}[The local-affine case]
\label{norm:local-affine}
If \(\mathcal H(\G,S)\) is \(\mathscr{L}\)-transformable, then
\(\Mono(\G)\) holds.
\end{lemma}

\begin{proof}
Let \(B\) be its common local-affine transformation. The transformed
boundary signatures are the columns of
\[
 L=B^{-1}S^{-1},\qquad R=B^{-1}S^T.
\]
Lemma~\ref{norm:local-support} makes all four columns computational-basis
rays. Since both matrices are invertible, \(L\) and \(R\) are monomial.
On an input port, a tensor coordinate change acts by its transpose
on the right of the gate matrix. Thus the transformed \(k\)-qubit gate
has the invertible matrix
\[
 W=L^{\otimes k}U(R^T)^{\otimes k}\in\mathscr{L}.
\]
Its coefficient tensor has arity \(2k\), so the preceding lemma bounds
its support by \(2^k\) points. Invertibility requires at least one
nonzero entry in each of its \(2^k\) rows and columns. There are
therefore exactly \(2^k\) nonzero entries, one per row and column.
Thus \(W\) is monomial. Multiplication on either side by an invertible
monomial matrix preserves monomiality, giving the result for \(U\).
\end{proof}

\subsection{Stabilizer tensors and the common Clifford frame}
\label{norm:affine-section}

Write \(\AffWit(\G)\) if there exists an invertible algebraic \(S\)
such that \(\mathcal H(\G,S)\) is \(\mathscr{A}\)-transformable.
Thus one transformation must work for all gate and boundary tensors,
including the transformed equality edge. We use the closure and
affine--Clifford results of
Cai, Guo, and Williams~\cite[Theorem~2.4 and Lemmas~3.1 and~4.1]{CGW18}:
affine signatures are closed under tensor products, port permutations,
and contractions, and an invertible square matrix with affine
coefficient tensor is a nonzero scalar multiple of a Clifford unitary,
with the converse also holding. The scalar is retained because it
contributes to the exact amplitude.

For arity one, the definition of \(\mathscr{A}\) gives exactly the six
nonzero rays
\begin{equation}
 [1,0],\ [0,1],\ [1,1],\ [1,-1],\ [1,i],\ [1,-i].
 \label{norm:six-rays}
\end{equation}
They are the Pauli eigenstate rays.

\begin{lemma}[Absorbing the equality edge]
\label{norm:absorb}
The condition \(\AffWit(\G)\) holds if and only if there is an
invertible algebraic matrix \(C\) such that, for every \(k\)-qubit
\(U\in\G\) and \(b\in\{0,1\}\),
\begin{equation}
 (C^{-1})^{\otimes k}UC^{\otimes k}\in\mathscr{A},
 \qquad C^{-1}e_b\in\mathscr{A},\qquad C^Te_b\in\mathscr{A}.
 \label{norm:affine-witness}
\end{equation}
\end{lemma}

\begin{proof}
Let \(S\) be the circuit basis and \(B\) a common affine
transformation, chosen algebraic by Theorem~\ref{hol:recognition}.
We combine these two changes of coordinates into one similarity
transformation using the affine equality-edge tensor \(E=B^TB\).
Put \(C=SB\). Applying \(B^{-1}\) to every gate port gives the affine matrix
\[
 W=(B^{-1}S^{-1})^{\otimes k}U(SB^{-T})^{\otimes k}.
\]
The output factor is \(C^{-1}\), whereas the input factor is
\(SB^{-T}\). The edge tensor converts the latter into \(C\) because
\(SB^{-T}E=SB=C\).
Contracting \(E\) on each input leg therefore preserves affinity and gives
\[
 WE^{\otimes k}=(C^{-1})^{\otimes k}UC^{\otimes k}.
\]
The transformed input pins are already \(C^{-1}e_b\).
For each transformed output pin, contract one copy of \(E\), obtaining
\[
 EB^{-1}S^Te_b=B^TS^Te_b=C^Te_b.
\]
These contractions establish membership in \(\mathscr{A}\) by its
closure properties and give the common structural witness
\eqref{norm:affine-witness}. Conversely, choose \(S=C\)
and \(B=I\). The gate tensors and boundary pins are affine by
\eqref{norm:affine-witness}, and \(B^TB=I\) is the affine equality
signature.
\end{proof}

\begin{theorem}[Affine unitary normal form]
\label{norm:affine}
For every finite algebraic unitary qubit library,
\[
 \DC(\G)\ \Longrightarrow\ \AffWit(\G)
 \ \Longrightarrow\ \ProdPerm(\G)\lor\DC(\G).
\]
If the library contains an entangling gate, then
\(\AffWit(\G)\) is equivalent to \(\DC(\G)\).
\end{theorem}

\begin{proof}
If \(\DC(\G)\) holds, choose \(S=D\). Every transformed gate is
Clifford and hence affine, and all four boundary tensors are scalar
computational-basis pins. Taking \(B=I\) gives an affine witness.

Conversely, fix the single matrix \(C\) supplied by
Lemma~\ref{norm:absorb} for the whole library. For each \(k\)-qubit
gate \(U\), put
\[
 Q_U=(C^{-1})^{\otimes k}UC^{\otimes k}.
\]
The affine--Clifford correspondence writes \(Q_U=\lambda V_U\),
where \(V_U\) is a Clifford unitary. Similarity to \(U\) gives
\(|\det Q_U|=1\), and hence \(|\lambda|^{2^k}=1\).
Thus \(|\lambda|=1\) and \(Q_U\) itself is a Clifford unitary under
our global-phase convention. We next restrict the common coordinate
matrix \(C\) to obtain the required diagonal unitary frame or the
product-permutation alternative. Unitarity of \(U\) gives
\begin{equation}
 Q_U^\dagger(C^\dagger C)^{\otimes k}Q_U
       =(C^\dagger C)^{\otimes k}.
 \label{norm:product-metric}
\end{equation}
The matrix \(C^\dagger C\) is the physical inner product expressed
in the \(C\) coordinates. Since \(Q_U\) is unitary in the standard
inner product as well, it must commute with this product metric.
This metric records the possible nonunitary part of the combined
change of coordinates $C=SB$. The boundary tensors restrict the rows
of this same \(C\). Nonorthogonal rows will force all gates to preserve
product states. Orthogonal rows will give a Clifford basis followed by
diagonal scalings, whose positive parts we can remove using the metric.

Write the rows of \(C\) as \(r_0,r_1\). By
\eqref{norm:affine-witness}, their transposes lie on the six rays in
\eqref{norm:six-rays}. Conjugation preserves these rays, so the columns
\(r_0^\dagger,r_1^\dagger\) lie on them as well. The rows are independent
because \(C\) is invertible. Suppose first that they are nonorthogonal.
Independent, nonorthogonal Pauli eigenstate rays belong to different
Pauli axes. Let \(a,b>0\) be the squared row norms and let \(n_0,n_1\)
be the Bloch vectors of the normalized columns
\(r_0^\dagger,r_1^\dagger\). Then
\[
 C^\dagger C=r_0^\dagger r_0+r_1^\dagger r_1
   =\frac{a+b}{2}I+\frac12(a n_0+b n_1)\cdot\sigma,
 \qquad \sigma=(X,Y,Z).
\]
The vectors \(n_0,n_1\) are signed coordinate vectors on distinct axes,
so \(C^\dagger C\) has at least two nonzero Pauli components.
Lemma~\ref{lem:metric} then forces the Clifford action of each \(Q_U\)
to send each one-qubit Pauli algebra to another one-qubit Pauli algebra.
Thus every \(Q_U\) is product-permutation, with local Clifford factors.

Conjugation by the same local \(C\) preserves this tensor structure:
a wire permutation commutes with \(C^{\otimes k}\), so the local
factors are simply conjugated individually. The resulting matrix is
the original unitary \(U\). The normalization argument in
Lemma~\ref{norm:balanced} makes its local factors unitary.
Thus \(\ProdPerm(\G)\) holds.

It remains to consider orthogonal rows. They form a Pauli eigenbasis.
There are therefore a one-qubit Clifford \(V\) and an invertible
diagonal matrix \(D_0\) such that \(C=D_0V\). The row norms and phases
are included in \(D_0\). Both \(V\) and \(D_0\) are chosen once from
\(C\) and are shared by all gates. Define
\[
 R_U=V^{\otimes k}Q_U(V^\dagger)^{\otimes k}.
\]
These are Clifford unitaries, and
\(U=D_0^{\otimes k}R_U(D_0^{-1})^{\otimes k}\).
Write the diagonal polar decomposition \(D_0=LP\), with \(L\)
diagonal unitary and \(P\) diagonal positive. These matrices also
depend only on \(C\). In this basis,
\eqref{norm:product-metric} gives
\[
 R_U^\dagger(P^2)^{\otimes k}R_U=(P^2)^{\otimes k}.
\]
Thus \(R_U\) commutes with \((P^2)^{\otimes k}\) and its unique
positive square root \(P^{\otimes k}\). This commutation removes
the positive factors from the similarity transformation:
\[
 \begin{aligned}
 U&=L^{\otimes k}P^{\otimes k}R_U
        (P^{-1})^{\otimes k}(L^\dagger)^{\otimes k}\\
  &=L^{\otimes k}R_U(L^\dagger)^{\otimes k}.
 \end{aligned}
\]
Divide \(L\) by its first diagonal entry. This does not change the
conjugation and produces \(D=\operatorname{diag}(1,z)\), \(|z|=1\),
common to the entire library. All entries of \(C,D_0,V\) are algebraic,
and the normalizations use positive square roots of algebraic numbers.
Hence \(z\) is algebraic. This proves \(\DC(\G)\).

If \(\G\) contains an entangling gate, the product-permutation outcome
is impossible, proving the last assertion.
\end{proof}

\begin{proof}[Proof of Theorem~\ref{norm:reduction}]
Suppose first that \(S\) witnesses
\(\ReTract(\mathcal H(\G,S))\). The tensor-closure and
product-transformable cases imply \(\ProdPerm(\G)\) by
Lemma~\ref{norm:tensor-cases}. The local-affine case implies
\(\Mono(\G)\) by Lemma~\ref{norm:local-affine}. The affine
alternative implies \(\ProdPerm(\G)\lor\DC(\G)\) by
Theorem~\ref{norm:affine}. This proves the forward implication,
including the explicit \(\Mono(\G)\) disjunct.

Conversely, if \(\ProdPerm(\G)\) holds, take \(S=I\).
Every gate tensor is a product of binary input/output factors and
every boundary signature is unary, so
\(\mathcal H(\G,I)\subseteq\langle\mathscr{T}\rangle\).
If \(\DC(\G)\) holds, take its diagonal \(D\) for \(S\).
The gates and boundaries are affine, giving the affine-transformable
case with \(B=I\). The monomial disjunct appears explicitly
on both sides of the equivalence.

To obtain Theorem~\ref{thm:normal}, assume that \(\G\) contains a
nonmonomial gate and an entangling gate. Then \(\Mono(\G)\) and
\(\ProdPerm(\G)\) are both false. The displayed equivalence therefore
reduces to
\[
 \exists S\in\mathrm{GL}_2(\Qbar)\;
       \ReTract(\mathcal H(\G,S))
 \quad\Longleftrightarrow\quad \DC(\G),
\]
which is the assertion of that theorem.
\end{proof}

\subsection{Recognition of the three normal forms}
\label{norm:recognition-section}

We now recognize the three structures directly from the gate matrices
and recover the data used by their simulators: local factors and wire
permutations, or a common diagonal frame and phased Clifford gates.
The normal forms give explicit tests and witnesses for the conditions
encoded by the general polynomial method of Theorem~\ref{hol:recognition}.

\begin{theorem}[Direct recognition]
\label{norm:recognition}
Each of \(\Mono(\G)\), \(\ProdPerm(\G)\), and \(\DC(\G)\) is decidable
from the exact algebraic gate matrices. Positive product-permutation
instances admit computable local factors and wire permutations.
A positive \(\DC\) instance admits a computable common diagonal
phase and Clifford representatives, including their global phases.
\end{theorem}

\begin{proof}
Monomiality is decided by checking whether every row and column of
each supplied matrix has exactly one nonzero entry.

For a \(k\)-qubit gate, enumerate the finitely many matchings of its
input ports with its output ports, equivalently the possible wire
permutations. Group each matched pair into an index of size four.
The resulting order-\(k\) tensor is a tensor product of \(k\)
four-dimensional vectors if and only if every one-index flattening
has rank one. One implication is immediate.
For the other, a rank-one flattening splits off its corresponding
factor. The remaining one-index ranks are still one, and induction
splits off all factors. Rank-one factorizations over algebraic number
fields are effective. The full gate's invertibility makes the resulting
local matrices invertible, and its unitarity normalizes them by
Lemma~\ref{norm:balanced}. Checking all matchings decides the
product-permutation condition and supplies its factors.

For \(\DC\), first enumerate the projective $k$-qubit Clifford group
for each $k$ occurring in the library, using their conjugation action
on the Pauli generators. For each of the \(2k\) generators
\(X_j,Z_j\), enumerate images from
\[
 \{\,\pm P_1\otimes\cdots\otimes P_k:
                         P_\ell\in\{I,X,Y,Z\}\,\},
\]
retaining tuples with the required commutation and anticommutation
relations. These images may act on several qubits. Solve the linear intertwining equations
\[
 QX_j=X'_jQ,\qquad QZ_j=Z'_jQ.
\]
Discard a choice with no nonzero solution. A nonzero solution is
invertible: its kernel is invariant under every input Pauli and hence
under the full matrix algebra, so its kernel is zero. Moreover,
\(Q^\dagger Q\) commutes with every input Pauli, because the chosen
images are Hermitian unitaries. It is therefore \(cI\) for \(c>0\).
Normalize by \(c^{-1/2}\) to obtain an algebraic Clifford representative.
Two invertible solutions for the same images differ by a scalar,
again because their quotient commutes with all Paulis. This finite
procedure enumerates all projective representatives.

Choose a tuple \((C_j)\) of these representatives for the supplied
gates \((U_j)\). We seek one phase \(z\) and a phase \(\lambda_j\)
for each gate such that
\[
 U_j=\lambda_jD(z)^{\otimes k_j}C_jD(z)^{-\otimes k_j}.
\]
Entrywise, this requires
\[
 U_j(y,x)=\lambda_jz^{|y|-|x|}C_j(y,x).
\]
Thus reject a tuple unless each \(C_j\) has the same zero pattern as
\(U_j\). At every nonzero position \(p=(y,x)\), define
\[
 r_{j,p}=\frac{U_j(y,x)}{C_j(y,x)},\qquad
 d_{j,p}=|y|-|x|.
\]
The ratio is \(r_{j,p}=\lambda_jz^{d_{j,p}}\). Choose one reference
position \(p_j\) for each gate and divide by its ratio to eliminate
\(\lambda_j\). Only the common phase \(z\) remains unknown.
A common diagonal frame with these representatives therefore requires
\begin{equation}
 z^{\,d_{j,p}-d_{j,p_j}}=\frac{r_{j,p}}{r_{j,p_j}}
       \quad\text{for every }j,p,
 \qquad |z|=1,\qquad |r_{j,p_j}|=1\quad\text{for every }j.
 \label{norm:phase-equations}
\end{equation}
Conversely, a solution of \eqref{norm:phase-equations} recovers
\(\lambda_j=r_{j,p_j}z^{-d_{j,p_j}}\), which has modulus one.
The Clifford representative including its actual global phase is
then \(\lambda_jC_j\).

Every zero exponent in \eqref{norm:phase-equations} requires its
right-hand side to equal one. If a nonzero exponent occurs, enumerate
the finitely many algebraic roots of that one equation and test
each candidate against all remaining equations and \(z\overline z=1\).
If all exponents are zero, check the constant equations and take
\(z=1\). Tests of absolute values and all other equalities are exact
algebraic-number tests. Accept if any tuple of Clifford representatives
and any candidate root succeeds. The enumerations are finite and
return the promised common phase and representatives. The procedure
need not run in time polynomial in the description of the library.
\end{proof}

\section{Arbitrary complex coefficients}
\label{app:coefficients}

For arbitrary fixed complex gate entries, we use the symbolic output
convention of Section~\ref{sec:classification}. Each amplitude is
represented by polynomials in the original real and imaginary gate
coordinates, reduced modulo all their rational relations. These
relations identify expressions with the same value. For example,
a unit phase times its conjugate reduces to one. We prove that the
resulting normal forms support the dichotomy in this exact-output model.

We prove two assertions needed to extend Theorem~\ref{thm:algebraic}.
First, these descriptions have polynomial length, and substituting suitable
algebraic constants into them transfers the established hardness result.
Second, the three easy simulations can produce the prescribed description,
even when a natural factorization introduces auxiliary constants. The
first part of this appendix develops the representation and specialization.
The second gives the conversion procedures and the simulations.

Fix a nonempty library $\G=(U_1,\ldots,U_t)$, in which $U_j$
acts on $k_j$ qubits, and put $k_* = \max_j k_j$. The circuit model is the
one in Section~\ref{sec:classification}. Write $m$ for the number of
gates and $N$ for the input length, including the two boundary strings.
Thus $m,n\leq N$. The numbers $k_j$ and the coefficient presentations are fixed
independently of the circuit. For an empty library, every circuit is
the identity and its amplitude is $\delta_{a,b}$. In this appendix,
$\Mono$, $\ProdPerm$, and $\DC$ allow arbitrary complex entries,
local unitary factors, and global phases. In $\DC$, the common diagonal
phase may be any $z\in\mathbb C$ with $|z|=1$. With these coefficient
ranges, the predicate from Appendix~\ref{app:prelim} is
\begin{equation}
 \QTract(\G)=\Mono(\G)\ \vee\ \ProdPerm(\G)\ \vee\ \DC(\G).
 \label{fg:easy-predicate}
\end{equation}

\subsection{Coefficient rings and exact output}
\label{fg:coefficient-rings}

Let $\alpha=(\alpha_1,\ldots,\alpha_s)\in\mathbb R^s$ list the real
and imaginary parts of the gate entries, with their matrix positions
recorded. Define
\begin{equation}
 I=\{p\in\mathbb Q[X_1,\ldots,X_s]:p(\alpha)=0\},
 \qquad
 R=\mathbb Q[\alpha]\cong\mathbb Q[X_1,\ldots,X_s]/I.
 \label{fg:ideal}
\end{equation}
The ideal $I$ is the full kernel of evaluation, and is prime because
$R\subset\mathbb R$ is a domain. The Hilbert basis theorem gives a
finite rational generating set. For the effective constructions, this set
is supplied with the promise that it generates the full ideal $I$.
The library, its coordinate positions, and this presentation are fixed
data. Preprocessing may depend on all these data, while the polynomial
bounds below concern the circuit input length. From the supplied generators,
compute a monic reduced Gr\"obner basis of $I$ in graded lexicographic
order with a fixed variable order.
Polynomial division gives the unique remainder $\NF_I(p)$ supported
on the standard monomials, namely those not divisible by a leading
monomial of the basis~\cite[Chapter~2]{CLO15}. In particular,
\[
 p(\alpha)=q(\alpha)
 \quad\Longleftrightarrow\quad
 \NF_I(p)=\NF_I(q).
\]
Every amplitude belongs to $R+iR$. We encode $p(\alpha)+iq(\alpha)$ by
\begin{equation}
 \bigl(\NF_I(p),\NF_I(q)\bigr),
 \label{fg:output}
\end{equation}
with each polynomial stored as a sorted list of monomials with binary
exponents and reduced rational coefficients. Write
$\Amp^{\NF_I}(\G)$ for the resulting function problem. The tuple
$\alpha$ and its relation ideal are fixed with the library, while the
input to this function problem is the circuit and its boundary strings.

\begin{theorem}[Dichotomy with symbolic exact outputs]
\label{fg:extension}
Fix a finite complex unitary qubit library $\G$ and the output
convention~\eqref{fg:output}. If $\QTract(\G)$ holds, then
$\Amp^{\NF_I}(\G)\in\FP$. Otherwise, $\Amp^{\NF_I}(\G)$ is
$\#\mathrm P$-hard.
The prescribed outputs have polynomial length. The tractable algorithms
and the overhead of the reductions have polynomial bit complexity,
with constants depending on the fixed library and presentation. All
oracle queries are legal circuits over the original library with its
prescribed computational-basis boundaries.

Given a finite rational generating set promised to equal the full
ideal $I$, together with the positions of the coordinate variables in
the unitary gate matrices, all three structural conditions are
decidable. The preprocessing for the tractable algorithms and the
hard algebraic specialization are effective from these data. The
algorithms compute every auxiliary presentation from $I$.
Their polynomial per-circuit bounds allow the fixed preprocessing
cost to depend on the library and its presentation.
\end{theorem}

\begin{lemma}[Compatibility with number-field output]
\label{fg:algebraic-encoding}
For a fixed algebraic library, the encoding~\eqref{fg:output} and the
number-field encoding of Theorem~\ref{thm:algebraic} are interconvertible
in polynomial time.
\end{lemma}
\begin{proof}
If every $\alpha_j$ is algebraic, $R$ is finite-dimensional over
$\mathbb Q$. Multiplication by a nonzero element is an injective
linear map on $R$, hence surjective, so $R$ is a number field. Its
standard monomials form a fixed rational basis. Let $K$ be the field
generated by the complex gate entries and their conjugates. Both $K$
and $R$ embed in the fixed number field $R(i)$. Expressing their fixed
bases in a basis of $R(i)$ gives rational linear maps. The inverse map
on the image of $K$ is also a fixed rational linear map. Thus an
amplitude can be converted in either direction by fixed-dimensional
rational linear algebra, with polynomial bit cost. The common field
$R(i)$ provides the comparison even when $i\notin K$.
\end{proof}

\subsection{Polynomial normal forms and bit complexity}
\label{fg:normal-form-arithmetic}

\begin{lemma}[Fixed-basis reduction]
\label{fg:gb-bit}
Fix a monic rational Gr\"obner basis in $r$ variables for graded
lexicographic order. A polynomial of degree at most $D$, represented
with a common positive denominator and with numerator and denominator
bit lengths at most $L$, can be reduced to normal form in
$\poly(D+L)$ bit operations. Its normal form has degree at most $D$
and polynomial bit length. The polynomial bound depends on the fixed
basis and $r$.
\end{lemma}
\begin{proof}
There are $M_r(D)=\binom{D+r}{r}$ monomials of degree at most $D$.
Cancelling a leading term by a basis element does not increase total
degree: every other monomial of the basis element is smaller than its
leading monomial and hence has no larger degree. After cancellation,
or after moving an irreducible leading term to the remainder, the
leading monomial of the working polynomial strictly decreases.
There are therefore at most $M_r(D)$ steps.

Choose a fixed positive integer $\Delta$ clearing all basis
coefficients, and a fixed $K\geq1$ bounding the multiplicative increase
in the coefficient $\ell_1$ norm in one step, including the remainder.
Such a $K$ exists because each cancellation replaces one term by a
scalar multiple of the fixed tail of a basis polynomial. Starting
with denominator $q$, after $u$ steps all denominators divide
$q\Delta^u$, and the coefficient norm is at most $K^u$ times its
initial value. The common-denominator numerators and denominator
therefore have bit lengths
\[
 O\bigl(L+\log M_r(D)+u(1+\log\Delta+\log K)\bigr).
\]
The numbers of monomials and steps are polynomial in $D$, since $r$
is fixed. All rational arithmetic has polynomial bit cost, including
conversion to reduced individual coefficients. If the input has
separate denominators, their product supplies a common denominator
with at most $M_r(D)L$ bits, giving the same conclusion.
\end{proof}

\begin{lemma}[Amplitude output bound]
\label{fg:amp-size}
For a circuit with $m$ gates, the real and imaginary parts of its
amplitude have representatives in $\mathbb Z[X]$ of degree at most
$m$ and coefficient bit length $O(m+1)$. Their normal forms have
length $\poly(N)$ and degree at most $m$.
\end{lemma}
\begin{proof}
Follow computational-basis states from the fixed input. At a
$k_j$-qubit gate there are at most $2^{k_j}$ choices of local output,
so there are at most $2^{k_*m}$ paths. Imposing the final boundary
only discards paths. Each path contributes a product of $m$ entries.
Expansion into real and imaginary parts produces at most $2^m$
signed monomials per path. The sum of the absolute values of all
coefficients is consequently at most $2^{(k_*+1)m}$. Combining
repeated monomials does not increase this bound. For $m=0$ the answer
is $\delta_{a,b}$. Lemma~\ref{fg:gb-bit} gives the normal-form bounds.
The path expansion supplies a size bound for the oracle output,
which will control the cost of the specialization reduction.
\end{proof}

The specialization and Clifford calculations below also use arithmetic
in fixed number fields. In a fixed rational basis, multiplication by
any fixed element is a fixed rational linear map. A product of $d$
such elements therefore has coordinate bit length $O(d+1)$.
Polynomially many such terms with polynomial-bit rational coefficients
can be added with polynomial-bit coordinates by using a common
denominator. Fixed rational basis changes have the same property.

\subsection{Clifford representatives and structural formulas}
\label{fg:structural-formulas}

The structural conditions must have rational first-order formulas
for algebraic specialization. To obtain such formulas for the Clifford
case, we retain Gaussian-integer representatives in the intertwining
construction of Theorem~\ref{norm:recognition}. These representatives
serve two purposes: they give rational formulas for specialization, and
they make the auxiliary Clifford amplitude a Gaussian integer that can
be combined with expressions in the original gate entries.

\begin{lemma}
\label{fg:cliff-table}
For every fixed $k$, one can effectively construct a finite list
$\mathcal W_k\subset\mathbb Z[i]^{2^k\times2^k}$ such that every
$W\in\mathcal W_k$ satisfies
\[
 W^\dagger W=h_WI,\qquad h_W\in\mathbb Z_{>0},
\]
and $W/\sqrt{h_W}$ is Clifford. Every $k$-qubit Clifford unitary,
including its global phase, is a nonzero scalar multiple of an
element of $\mathcal W_k$.
\end{lemma}
\begin{proof}
Enumerate the possible signed Hermitian Pauli images $P_r,Q_r$ of
$X_r,Z_r$, for $1\leq r\leq k$, and solve
\[
 MX_r=P_rM,\qquad MZ_r=Q_rM
\]
as a homogeneous linear system over $\mathbb Q(i)$. Discard a tuple
with zero solution space. For a nonzero solution $M$, its kernel is
invariant under all $X_r,Z_r$, hence under the full matrix algebra
they generate. The kernel is therefore zero, and $M$ is invertible.
Hermiticity and the intertwining equations imply that $M^\dagger M$
commutes with every input Pauli. Thus $M^\dagger M=hI$ for $h>0$.
Clear the rational denominators of $M$ to obtain $W\in\mathbb Z[i]^{2^k\times2^k}$.
Then $h_W=\sum_y|W(y,x)|^2$ is a positive integer, independent of $x$.
The normalized matrix is unitary and sends the Pauli generators to
Paulis, so it is Clifford.

Conversely, a Clifford unitary yields one of the enumerated tuples
and a nonzero complex solution of its equations. Matrix rank over
$\mathbb Q(i)$ is unchanged by extension to $\mathbb C$, so this
tuple has a nonzero solution over $\mathbb Q(i)$. Two invertible
solutions differ by a scalar, since their quotient commutes with
all input Paulis. Hence the selected $W$ represents every Clifford
with that tuple of images. All enumerations and linear systems are
finite and depend only on $k$.
\end{proof}

The normalized matrices in this lemma form a fixed algebraic
Clifford library, for which exact simulation retains the overall
phase~\cite{DH22}. The integer representatives also express $\QTract$ by
a rational first-order formula on the unitary locus:
\begin{itemize}[leftmargin=*,itemsep=3pt]
\item For $\Mono$, enumerate the matrix permutations and set all
entries outside the selected positions to zero. Unitarity forces
the selected entries to be nonzero.
\item For $\ProdPerm$, enumerate wire permutations and quantify over
the real and imaginary parts of the local $2\times2$ matrices.
Their unitarity and the tensor-factorization equations are polynomial.
\item For $\DC$, enumerate tuples $W_j\in\mathcal W_{k_j}$ and
quantify over $z,s_1,\ldots,s_t\in\mathbb C$, each split into real
and imaginary parts. Require
\begin{equation}
 |z|^2=1,\qquad h_{W_j}|s_j|^2=1,\qquad
 U_j=s_jD(z)^{\otimes k_j}W_jD(z)^{-\otimes k_j},
 \label{fg:dc-firstorder}
\end{equation}
where $D(z)=\diag(1,z)$. Negative powers of $z$ are replaced by
powers of $\bar z$, making every entry equation polynomial. The
scalar $s_j$ includes the normalization and the Clifford global phase.
\end{itemize}
Unitarity is itself a rational polynomial condition. Thus both $\QTract$
and its negation are first-order conditions over real closed fields.

\subsection{Specialization and exact oracle reduction}
\label{fg:specialization}

The preceding formulas let us choose algebraic gate entries that retain
unitarity, failure of the three tractable conditions, and every relation
in $I$. The transfer principle used for this choice is the following
consequence of quantifier elimination: a rational
first-order sentence has the same truth value in $\mathbb R$ and
in the real algebraic numbers $\mathbb R_{\mathrm{alg}}$.
The same holds for formulas with parameters in a common ordered
coefficient field~\cite[Theorem~4.4 and Corollary~4.5]{Pla23}.
In particular, a formula at a real algebraic tuple has the same truth
value in $\mathbb R_{\mathrm{alg}}$ and $\mathbb R$, including when
it contains negated existential conditions.

\begin{lemma}[Hard algebraic specialization]
\label{fg:specialize}
If $\neg \QTract(\G)$, there is a fixed
$\beta\in\mathbb R_{\mathrm{alg}}^s$ satisfying every polynomial in
$I$ such that the specialized library $\G_\beta$ is unitary and
satisfies $\neg \QTract(\G_\beta)$. Such a tuple can be found effectively
from a finite generating set of $I$.
\end{lemma}
\begin{proof}
Let $g_1,\ldots,g_v$ generate $I$, and let $U_j(X)$ be the formal
matrix of coordinate entries. With
$\G(x)=(U_1(x),\ldots,U_t(x))$, the rational first-order sentence
\begin{equation}
 \exists x\;\left[
 \bigwedge_{\nu=1}^{v}g_\nu(x)=0\ \wedge\
 \bigwedge_{j=1}^{t}U_j(x)^\dagger U_j(x)=I\ \wedge\
 \neg \QTract\bigl(\G(x)\bigr)\right]
 \label{fg:sentence}
\end{equation}
holds in $\mathbb R$, witnessed by $\alpha$. Transfer to
$\mathbb R_{\mathrm{alg}}$ gives a tuple $\beta$ satisfying the
bracketed formula in that field. Since the $g_\nu$ generate $I$,
every polynomial in $I$ vanishes at $\beta$.

Now fix this algebraic tuple and transfer the bracketed formula back
to $\mathbb R$. This second transfer ensures that $\G_\beta$ is
unitary and fails all three tractable conditions with witnesses ranging
over arbitrary complex entries. In particular, the diagonal-frame
condition fails for every complex unit phase.

For effectiveness, enumerate tuples of real algebraic numbers in
minimal-polynomial and isolating-interval encodings. Decide the
bracketed formula at each tuple by effective real quantifier
elimination. A witness exists, so the search terminates. This is
preprocessing for the fixed library. The resulting tuple is shared
by all queried circuits, so the search cost is independent of their size.
\end{proof}

\begin{lemma}[One-query specialization reduction]
\label{fg:hard-reduction}
For $\beta$ as in Lemma~\ref{fg:specialize},
\[
 \Amp(\G_\beta)\leq_T^{\poly}\Amp^{\NF_I}(\G),
\]
where the left-hand problem uses number-field coordinates.
The reduction makes one oracle query per input circuit, keeps its
placements and boundary strings, and has polynomial bit overhead.
\end{lemma}
\begin{proof}
There is a well-defined ring homomorphism
\begin{equation}
 \phi:R\longrightarrow\mathbb R_{\mathrm{alg}},\qquad
 p(\alpha)\longmapsto p(\beta),
 \label{fg:specialization-map}
\end{equation}
because $I\subseteq I_\beta$, where
$I_\beta=\{p\in\Q[X]:p(\beta)=0\}$. This inclusion allows
$\beta$ to satisfy additional relations and suffices to define $\phi$.
Extend $\phi$ to $R+iR$ by fixing $i$.
Given a circuit over $\G_\beta$, query the identical sequence of
labels and placements over $\G$. If the answer is $(p_0,q_0)$,
output $p_0(\beta)+iq_0(\beta)$. An amplitude is a polynomial
contraction of the gate entries, so applying $\phi$ is the same as
specializing every gate. Reduction modulo $I$ before specialization
does not change the answer.

Lemma~\ref{fg:amp-size} bounds the oracle answer's length by
$\poly(N)$ and its degree by $m$. Evaluate it in the fixed number
field $L=\mathbb Q(\beta_1,\ldots,\beta_s,i)$. There are polynomially
many monomials, each a product of at most $m$ fixed elements, with
polynomial-bit rational coefficients. The number-field bounds above
give polynomial-bit evaluation. The resulting value belongs to the
subfield generated by the specialized gate entries and their
conjugates. Conversion from $L$ to a fixed basis of that subfield
is a fixed-dimensional rational linear calculation, giving the
required number-field output.

The query keeps the circuit's directed structure and boundaries,
with gate labels interpreted over $\G$. Composing this one-query
specialization with the hardness reduction of
Theorem~\ref{thm:algebraic} gives a polynomial-time Turing reduction
from $\#\mathrm{SAT}$ to $\Amp^{\NF_I}(\G)$.
\end{proof}

\subsection{Localizations and conversion of exact outputs}
\label{fg:easy-conversion}

We now turn to the tractable algorithms. Their intermediate expressions
can involve ratios of gate entries, whereas the output must be a pair
of polynomials in the original coordinates. To convert an expression
containing inverses into this output, we use the representative of degree
at most $m$ guaranteed by Lemma~\ref{fg:amp-size}. The degree bound lets us recover
its normal-form coefficients by rational linear algebra in a space of
polynomial dimension. We first prove this conversion for an injective
map between fixed finitely generated $\Q$-algebras. We then apply it to
a localization of $R=\Q[\alpha]$, a ring obtained by allowing inverses
of specified nonzero elements. This localization will contain the ratios
used by the product-permutation and diagonal-Clifford simulations.

\begin{lemma}[Descent through a fixed polynomial embedding]
\label{fg:embedding-descent}
Let
\[
 A=\Q[X_1,\ldots,X_r]/I_A,\qquad
 B=\Q[Y_1,\ldots,Y_s]/I_B
\]
be fixed finitely generated \(\Q\)-algebras with given defining ideals,
and fix graded-lexicographic Gr\"obner normal forms for both. Suppose a
fixed injective homomorphism \(\iota:A\longrightarrow B\) is specified
by polynomials \(H_j(Y)\) representing \(\iota(X_j)\).
Given the \(B\)-normal form \(h(Y)\) of \(\iota(v)\), of degree at most
\(D\) and coefficient bit length at most \(L\), and the promise that
\(v\in A\) has a polynomial representative of degree at most \(m\),
one can compute \(\NF_{I_A}(v)\) in
\(\poly(m+D+L)\) bit operations. The polynomial depends on the fixed
presentations and embedding.
\end{lemma}

\begin{proof}
Let \(S_m\) be the \(I_A\)-standard monomials of degree at most \(m\).
There are at most \(\binom{m+r}{r}\) of them. Degree-compatible
reduction shows that the desired answer is a rational linear combination
of these monomials. Set
\(\kappa=\max(1,\deg H_1,\ldots,\deg H_r)\).
For each \(\mu=X_1^{a_1}\cdots X_r^{a_r}\in S_m\), compute
\[
 v_\mu(Y)=\NF_{I_B}
       \bigl(H_1(Y)^{a_1}\cdots H_r(Y)^{a_r}\bigr).
\]
Before reduction, this polynomial has degree at most \(\kappa m\).
The factors \(H_j\) are fixed: a fixed denominator raised to \(O(m)\)
clears its coefficients, and its coefficient norm grows at most
exponentially in \(m\). Thus its coefficient bit length is \(O(m+1)\).
Dense polynomial arithmetic and Lemma~\ref{fg:gb-bit} compute all
\(v_\mu\) with polynomial bit cost.

The unknown coefficients of the desired normal form are the
$c_\mu$ in $\sum_{\mu\in S_m}c_\mu\mu$. Mapping this expression
into $B$ and equating it to the given $h$ gives the linear system:
\begin{equation}
 \sum_{\mu\in S_m}c_\mu v_\mu(Y)=h(Y)
 \label{fg:embedding-system}
\end{equation}
Expand both sides in the \(I_B\)-standard monomials of degree at most
\(M=\max(D,\kappa m)\). Each such monomial gives one coefficient
equation, and each \(v_\mu\) gives one column. There are at most
\(\binom{M+s}{s}\) rows,
\(\binom{m+r}{r}\) columns, and polynomial-bit rational entries.
The columns are independent. A relation among them would give
\(\iota(\sum_\mu c_\mu\mu)=0\). Injectivity and independence of the
\(I_A\)-standard monomials force all coefficients to vanish.
The degree promise guarantees consistency, so the solution is unique
and \(\sum_\mu c_\mu\mu\) is \(\NF_{I_A}(v)\).

This linear solving has polynomial bit complexity. Clear the
denominators in the entire polynomial-size system. Their product still
has polynomial bit length. Fraction-free elimination uses polynomially
many integer operations, whose relevant minors have polynomial bit
length: for an \(u\times u\) integer matrix with entry bit length
\(H\), the determinant has absolute value at most
\(u!2^{uH}\). The resulting unique rational solution consequently has
polynomial bit length and is computable within the claimed bound.
\end{proof}

Return to the full evaluation ideal \(I\) of the gate coordinates.
List the nonzero complex entries of the fixed library as
\(u_1,\ldots,u_\ell\), retaining their positions, and write
\(u_r=u_{r,R}+iu_{r,I}\) in coordinate polynomials. We encode all
their inverses using one new real variable \(T\). Choose \(q\) as
the product of their squared moduli and impose \(T=1/q\):
\begin{equation}
 q(X)=\prod_{r=1}^{\ell}
          \bigl(u_{r,R}(X)^2+u_{r,I}(X)^2\bigr),
 \qquad
 J=I\Q[X,T]+(Tq(X)-1).
 \label{fg:localization}
\end{equation}
The empty product is one. An entry is zero exactly when both coordinate
polynomials reduce to zero modulo \(I\), so the list is computable from
\(I\). Since \(q(\alpha)>0\), we have \(q\notin I\).

\begin{lemma}[Presentation and descent]
\label{fg:descent}
The ideal \(J\) in~\eqref{fg:localization} is the full relation ideal
of \((\alpha,1/q(\alpha))\), and
\(J\cap\Q[X]=I\). Fix a graded-lexicographic Gr\"obner basis for
\(J\). Suppose \(v\in R\) has a representative of degree at most
\(m\) in the original variables. Given a polynomial \(h(X,T)\)
representing \(v\) in the localization, of degree at most \(D\) and
coefficient bit length at most \(L\), one can compute \(\NF_I(v)\)
in \(\poly(m+D+L)\) bit operations. The same conclusion holds
componentwise for \(R+iR\).
\end{lemma}

\begin{proof}
Let \(\bar q\) be the class of \(q\) in \(R\). There are natural
isomorphisms
\[
 \Q[X,T]/J
 \cong R[T]/(T\bar q-1)
 \cong R[\bar q^{-1}].
\]
The second sends \(T\) to \(\bar q^{-1}\). Its inverse sends
\(r/\bar q^d\) to \(rT^d\). Since \(R\) is a domain and
\(\bar q\ne0\), this localization embeds into
\(\Frac(R)\subset\R\) by actual evaluation. The evaluation kernel
of the quotient is therefore zero. Thus \(J\) is the full relation
ideal, and restricting the evaluation map to \(\Q[X]\) gives
\(J\cap\Q[X]=I\).

Reduce \(h\) modulo \(J\) using Lemma~\ref{fg:gb-bit}, and apply
Lemma~\ref{fg:embedding-descent} to the injective inclusion
\(R\longrightarrow\Q[X,T]/J\). Its generator images are simply
the variables \(X_j\). Explicitly, the system is
\begin{equation}
 \sum_{\mu\in S_m}c_\mu\NF_J(\mu)=\NF_J(h),
 \label{fg:descent-system}
\end{equation}
where \(S_m\) consists of the \(I\)-standard monomials of degree at
most \(m\). There are at most \(\binom{m+s}{s}\) columns and
\(\binom{\max(m,D)+s+1}{s+1}\) rows. The preceding lemma proves
the claimed bound. For a complex expression, apply the procedure to
its real and imaginary parts separately.
\end{proof}

Every inverse gate entry belongs to this one localization:
\begin{equation}
 u_r^{-1}=\overline{u_r}\,T
                  \prod_{h\ne r}|u_h|^2,
 \label{fg:inverse-entry}
\end{equation}
where the product ranges over the other positions in the list and all
expressions are evaluated at \((\alpha,1/q(\alpha))\).
The product counts positions, including repeated entries. Hence every fixed rational
expression in gate entries and their inverses is a fixed polynomial in
\(X,T\) with coefficients in \(\Q(i)\). Complex conjugation acts
only on these rational complex coefficients, since \(X,T\) evaluate
to real numbers.

We will use elementary size bounds before applying any Gr\"obner
reduction. Each fixed polynomial factor has constant degree and
coefficient norm, and its rational denominators divide a fixed integer.
A product of \(O(N)\) such factors has degree \(O(N)\), and its
common-denominator coefficients have bit length \(O(N+1)\).
The same bounds hold for a sum over exponentially many paths of length
\(O(N)\) with a fixed number of choices per step. Because the number
of variables is fixed, there are only polynomially many monomials.
The algorithms below construct the required polynomials by polynomially
many dense additions and multiplications. The path estimates bound
their coefficient sizes.

\subsection{Monomial and product-permutation libraries}
\label{fg:easy-product}

\begin{lemma}
\label{fg:mono-prod}
If \(\Mono(\G)\) or \(\ProdPerm(\G)\) holds, then
\(\Amp^{\NF_I}(\G)\in\FP\). Recognition and the fixed
preprocessing for these algorithms are effective from \(I\).
\end{lemma}

\begin{proof}
For \(\Mono\), use zero tests modulo \(I\) to find the unique
nonzero entry in every column. Starting at the input string \(b\),
follow the unique basis trajectory and multiply its \(m\) transition
entries. Return zero if the final string differs from \(a\).
The product has degree at most \(m\) and polynomial-bit coefficients.
Lemma~\ref{fg:gb-bit} computes its output normal form. The same zero
tests determine whether every gate is monomial.

For \(\ProdPerm\), enumerate the wire permutations of each
\(k\)-qubit gate. Undo a proposed output permutation and group the
input-output pairs as indices \(v_r\in\{0,1\}^2\). This gives a
\(k\)-way tensor \(F(v_1,\ldots,v_k)\) with four-valued indices.
A nonzero \(F\) is a tensor product of \(k\) four-dimensional
vectors if and only if all one-index flattenings have rank one.
For sufficiency, factor off the first rank-one flattening.
Each remaining flattening has the same rank as the corresponding
flattening of the remaining tensor, because the removed factor is
nonzero. Induction gives the full product. Rank at most one is tested
by the \(2\times2\) minors, and the gate's invertibility ensures
that the tensor is nonzero. These are equality tests in \(R+iR\).

Choose a nonzero reference entry
\(v^0=(v_1^0,\ldots,v_k^0)\), and define
\begin{equation}
 \rho=F(v^0),\qquad
 B_r(v_r)=
 \frac{F(v_1^0,\ldots,v_{r-1}^0,v_r,v_{r+1}^0,\ldots,v_k^0)}
      {\rho}.
 \label{fg:tensor-ratios}
\end{equation}
For such a tensor product, cancellation of the nonzero reference factors gives
\[
 F(v_1,\ldots,v_k)=\rho\prod_{r=1}^kB_r(v_r).
\]
The identity also covers zero entries. All coefficients belong to the
localization~\eqref{fg:localization}, by~\eqref{fg:inverse-entry}.
The only nonconstant denominators used here are nonzero original entries.

To show that this test recognizes precisely \(\ProdPerm\), we
normalize the local factors to unitaries. This normalization establishes
the structural condition, while the simulation below uses the matrices
\(B_r\) from~\eqref{fg:tensor-ratios} directly.
Every local matrix \(B_r\) is invertible, since their tensor product
is invertible, and unitarity gives
\[
 |\rho|^2\bigotimes_r(B_r^\dagger B_r)=I.
\]
Compare products of positive eigenvalues that differ in just one
factor. The two eigenvalues of \(B_r^\dagger B_r\) must be equal,
so \(B_r^\dagger B_r=c_rI\) with \(c_r>0\).
Thus \(B_r/\sqrt{c_r}\) is unitary, and
\(\rho\prod_r\sqrt{c_r}\) is a unit phase that can be absorbed
into one local factor. This proves that the factorization test
recognizes exactly the product-preserving unitaries.

For simulation, use the unnormalized matrices $B_r$ directly, keeping
a two-dimensional vector per wire, initialized
to its basis input, and a global scalar. At each gate apply its local
matrices \(B_r\), multiply the scalar by \(\rho\), and perform
the wire permutation. The final amplitude is the scalar times the
product of the vector coordinates selected by \(a\).
There are at most \(k_*m\) local matrix applications. Across all
final wire histories, these applications use \(O(m)\) factors.
Permutations only relabel the histories. Each coordinate is a sum
over at most exponentially many local choices. Substituting the fixed
polynomials in~\eqref{fg:tensor-ratios} therefore gives degree
\(O(m)\) and polynomial coefficient bit length throughout.
Dense arithmetic has polynomial cost. The final value has an original
degree-\(m\) representative by Lemma~\ref{fg:amp-size}, so
Lemma~\ref{fg:descent} converts it to the required pair of normal forms.
\end{proof}

\subsection{Eliminating the common diagonal phase}
\label{fg:easy-diagonal}

The common diagonal phase need not belong to the original coefficient
ring. We therefore express its contribution to the amplitude through
powers determined by gate-entry ratios. When these ratios impose a
nonzero phase exponent, they determine a power \(w=z^d\) directly
in the localization already constructed. The total phase exponent of
every nonzero amplitude is then divisible by \(d\), which expresses
the amplitude using \(w\) and the original gate entries. If all local
exponents vanish, the ratio equations allow the frame \(z=1\).

Choose a tuple \(W_j\in\mathcal W_{k_j}\) from
Lemma~\ref{fg:cliff-table}, with the same support as \(U_j\).
For each nonzero position \(p=(y,x)\) of gate \(j\), put
\begin{equation}
 r_{j,p}=\frac{U_j(y,x)}{W_j(y,x)},\qquad
 d_{j,p}=\wt(y)-\wt(x).
 \label{fg:ratios}
\end{equation}
Fix a reference position \(p_j\) for each gate and set
\begin{equation}
 r_j=r_{j,p_j},\quad \delta_j=d_{j,p_j},\quad
 c_{j,p}=\frac{r_{j,p}}{r_j},\quad
 n_{j,p}=d_{j,p}-\delta_j.
 \label{fg:ratio-data}
\end{equation}
Here $\delta_j$ records the reference entry's change in Hamming weight,
and $n_{j,p}$ is the difference from that reference. Dividing the two
gate-entry equations will leave the phase power $z^{n_{j,p}}$.
The scalars \(r_j\) and \(c_{j,p}\) are nonzero. They and their inverses are fixed
polynomial expressions in the localization: nonzero entries of
\(W_j\) are constants in \(\Q(i)\), and every nonconstant
denominator is a product of nonzero original entries.

The local equations \(z^{n_{j,p}}=c_{j,p}\) must hold for one
common phase. When some exponent is nonzero, B\'ezout's identity
combines the exponents into their greatest common divisor \(d\),
so their right-hand sides determine
the shared quantity \(w=z^d\). The following test checks consistency
using this computable quantity.

\begin{lemma}[Phase test without adjoining roots]
\label{fg:phase-test}
For a fixed tuple \((W_j)\), a common diagonal Clifford frame with
these projective representatives exists if and only if the following
test accepts.
If every \(n_{j,p}\) is zero, require every \(c_{j,p}=1\).
Otherwise choose fixed B\'ezout coefficients satisfying
\[
 d=\gcd\{|n_{j,p}|:n_{j,p}\ne0\}>0,\qquad
 \sum_{j,p}b_{j,p}n_{j,p}=d,
 \qquad w=\prod_{j,p}c_{j,p}^{\,b_{j,p}},
\]
and require
\begin{equation}
 w\overline w=1,\qquad
 c_{j,p}=w^{n_{j,p}/d}\quad\text{for every }j,p.
 \label{fg:gcd-test}
\end{equation}
The test is effective from \(I\), using the relation ideal \(J\)
computed in Lemma~\ref{fg:descent}.
In the second case, every \(d\)-th root \(z\) of \(w\) has
\(|z|=1\) and satisfies
\begin{equation}
 U_j=r_jz^{-\delta_j}
        D(z)^{\otimes k_j}W_jD(z)^{-\otimes k_j}.
 \label{fg:gate-phase-form}
\end{equation}
In the first case the same identity holds with \(z=1\).
\end{lemma}

\begin{proof}
A common frame for these projective representatives is equivalent to
the existence of a unit phase \(z\) satisfying
\[
 z^{n_{j,p}}=c_{j,p}\quad\text{for every }j,p.
\]
One direction follows by dividing each entry equation by its reference
equation. Conversely, the ratios reconstruct
\eqref{fg:gate-phase-form} on the common support. Since \(U_j\)
is unitary and \(W_j^\dagger W_j=h_{W_j}I\), the scalar
\(r_jz^{-\delta_j}\) in that identity has modulus
\(1/\sqrt{h_{W_j}}\). Its product with \(\sqrt{h_{W_j}}\)
is exactly an allowed Clifford global phase. Thus unitarity supplies
the scalar normalization as soon as the entry-ratio equations hold.

If all exponents vanish, the system is precisely \(c_{j,p}=1\),
and \(z=1\) is a solution. Otherwise any solution gives
\(w=z^d\) by B\'ezout's identity and therefore passes
\eqref{fg:gcd-test}. In the other direction, if the test accepts,
choose a complex \(d\)-th root \(z\) of \(w\). The equality
\(w\overline w=1\) makes \(|z|=1\), and divisibility by \(d\)
gives
\[
 z^{n_{j,p}}=(z^d)^{n_{j,p}/d}
            =w^{n_{j,p}/d}=c_{j,p}.
\]
The root \(z\) establishes the existence of a common frame. The
test itself computes \(w\) and checks~\eqref{fg:gcd-test}.
All powers, inverses, and conjugates appearing in these checks are fixed
localized expressions. Equality is decided modulo the explicitly
computed full ideal \(J\) of Lemma~\ref{fg:descent}.
Enumerating the finite tuples \((W_j)\) and checking their supports
therefore decides \(\DC\).
\end{proof}

When the gate entries are algebraic, so is \(w\), and its roots can
be chosen algebraic. The arbitrary-phase definition of \(\DC\)
thus agrees on algebraic libraries with the algebraic-phase condition
in Theorem~\ref{thm:algebraic}.

\begin{lemma}[Exact simulation in the original coefficient ring]
\label{fg:dc-simulation}
If \(\DC(\G)\) holds, then \(\Amp^{\NF_I}(\G)\in\FP\).
Its fixed preprocessing is effective from \(I\).
\end{lemma}

\begin{proof}
Use Lemma~\ref{fg:phase-test} to find an accepted tuple \((W_j)\)
and its fixed ratio data. In an input circuit \(C\), let \(m_j\)
count occurrences of gate \(j\). Replace each occurrence by \(W_j\)
on the same ordered wires and define
\[
 A_W=\bra a W_C\ket b\in\Z[i],\qquad
 e=\wt(a)-\wt(b)-\sum_jm_j\delta_j.
\]
Equation~\eqref{fg:gate-phase-form} gives
\begin{equation}
 \bra aU_C\ket b
       =\left(\prod_jr_j^{m_j}\right)z^eA_W.
 \label{fg:amplitude-phase}
\end{equation}
To see the cancellation for arbitrary gate placements, let \(\mathcal D=D(z)^{\otimes n}\). An embedded framed gate is
the conjugate by \(\mathcal D\) of its embedded \(W_j\), times
the scalar \(r_jz^{-\delta_j}\). On other wires the two diagonal
factors cancel. Multiplying the circuit cancels all internal copies
of \(\mathcal D^{-1}\mathcal D\), leaving the boundary factor
\(z^{\wt(a)-\wt(b)}\) and the stated scalar factors.

Thus \(e\) is the total phase exponent left after the internal
changes of basis and the reference-entry phases have been accounted for.
We first compute \(A_W\) exactly. For a nonzero result, the phase
test will let us evaluate \(z^e\) using either \(z=1\) or a power
of \(w\).

To compute \(A_W\), simulate the Gaussian-integer matrices through
their normalized Clifford representatives. Set \(\widehat W_j=W_j/\sqrt{h_{W_j}}\). This is a fixed
algebraic Clifford library, so the tractable part of
Theorem~\ref{thm:algebraic} computes its amplitude exactly in
polynomial time, with its gate phases retained. Then
\begin{equation}
 A_W=\left(\prod_jh_{W_j}^{m_j/2}\right)
                  \bra a\widehat W_C\ket b.
 \label{fg:w-simulation}
\end{equation}
This calculation takes place entirely in the known fixed number field
\(\Q(i,\sqrt{h_{W_1}},\ldots,\sqrt{h_{W_t}})\), or a fixed
algebraic extension used by the simulator. The prefactor has coordinate
bit length \(O(m+1)\). Fixed rational linear algebra recovers the
two integer coordinates of \(A_W\). Their bit length is
\(O(m+1)\) directly from the path expansion with fixed
Gaussian-integer gate entries. In particular, the test \(A_W=0\)
is exact. This auxiliary calculation is completed in \(\Z[i]\)
before it is combined with the original coefficient expressions.
Only the resulting Gaussian integer enters that symbolic calculation,
so the auxiliary square roots require no new symbolic coordinates.
If \(A_W=0\), return the zero pair of normal forms.

Assume now that \(A_W\ne0\). If every \(n_{j,p}\) is zero, use
\(z=1\) in~\eqref{fg:amplitude-phase}. Otherwise take \(d,w\)
from Lemma~\ref{fg:phase-test}. At least one computational-basis path has a
nonzero contribution to \(A_W\). Along that path, summing the
local weight changes gives the total weight change of the boundary
strings. Hence
\begin{equation}
 e=\sum_{\text{occurrences on the path}}(d_{j,p}-\delta_j)
   =\sum_{\text{occurrences}}n_{j,p}\in d\Z.
 \label{fg:divisibility}
\end{equation}
This path establishes the divisibility of \(e\). The algorithm
computes \(e\) directly from the boundary weights, the gate counts,
and the fixed \(\delta_j\). The desired answer is consequently
\begin{equation}
 \bra aU_C\ket b
       =A_W\left(\prod_jr_j^{m_j}\right)w^{e/d}.
 \label{fg:root-free-amplitude}
\end{equation}
This formula is independent of the chosen \(d\)-th root.
For negative \(e/d\), use the corresponding positive power of
\(\overline w\), since \(w\overline w=1\).

The expressions \(r_j,w,\overline w\) are fixed polynomials in
\(X,T\), and \(|e|\le n+k_*m=O(N)\).
Equation~\eqref{fg:root-free-amplitude} can therefore be computed
densely with degree \(O(N)\) and polynomial coefficient bit length.
The all-zero-exponent branch has the same bound. The result is an
original circuit amplitude, which has a degree-\(m\) representative
by Lemma~\ref{fg:amp-size}. Applying Lemma~\ref{fg:descent} gives
the prescribed pair of normal forms using the computed ratio data
and the presentation $I$.
\end{proof}

\subsection{Completion of the coefficient extension}
\label{fg:completion}

\begin{proof}[Proof of Theorem~\ref{fg:extension}]
Lemma~\ref{fg:amp-size} gives polynomial output length for every
circuit. If \(\Mono\) or \(\ProdPerm\) holds,
Lemma~\ref{fg:mono-prod} supplies a polynomial-time algorithm.
If \(\DC\) holds, Lemma~\ref{fg:dc-simulation} supplies one.
Their recognition procedures are finite and use only \(I\), the
explicit ideal \(J\), and the Clifford tables of
Lemma~\ref{fg:cliff-table}. The required fixed Gr\"obner bases are
effectively computable from these finite rational presentations.

If all three conditions fail, Lemma~\ref{fg:specialize} gives a
fixed algebraic unitary library \(\G_\beta\) for which they still
fail. The algebraic-phase and arbitrary-phase conditions agree there,
as shown after Lemma~\ref{fg:phase-test}. Theorem~\ref{thm:algebraic}
and Lemma~\ref{fg:hard-reduction} give
\[
 \SAT\leq_T^{\poly}\Amp(\G_\beta)
       \leq_T^{\poly}\Amp^{\NF_I}(\G).
\]
The second reduction preserves each queried circuit's placements,
ordering, and basis boundaries. Thus every query in the composed
reduction remains a legal acyclic circuit over the original library.
Its conversion cost is polynomial in circuit size, and the
specialization is effective from the promised presentation of \(I\).
This proves Theorem~\ref{fg:extension}.

Under the exact-output convention in the main text, the same
algorithms, recognition procedures, and reductions prove
Theorem~\ref{thm:intro-dichotomy}. For algebraic libraries,
Lemma~\ref{fg:algebraic-encoding} gives polynomial-time equivalence
with the number-field encoding, so this also retains the original
algebraic statement.
\end{proof}

\section{Consequences for specific gate libraries}
\label{sec:applications}

We now apply Theorem~\ref{thm:intro-dichotomy} to the gate-library
consequences of Section~\ref{sec:consequences}, retaining computational-basis
boundaries and unrestricted gate placements. A frame-uniqueness lemma
gives the three-gate bound and the compatibility tests for specific
libraries. A column-magnitude criterion then gives singleton hardness.
The final paragraphs connect these tests to the further applications
in the next appendix and to the parameter and closure results in
Appendices~\ref{app:parameters} and~\ref{app:closure}.

\subsection{Gates that constrain the Clifford frame}
\label{app:frame-tests}

For a nonmonomial gate, any diagonal Clifford frame determines all the
others up to a fourth root of unity. This restricts the possible common
frames of a library once one nonmonomial gate has been selected.

\begin{lemma}[Uniqueness of a diagonal Clifford frame]
\label{lem:frame-uniqueness}
Let $U$ be a nonmonomial $k$-qubit unitary. If
$\bigl(D(z)^\dagger\bigr)^{\otimes k}UD(z)^{\otimes k}$ is Clifford for a unit
phase $z$, then the phases giving a diagonal Clifford frame for $U$
are exactly $z\{1,i,-1,-i\}$.
\end{lemma}
\begin{proof}
Write $\mu_4=\{1,i,-1,-i\}$.
Set $C=\bigl(D(z)^\dagger\bigr)^{\otimes k} U D(z)^{\otimes k}$. Since $C$ is Clifford, each
$CZ_jC^\dagger$ is a Pauli. For some $j$, it has an $X$ or $Y$
factor. Indeed, if all these Paulis were diagonal, conjugation by $C$
would map the diagonal algebra generated by $Z_1,\ldots,Z_k$ onto
itself. It would permute its rank-one computational-basis projectors,
making $C$, and hence $U$, monomial.

Fix such a Pauli $P=CZ_jC^\dagger$. If $D(z')$ is another frame,
put $w=z'/z$. The Clifford $\bigl(D(w)^\dagger\bigr)^{\otimes k} C D(w)^{\otimes k}$ conjugates
$Z_j$ to $\bigl(D(w)^\dagger\bigr)^{\otimes k} P D(w)^{\otimes k}$. For a unit phase
$w\notin\mu_4$, conjugating an $X$ or $Y$ factor by $D(w)$ gives
a linear combination of $X$ and $Y$ with both coefficients nonzero.
Expanding the tensor product therefore gives at least two distinct
Pauli strings with nonzero coefficients. Linear independence of Pauli
strings shows that this conjugate is not a Pauli, a contradiction.
Thus $w\in\mu_4$.

Conversely, $D(w)$ is Clifford for every $w\in\mu_4$.
Conjugation by its tensor power preserves Clifford membership, so
every phase in $z\mu_4$ gives a frame for $U$.
\end{proof}

\begin{proof}[Proof of Corollary~\ref{cor:three-gate-witness}]
Choose a nonmonomial gate $U\in\G$ and a gate $V\in\G$ that
entangles a product state, taking $U=V$ if one gate has both properties.
The sublibrary $\{U,V\}$ already fails \textup{(i)} and \textup{(ii)}.
If $U$ has no diagonal Clifford frame, this sublibrary fails
\textup{(iii)} as well.

Otherwise choose a frame phase $z$ for $U$. Since $\G$ has no common
frame, some $W\in\G$ is not Clifford in the frame $D(z)$.
Lemma~\ref{lem:frame-uniqueness} restricts every frame for $U$ to
$z\mu_4$, where $\mu_4=\{1,i,-1,-i\}$. Multiplication by an element
of $\mu_4$ is a Clifford change of basis, so $W$ fails in every one
of these frames. Hence $\{U,V,W\}$ fails all three conditions.
It contains at most three gates, or at most two when $U=V$.
Theorem~\ref{thm:intro-dichotomy} gives the claimed hardness.
The library $\{H,T,\CNOT\}$ from the introduction fails all three
conditions while every pair satisfies one, proving sharpness.
\end{proof}

\begin{corollary}[Libraries containing CNOT]
\label{cor:cnot}
Suppose $\CNOT\in\G$. If every gate in $\G$ is monomial, or every gate
is Clifford, then $\Amp(\G)\in\FP$. Every other such library is
$\#\mathrm P$-hard. Each other gate may act on any fixed number of qubits.
\end{corollary}
\begin{proof}
CNOT entangles a product input. If $D=\diag(1,z)$ gives a common
diagonal Clifford frame, then $(D^\dagger)^{\otimes2}\CNOT D^{\otimes2}$
is Clifford. Its nonzero entry in column $(a,b)$ is $z^{-a+2ab}$.
Ratios of nonzero entries of a Clifford matrix are fourth roots of
unity~\cite{CGW18}. Comparing inputs $00$ and $10$ gives $z^4=1$.
Hence $D$ is Clifford, and every gate in the common frame is an ordinary
Clifford gate. Apply Theorem~\ref{thm:intro-dichotomy}.
\end{proof}

Thus a library containing CNOT is hard as soon as it contains a
nonmonomial gate and a non-Clifford gate. These two properties may
come from different gates.
Hadamard gives a complementary test for diagonal interactions:
$\{H,D_1,\ldots,D_s\}$ is tractable if all $D_j$ are products of
one-qubit diagonal gates, or all are Clifford, and is hard otherwise
(Corollary~\ref{app:hadamard}). For
$D_k(t)\ket x=t^{x_1\cdots x_k}\ket x$, $|t|=1$, the library
$\{H,D_k(t)\}$ is tractable for all phases when $k=1$, exactly
$t=\pm1$ when $k=2$, and only $t=1$ when $k\ge3$.

To test compatibility for a family of rotated Hadamard gates, write
$H_z=D(z)HD(z)^\dagger$, where $D(z)=\diag(1,z)$ and $|z|=1$.

\begin{corollary}[Compatibility of phase frames]
\label{cor:phase-compatibility}
Let $\G=\{\CZ,H_{z_1},\ldots,H_{z_s}\}$, where $s\ge1$ and $|z_j|=1$.
If $z_j/z_1\in\{1,-1,i,-i\}$ for every $j$, then
$\Amp(\G)\in\FP$; otherwise $\Amp(\G)$ is $\#\mathrm P$-hard.
\end{corollary}
\begin{proof}
CZ excludes product preservation, and each $H_{z_j}$ excludes
monomiality. In a common frame $D(w)$, CZ is unchanged and $H_{z_j}$
becomes $H_{z_j/w}$. The matrix $H_t$ is Clifford exactly when $t^4=1$.
This forces every $z_j/z_1$ to be a fourth root of unity. Conversely,
under this condition $w=z_1$ supplies a common frame.
Apply Theorem~\ref{thm:intro-dichotomy}.
\end{proof}

The same frame constraint gives the parameter classifications for the
MS pulse defined in Section~\ref{sec:consequences}.

\begin{corollary}[MS pulses and their combinations]
\label{cor:ms-regimes}
For fixed real angles, the following libraries have exact amplitudes
in $\FP$ under the listed conditions and are $\#\mathrm P$-hard otherwise:
\begin{enumerate}[label=\textup{(\roman*)},leftmargin=2em]
\item $\{M_\phi(\theta)\}$, when $\theta\in(\pi/2)\Z$.
\item $\{M_{\phi_1}(\pi/2),M_{\phi_2}(\pi/2)\}$, when
$\phi_1-\phi_2\in(\pi/2)\Z$.
\item $\{M_\phi(\pi/2),R_z(\lambda)\}$, where
$R_z(\lambda)=e^{-i\lambda Z/2}$, when $\lambda\in(\pi/2)\Z$.
\end{enumerate}
\end{corollary}
\begin{proof}
Since $D(e^{i\phi})XD(e^{i\phi})^\dagger=\sigma_\phi$,
the frame $D(e^{i\phi})$ sends $M_\phi(\theta)$ to
$e^{-i\theta X\otimes X/2}$. At $\theta\in(\pi/2)\Z$ this is
Clifford, giving tractability in \textup{(i)}. At every other angle,
\[
 M_\phi(\theta)\ket{00}
 =\cos(\theta/2)\ket{00}
   -ie^{2i\phi}\sin(\theta/2)\ket{11}.
\]
Both coefficients are nonzero and have unequal magnitudes.
The output is entangled, excluding product preservation, and it
branches, excluding monomiality. Every Clifford column is a stabilizer
state and has constant magnitude on its nonzero support~\cite{CGW18}.
Diagonal conjugation preserves entry magnitudes, so these unequal
coefficients also exclude a diagonal Clifford frame. The dichotomy
gives hardness at all remaining angles.

For \textup{(ii)}, each pulse at $\theta=\pi/2$ is nonmonomial and
entangles $\ket{00}$. It therefore excludes the first two tractable
conditions. Lemma~\ref{lem:frame-uniqueness} shows that the frame
phases for $M_\phi(\pi/2)$ are exactly $e^{i\phi}\mu_4$, where
$\mu_4=\{1,i,-1,-i\}$. The two sets of phases intersect exactly when
$e^{i(\phi_1-\phi_2)}\in\mu_4$, which proves the stated condition.
Equivalently, in the frame $D(e^{i\delta})$ the pulse becomes
$M_{\phi-\delta}(\pi/2)$, Clifford precisely when
$\phi-\delta\in(\pi/2)\Z$.

For \textup{(iii)}, the MS gate again excludes the first two conditions.
The rotation $R_z(\lambda)$ commutes with every diagonal frame, so
it can share such a frame with the pulse exactly when it is already
Clifford. Its Pauli conjugation rotates $X$ into
$\cos\lambda\,X+\sin\lambda\,Y$, which is a Pauli exactly at
$\lambda\in(\pi/2)\Z$. At these angles it remains Clifford in
the frame $D(e^{i\phi})$, proving sufficiency as well.
\end{proof}

\subsection{Single gates, combinations and parameter families}

\begin{corollary}[Different column-magnitude multisets]
\label{cor:column-magnitudes}
If two columns of a unitary qubit gate $U$ have different multisets of
squared entry magnitudes, then $\Amp(\{U\})$ is $\#\mathrm P$-hard.
The multisets include zeros with their multiplicities.
\end{corollary}
\begin{proof}
We show that all columns of a gate in any of the three tractable classes
have the same multiset of squared magnitudes. For a monomial gate,
each column has one entry of modulus one and all other entries zero.
The two columns of a one-qubit unitary have squared magnitudes
$(p,1-p)$ and $(1-p,p)$. A tensor product of such matrices has column
multisets obtained by multiplying the same local multisets, independently
of the input string. A wire permutation only permutes columns or rows.
Thus product-preserving gates also have identical column multisets.

For a Clifford gate $C$, write $\ket x=X^x\ket{0^k}$ with
$X^x=\bigotimes_jX^{x_j}$. Then
$C\ket x=(CX^xC^\dagger)C\ket{0^k}$, so all columns are Pauli
translates of the first. A Pauli only permutes coordinates and changes
their phases, and diagonal conjugation also preserves all entry
magnitudes. This proves the required equality for the third class.
Different column multisets therefore exclude all three conditions,
and Theorem~\ref{thm:intro-dichotomy} gives hardness.
\end{proof}

\begin{corollary}[A deterministic column and a branching column]
\label{cor:deterministic-column}
If a nonmonomial unitary $U$ maps even one computational-basis state to
a computational-basis state up to phase, then $\Amp(\{U\})$ is
$\#\mathrm P$-hard. In particular, every nonmonomial particle-number-conserving
qubit gate has this property.
\end{corollary}
\begin{proof}
A deterministic column has squared-magnitude multiset
$\{1,0,\ldots,0\}$. A nonmonomial unitary has a branching column,
whose multiset differs, so Corollary~\ref{cor:column-magnitudes}
applies. A particle-number-conserving gate has a deterministic column
at the one-dimensional vacuum sector.
\end{proof}

For example, a controlled-$V$ has a deterministic column at control
value zero and a branching column at control value one whenever $V$
is nonmonomial. This gives singleton hardness for controlled-Hadamard.
For the nonmonomial exchange and fSim gates of
Section~\ref{sec:consequences}, Corollary~\ref{app:fixed-filling}
shows that hardness persists at each fixed rational filling density
strictly between zero and one.

Tensoring separately easy gates can produce a hard gate.
For any nonmonomial one-qubit $R$, both $\{\mathrm{Toffoli}\}$
and $\{R\}$ are tractable, whereas the single four-qubit gate
$\mathrm{Toffoli}\otimes R$ is hard. Different placements can
interleave its two factors (Proposition~\ref{app:packing}).
Boundary changes can also produce hardness: allowing both
$\ket+$ inputs and $\bra+$ outputs makes Toffoli amplitudes hard
(Proposition~\ref{app:boundaries}).

With the number of qubits fixed for each gate and at least one multiqubit gate, the complement of the
three classes is open, dense, and of full Haar measure
(Theorem~\ref{param:geometry}). Along a rational one-phase family with algebraic coefficients that is
unitary at every unit phase away from its poles, either every
admissible phase is tractable or only finitely many are;
the exceptional phases in the latter case are algebraic and
effectively computable (Theorem~\ref{param:finite}).
These are pointwise statements about exact amplitudes, with each
coefficient presentation fixed separately. The next appendix develops
the circuit applications, including the fixed-angle exchange compiler
(Appendix~\ref{app:exchange-compiler}) and the comparison between finite
commuting dynamics and amplitude complexity
(Appendix~\ref{app:generated-groups}). Appendix~\ref{app:parameters}
proves the parameter results above, and Appendix~\ref{app:closure}
relates algebraic exact evaluation to fixed-width gate closures.

\section{Further consequences}
\label{app:applications}

Gates in this appendix may have arbitrary fixed complex entries. Exact
answers use the polynomial normal forms of Appendix~\ref{app:coefficients},
with the full rational relation ideal of the fixed coefficient tuple.
Whenever additional boundary vectors are allowed, their real and imaginary
coordinates are included in that tuple. Each library and its coefficient
presentation are fixed before the circuit input is given.

We first classify concrete libraries and examine particle-number
restrictions. We then turn to approximate exchange compilation, where
the angle is assumed polynomial-time computable, and finally study
tensor packing, boundary states, and generated groups. The exact
classifications use the fixed symbolic presentations above.

\subsection{Diagonal gates and nonuniform columns}

For a one-qubit gate, Clifford membership has a short entrywise test.
A diagonal or antidiagonal unitary is Clifford precisely when its two
nonzero entries have fourth-root ratio. A unitary with four nonzero
entries $\left(\begin{smallmatrix}a&b\\c&d\end{smallmatrix}\right)$
is Clifford precisely when
\[
 |a|^2=\tfrac12,\qquad (b/a)^4=(c/a)^4=1.
\]
Orthogonality then forces $d=-a(b/a)(c/a)$, giving the affine quadratic
phase relation~\cite{CGW18}. Thus Corollary~\ref{cor:cnot} is directly
testable for one-qubit additions to CNOT.

\begin{corollary}[Hadamard and diagonal gates]
\label{app:hadamard}
For $\G=\{H,D_1,\ldots,D_s\}$ with each diagonal gate $D_j$ acting on any fixed
number of qubits, the tractable condition is that all $D_j$ are products of
one-qubit diagonal gates, or that all $D_j$ are Clifford.
Every other such library has $\#\mathrm P$-hard exact amplitudes.
\end{corollary}
\begin{proof}
Hadamard excludes monomiality. A diagonal product-permutation unitary
must be a tensor product of diagonal one-qubit gates. Indeed, after a
wire permutation, output $j$ depends on one input bit. To equal
$\ket{x_j}$ up to phase for every independently chosen input string,
that bit must be $x_j$ and the local factor must be diagonal.
If a common frame $D=\diag(1,z)$ exists, $D^\dagger HD$ is Clifford.
Its entry ratios include $z$, so $z^4=1$ and $D$ is itself Clifford.
The remaining gates must then be ordinary Clifford gates.
Theorem~\ref{thm:intro-dichotomy} proves both implications.
\end{proof}

A diagonal gate $D_f=\diag(f)$ is Clifford precisely when
\begin{equation}
 f(x)=\lambda i^{\sum_j a_jx_j+2\sum_{j<\ell}b_{j\ell}x_jx_\ell},
 \qquad a_j\in\mathbb Z_4,\quad b_{j\ell}\in\mathbb F_2,
 \quad |\lambda|=1.
 \label{app:diagonal-phase}
\end{equation}
For $D_k(t)\ket x=t^{x_1\cdots x_k}\ket x$, $|t|=1$, the resulting
classification is
\begin{center}
\begin{tabular}{ll}
\toprule
Library & Parameters giving polynomial-time amplitudes\\
\midrule
$\{H,D_1(t)\}$ & Every $t$\\
$\{H,D_2(t)\}$ & $t\in\{1,-1\}$\\
$\{H,D_k(t)\}$, $k\ge3$ & $t=1$\\
\bottomrule
\end{tabular}
\end{center}
The parameter $t$ may be any fixed complex phase. All remaining parameters
give $\#\mathrm P$-hard amplitudes. For $k\ge2$, a nontrivial $D_k(t)$
is not a tensor product of one-qubit gates: the phase ratios along one input coordinate
depend on the others. Equation~\eqref{app:diagonal-phase} permits only
a sign on a quadratic interaction and no isolated interaction of degree
at least three, which proves the table.

\begin{corollary}[A nonuniform column]
\label{app:biased}
Suppose $\G$ contains an entangling gate and some circuit over $\G$ has
a computational-basis column with two unequal nonzero magnitudes.
Then $\Amp(\G)$ is $\#\mathrm P$-hard. In particular, a one-qubit
unitary $B$ satisfying $0<|B_{00}|^2<1$ and $|B_{00}|^2\ne1/2$,
together with any entangling gate, gives a hard library.
\end{corollary}
\begin{proof}
The entangler excludes product preservation, and the branching column
excludes monomiality. Clifford columns have constant magnitude on their
nonzero support. A common diagonal frame preserves these magnitudes by
the boundary-phase identity in Section~\ref{sec:classification}, so a common diagonal Clifford frame is also impossible.
Theorem~\ref{thm:intro-dichotomy} applies.
\end{proof}

Corollary~\ref{cor:column-magnitudes} gives a complementary singleton
test: two columns with different multisets of squared magnitudes make
the gate hard by itself. The qubit
assumption matters. In a higher local dimension, a one-system unitary
can have both a deterministic column and a branching column while
never creating inter-system entanglement.

\subsection{Number conservation at extensive filling}
\label{app:number-conservation}

\begin{corollary}[Prescribed filling]
\label{app:half-filling}
\label{app:fixed-filling}
Fix a rational density $\rho=p/q$ with $0<p<q$. For a nonmonomial
number-conserving gate $U$, singleton amplitude hardness persists on
$N=q\ell$ qubits when the input is fixed to
$1^{p\ell}0^{(q-p)\ell}$ and the output has weight $p\ell$.
In particular, hardness holds with a fixed half-filled input and a
half-filled output.
\end{corollary}
\begin{proof}
Corollary~\ref{cor:deterministic-column} proves unrestricted singleton
hardness. A query whose boundary weights differ is known to have value
zero. For any remaining query on $n\ge1$ wires, let the common weight be
$w$ and choose
\[
 \ell=\max\left\{
       \left\lceil\frac{w}{p}\right\rceil,
       \left\lceil\frac{n-w}{q-p}\right\rceil\right\},
 \qquad N=q\ell.
\]
Add $p\ell-w$ idle wires fixed to $\ket1$ at both boundaries and
$(q-p)\ell-(n-w)$ idle wires fixed to $\ket0$. Both numbers are
nonnegative, the new width is $N$, and both boundary weights are
$p\ell$. Since $p,q$ are fixed, $N=O(n)$. Relabel the wires to place
the occupied input positions first, carrying the output string and all
ordered gate placements along with the relabelling. The amplitude
is unchanged, and the added idle wires use only the existing basis
boundary coefficients. Thus the reduction also
preserves the output presentation. The output string may be any string
of the required weight. Only the input is fixed in canonical order.
\end{proof}

\begin{corollary}[Exchange and fSim gates]
\label{cor:exchange-fsim}
For any fixed real $\theta,\varphi$, the singleton library
$\{U_\theta\}$ or $\{\operatorname{fSim}(\theta,\varphi)\}$ has
exact amplitudes in $\FP$ when $\theta\in(\pi/2)\Z$ and is
$\#\mathrm P$-hard otherwise. At every hard angle, hardness persists
at each prescribed rational filling density $0<\rho<1$.
\end{corollary}
\begin{proof}
The exchange pulse is
$U_\theta=e^{i\theta\SWAP}=\cos\theta\,I+i\sin\theta\,\SWAP$.
In the computational basis ordered as $00,01,10,11$, the fSim gate is
\[
 \operatorname{fSim}(\theta,\varphi)
   =1\;\oplus\!
     \begin{pmatrix}\cos\theta&-i\sin\theta\\
                     -i\sin\theta&\cos\theta\end{pmatrix}
       \!\oplus\;e^{-i\varphi}.
\]
Both gates preserve particle number. At $\theta\in(\pi/2)\Z$
they are monomial. At every other angle, their $\ket{01}$ columns
have two nonzero entries, of magnitudes $|\cos\theta|$ and
$|\sin\theta|$, while their $\ket{00}$ columns are deterministic.
Corollary~\ref{cor:deterministic-column} therefore gives
singleton hardness, and Corollary~\ref{app:fixed-filling} preserves
it at each fixed rational density strictly between zero and one.
\end{proof}

For fSim the coefficient presentation contains the full relations
among both phases, including any relations coupling the parameters.
A higher-body example is a four-qubit gate equal to the identity
except for a nonmonomial rotation of
$\operatorname{span}\{\ket{0011},\ket{1100}\}$.

For a length-$m$ circuit with gates on at most $k$ qubits, direct evolution in
the weight-$w$ sector costs $O(m2^k\binom nw)$ ring operations.
Every intermediate entry is an amplitude of a circuit prefix, so it
has a polynomial representative of degree at most $m$ and polynomial
coefficient bit length. Reduction by the fixed relation ideal therefore
gives polynomial bit cost when the particle number or the hole number
is constant. The hardness statement permits extensive filling.
Fermionic-mode interactions have a different placement convention:
their qubit representations include Jordan--Wigner parity strings.
Applying a fixed two-qubit matrix to arbitrary nonadjacent ordinary
qubits generally omits these strings. The number-conserving conclusion
is therefore consistent with noninteracting-fermion and nearest-neighbour
matchgate simulation~\cite{TD02,JM08}.

For example, the ideal Hartree--Fock circuits in~\cite[Fig.~1]{Arute20}
compile fermionic Givens rotations into nearest-neighbour
$\sqrt{\mathrm{iSWAP}}$ gates and single-qubit $Z$ rotations. In our
convention, $\sqrt{\mathrm{iSWAP}}=\operatorname{fSim}(-\pi/4,0)$.
The nearest-neighbour geometry preserves the free-fermion structure
underlying classical simulation~\cite{Val02,TD02}, although unrestricted placements of
$\sqrt{\mathrm{iSWAP}}$ give a hard singleton library by
Corollary~\ref{cor:exchange-fsim}. Other simulation algorithms exploit
polynomially bounded Schmidt ranks of the evolving state~\cite{Vid03}
or a circuit graph whose treewidth is logarithmic in the circuit
size~\cite{MarkovShi08}. These guarantees use restrictions on the
circuit instances beyond the fixed gate library.

The preceding exact classification also covers transcendental phases
under the symbolic output convention. For a single exchange phase,
the next lemma gives the full relation ideal explicitly, supplying the
presentation needed to apply that classification.

\begin{lemma}[A transcendental phase presentation]
\label{app:circle-presentation}
Let $z=c+is$ be a unit phase that is transcendental over $\mathbb Q$,
with $c,s\in\mathbb R$. Its full rational relation ideal is
\[
 \{p\in\mathbb Q[C,S]:p(c,s)=0\}
       =(C^2+S^2-1).
\]
Consequently, the coefficients of the exchange pulse
$U_\theta=\cos\theta I+i\sin\theta\,\SWAP$, with
$z=e^{i\theta}$ transcendental, have an explicit finite presentation.
In the coordinates $C=\cos\theta$, $S=\sin\theta$, an exact amplitude
of a length-$m$ circuit has a unique representative
$P(S)+CQ(S)$ with $P,Q\in\mathbb Q(i)[S]$ and total degree at most $m$.
\end{lemma}
\begin{proof}
There is an isomorphism
\[
 \mathbb Q(i)[C,S]/(C^2+S^2-1)
       \cong\mathbb Q(i)[Z,Z^{-1}],
 \qquad
 C\longmapsto\frac{Z+Z^{-1}}2,
 \quad S\longmapsto\frac{Z-Z^{-1}}{2i}.
\]
Its inverse sends $Z$ to $C+iS$ and $Z^{-1}$ to $C-iS$.
Since $z$ is transcendental over $\mathbb Q$, it is also transcendental
over the algebraic extension $\mathbb Q(i)$. Evaluation at $Z=z$ is
therefore injective on the Laurent polynomial ring. Its kernel in
$\mathbb Q(i)[C,S]$ is exactly the displayed circle ideal. Contracting
this ideal to $\mathbb Q[C,S]$ gives the same principal ideal: division
by the monic polynomial $C^2+S^2-1$, viewed as a polynomial in $C$,
has rational coefficients and a unique remainder of $C$-degree at most
one.

Every real or imaginary entry of $U_\theta$ is zero, $C$, or $S$,
and both $C,S$ occur. Thus the relation ideal in the original entry
coordinates is obtained by adjoining their explicit linear
identifications to the circle relation. With graded lexicographic
order $C>S$, the leading monomial of that relation is $C^2$.
Reduction gives precisely $P(S)+CQ(S)$ without increasing total
degree. A circuit amplitude is a polynomial of degree at most $m$
in the entries, giving the asserted bound. The linear identifications
and Lemma~\ref{fg:embedding-descent} give polynomial-time conversion
between this presentation and the entry-coordinate normal form.
\end{proof}

For every transcendental exchange phase, the lemma therefore supplies
the full relation ideal required by the output convention. With several
phases, that convention also includes all rational relations coupling
their coordinates.

\subsection{Compiling exchange pulses at a fixed angle}
\label{app:exchange-compiler}

Burkat, Strelchuk, and Studzi\'nski proved encoded universality for
$\pi/4$ exchange pulses and asked which other fixed angles suffice,
including $\pi/6$~\cite[Theorem~5 and Section~VIII]{BSS26}.
At $\theta\in(\pi/2)\Z$ each pulse is the identity or $\SWAP$ up to
phase. At every other efficiently computable angle, the following
construction compiles the $\pi/4$ pulse.

\begin{theorem}[Fixed-angle exchange compilation]
\label{thm:exchange-compiler}
Fix a polynomial-time computable $\theta\notin(\pi/2)\Z$.
Using only forward $e^{i\theta\SWAP}$ pulses on arbitrary pairs of
wires and two reusable auxiliary qubits initialized to $\ket{00}$,
one can implement $e^{i(\pi/4)\SWAP}$ on any data pair up to a global
phase and operator-norm error $\epsilon$, with compilation time and
gate count polynomial in $1/\epsilon$. Consequently, this fixed pulse
suffices for encoded universal quantum computation with computational-basis
preparation and readout.
\end{theorem}

More precisely, write $E_{ij}=\SWAP_{ij}$ and
$U_{ij}(t)=e^{itE_{ij}}$. The compiler constructs a four-wire unitary
$V_\epsilon$ satisfying
\[
 \bigl\|V_\epsilon-U_{12}(\pi/4)U_{34}(-\pi/4)\bigr\|\le\epsilon.
\]
This operator norm compares the full four-wire unitaries. On auxiliary
input $\ket{00}_{34}$, the target applies $U_{12}(\pi/4)$ to the data
and returns the auxiliary pair with phase $e^{-i\pi/4}$.
The polynomial bounds may depend on the fixed angle $\theta$.

\begin{proof}
As in~\cite[Lemma~5]{BSS26}, we construct the four-wire target from
two overlapping three-wire circuits. On three wires, the construction
must match the action on two identical two-dimensional blocks as well
as the phase on their complement. We first compile the action on the blocks,
then cancel the complementary phase and combine the two circuits.
Finally, we replace inverse pulses by forward pulses. All errors below
are in operator norm.

Let $\ket{s}=(\ket{01}-\ket{10})/\sqrt2$ and let $P_{ij}$ project
qubits $i,j$ onto $\ket{s}$, so $E_{ij}=I-2P_{ij}$. Contracting two
singlets gives $(\bra{s}_{12}\otimes I_3)(I_1\otimes\ket{s}_{23})=-I/2$,
where the remaining qubit spaces are identified in the computational
basis. Thus $P=P_{12}$ and $Q=P_{23}$ on three wires satisfy
\[
       PQP=P/4,\qquad QPQ=Q/4.
\]
Take an orthonormal basis $u_1,u_2$ of $\operatorname{im}P$ and set
$v_j=2Qu_j$. The $v_j$ form an orthonormal basis of
$\operatorname{im}Q$, with $\langle u_i,v_j\rangle=\delta_{ij}/2$.
The planes $L_j=\operatorname{span}(u_j,v_j)$ are orthogonal. On
each, $P$ and $Q$ are the rank-one projectors onto $u_j$ and $v_j$,
and on $(L_1\oplus L_2)^\perp$ both vanish. In particular the two
planes carry identical representations, and each physical pulse acts
on the complement as the scalar $e^{i\theta}$.

On either plane write $R=I-2P$, $S=I-2Q$ for the restricted
reflections and put $A=e^{i\theta R}$, $B=e^{i\theta S}$. We show
that $A,B$ generate a dense subgroup of $SU(2)$ whenever
$\theta\notin(\pi/2)\Z$. The reflections are traceless and
$\Tr(RS)=-1$ because $\Tr(PQ)=1/4$ on the plane. We first
exclude a finite generated group using the trace identity
\[
        \Tr(AB)=2\cos^2\theta+\sin^2\theta=1+\cos^2\theta.
\]
If $\theta/\pi$ is irrational, $A$ has infinite order. Otherwise
$\zeta=e^{2i\theta}$ is a root of unity of order greater than two.
Every conjugate of $x=\cos^2\theta=(2+\zeta+\zeta^{-1})/4$ lies
strictly between zero and one, since every conjugate of $\zeta$ has
the same order. If $x$ were an algebraic integer, its norm from
$\Q(x)$ to $\Q$ would be a positive integer less than one. Thus
$1+x$ is not an algebraic integer, whereas the trace of a
finite-order matrix is a sum of roots of unity. In this case $AB$
has infinite order.

The closure of the group is therefore a positive-dimensional compact
Lie subgroup of $SU(2)$. A proper such subgroup has a one-dimensional
Lie algebra. Under $\mathfrak{su}(2)\cong\R^3$ the bracket is a
multiple of the cross product, so there is no two-dimensional
subalgebra. The subgroup must then preserve its Lie algebra as an
unoriented line under conjugation. The Bloch rotations for $A,B$ have distinct axes since
$\Tr(RS)/2=-1/2$. Each has rotation angle $2\theta$ up to orientation,
so neither is the identity or a half-turn. Each therefore preserves
only its own axis as a line. This rules
out a proper closure. The rationality split establishes density.
The algorithm uses the fixed pulse matrices to compile, with no
rationality test on $\theta/\pi$.

First allow inverse pulses. Fix $t=\pi/4$. By the Solovay--Kitaev
algorithm~\cite{DN06}, for $0<\delta<1/4$ one can compile
$e^{itR}$ on the plane to error $\delta$ using a word of length
$O_\theta(\log^c(1/\delta))$ in $A^{\pm1},B^{\pm1}$ for a constant $c$.
Use the same word in the physical pulses on pairs $12$ and $23$,
and call it $w$. It approximates $U_{12}(t)$ on $L_1\oplus L_2$
and acts as some scalar $\lambda$ on the complement. To cancel this
complementary phase, relabel wires $1$ and $3$ in the word, obtaining
$\widetilde w$. At the operator level this is conjugation by the
wire permutation that swaps $1$ and $3$. It interchanges $P$ and $Q$,
preserves the sum of their images, and leaves the scalar $\lambda$
unchanged. Conjugating the approximation and using invariance of the
operator norm shows that $\widetilde w$ approximates $U_{23}(t)$
on that sum with the same error $\delta$.

On the complement, $w\widetilde w^\dagger$ acts as
$\lambda\overline\lambda=1$. Both target pulses act there as
$e^{it}$, so $U_{12}(t)U_{23}(-t)$ also acts as the identity.
The block errors therefore give the full three-wire estimate
\[
    \|w\widetilde w^\dagger-U_{12}(t)U_{23}(-t)\|\le 2\delta .
\]
To cancel the intermediate pulse on wires $2,3$, place a second copy
of this construction on wires $2,3,4$. The two ideal operations compose to
\[
   [U_{12}(t)U_{23}(-t)] [U_{23}(t)U_{34}(-t)]
       = U_{12}(t)U_{34}(-t).
\]
The four-wire error is at most $4\delta$.

To obtain error $\epsilon$ with forward pulses only, allocate half
the error to block compilation and half to inverse-pulse replacement.
Choose $\delta=\epsilon/8$, with $0<\epsilon<1$, and let $m$ be the
length of the resulting four-wire word. Set $\eta=\epsilon/(2m)$.
Among
$1,e^{i\theta},\ldots,e^{iN\theta}$, two points lie in the same one
of $N=\lceil 8\pi/\eta\rceil$ equal arcs. Their index difference
$q$ satisfies $1\le q\le N$ and $|e^{iq\theta}-1|\le\eta/4$.
Scanning these indices with numerical error at most $\eta/4$ finds a
$q$ with $|e^{iq\theta}-1|\le\eta$ (accept a computed distance at
most $\eta/2$). The eigenvalues of $E_{ij}$ are $\pm1$, so replacing
$U_{ij}(\theta)^{-1}$ by $U_{ij}(\theta)^{q-1}$ costs at most $\eta$
in operator norm. The total error is at most $4\delta+m\eta=\epsilon$.
Since $\theta$ is fixed and computable to $b$ bits in time polynomial
in $b$, the search, the compilation
on the plane, and the output length are all polynomial in $1/\epsilon$.

Finally $U_{34}(-t)\ket{00}=e^{-it}\ket{00}$. The ideal four-wire
unitary therefore applies $U_{12}(t)$ to arbitrary data up to a global
phase and returns the auxiliary pair exactly to $\ket{00}$.
The error estimate holds on the full four-wire space, so it also
applies to the perturbed auxiliary state passed to
the next use. To simulate a circuit of $L$ such gates, use the same
auxiliary pair throughout and choose error $\epsilon/L$ per replacement.
A telescoping sum on the full data-and-auxiliary space bounds the
total operator error by $\epsilon$, even for data entangled with
other wires. Thus the auxiliary pair can be reused without intermediate
reset. The simulation preserves the computational-basis preparation
and readout of~\cite[Theorem~5]{BSS26}, with bounded error and
polynomial overhead.
\end{proof}

\subsection{Tensor products and boundary states}

\begin{proposition}[Packing a library into one gate]
\label{app:packing}
For $\G=\{U_1,\ldots,U_s\}$, put $W=U_1\otimes\cdots\otimes U_s$.
Then $\Amp(\{W\})\equiv_T^{\poly}\Amp(\G)$. Each direction uses one
amplitude query, linear circuit overhead, and polynomial-time conversion
to the requested output normal form. The reduction from $\G$ to
$\{W\}$ also uses a known nonzero scalar correction. From the full
relation ideal of $\G$, that of $W$ can be computed during fixed
preprocessing.
\end{proposition}
\begin{proof}
Expanding the tensor product simulates $W$ over $\G$, and placing
the unused factors on private wires simulates a gate of $\G$ over
$\{W\}$ up to a nonzero scalar. To turn these circuit constructions
into reductions between the stated function problems, we also convert
the oracle answer to the coefficient normal form requested on each side.

Let $X$ denote the real and imaginary entry coordinates of $\G$,
and let $Y$ denote those of $W$. Tensor multiplication expresses
$Y=H(X)$ for a fixed tuple of rational polynomials. Write $I_\G,I_W$
for the two full relation ideals. Then
\[
 I_W=\bigl(I_\G+(Y-H(X))\bigr)\cap\mathbb Q[Y],
\]
where the ideal on the right is formed in $\mathbb Q[X,Y]$.
Indeed, substitution identifies its quotient with
$\mathbb Q[X]/I_\G$, and its intersection with $\mathbb Q[Y]$ is
exactly the kernel of evaluation at the entries of $W$. This computes
$I_W$ by elimination and defines a fixed polynomial injection
\[
 \eta:\mathbb Q[Y]/I_W\hookrightarrow\mathbb Q[X]/I_\G,
       \qquad Y\longmapsto H(X).
\]
This elimination is fixed preprocessing, whose cost may depend on
the library and its supplied presentation.

To evaluate a circuit with $m$ occurrences of $W$, expand each occurrence
into its factors on disjoint wires and query $\Amp(\G)$. The returned
normal form represents the desired value in the larger coefficient
ring. To recover its coordinates in the smaller $W$ ring, use the
fact that a circuit with $m$ occurrences of $W$ has a polynomial
representative of degree at most $m$ in those coordinates.
Lemma~\ref{fg:embedding-descent}, applied to the inclusion $\eta$
and separately to the real and imaginary parts, recovers the normal
form modulo $I_W$ with polynomial bit cost. Thus this direction
uses only one amplitude query.

Conversely, choose a nonzero entry
$u_j=\bra{a_j}U_j\ket{b_j}$ of every factor, using exact zero tests
modulo $I_\G$. To simulate an occurrence of $U_i$, place its block
of $W$ on the desired wires and put every other block on fresh
private wires with input $b_j$ and output $a_j$. Retire those wires
until the final boundary. This implements
$(\prod_{j\ne i}u_j)U_i$. For an original circuit of length $m$,
the single resulting $W$ query is its amplitude multiplied by a
known nonzero product of these fixed entry factors, with exponents
$O(m)$. This direction first removes the scalar correction, then
converts the resulting expression back to the original coefficient ring.
Substitute $Y=H(X)$ in the oracle answer. Inverses of all
the factors $u_j$ belong to the fixed localization used in
Lemma~\ref{fg:descent}. Division by the correction therefore gives
a localized polynomial of degree and coefficient bit length
polynomial in the circuit size. The requested amplitude has a
degree-$m$ representative in the original $X$ coordinates, so that
lemma converts the result to the prescribed normal form modulo
$I_\G$. Both circuit constructions have linear overhead and preserve
acyclicity and distinct-wire placements.
\end{proof}

Proposition~\ref{app:packing} transfers the exact-amplitude complexity
of a whole library to its tensor-product gate. In particular, the union
of the three easy families is not closed under tensor products of
separately tractable gates. If $R$ is any nonmonomial
one-qubit rotation, $\{\mathrm{Toffoli}\}$ and $\{R\}$ are separately
easy, whereas $\{\mathrm{Toffoli}\otimes R\}$ is hard. Toffoli with
a basis-one control supplies CNOT, and Toffoli itself is non-Clifford.
Corollary~\ref{cor:cnot} and Proposition~\ref{app:packing} apply.
Across different placements, the two tensor factors can act on
overlapping sets of wires, allowing their operations to interact.
The monomial and product-preserving families are individually tensor
closed, as is any fixed diagonal Clifford frame.

\begin{proposition}[One-sided and two-sided coherent boundaries]
\label{app:boundaries}
For monomial libraries, product inputs from any fixed finite set of
complex one-qubit states and computational-basis outputs admit polynomial-time
exact evaluation. The same holds with the input and output conventions
reversed. The output presentation includes the boundary coefficients.
Allowing both product inputs and outputs makes Toffoli
amplitudes $\#\mathrm P$-hard as soon as $\ket+$ and $\bra+$ are
allowed alongside basis pins.
\end{proposition}
\begin{proof}
A monomial circuit defines a permutation $p$ of basis strings and a
path phase. Given the output $a$, compute $x=p^{-1}(a)$ by classical
reverse lookup through the fixed gate tables. For input
$\bigotimes_j\psi_j$, the amplitude is its path phase multiplied by
$\prod_j\bra{x_j}\psi_j\rangle$. The reverse convention follows the
unique forward path and evaluates the product effect. Both directions
use classical lookup in the fixed monomial gate matrices.
The resulting formula multiplies at most $m+n$ coefficients from a fixed finite
set, so its degree, coefficient bit length, and normal-form reduction
cost are polynomial in the circuit description. The full relation
ideal of gates and boundaries supplies the required exact zero tests
and output encoding.

For hardness, reversibly compute a Boolean formula $F(x)$ in work bits,
copy its value into a fresh zero result bit, and uncompute the work bits.
Basis-one controls let Toffoli implement X and CNOT. Prepare the $n$
variables in $\ket+^{\otimes n}$, select the result bit as one and the
clean workspace as zero, and apply $\bra+^{\otimes n}$ to the variables.
The amplitude is $2^{-n}\#\mathrm{SAT}(F)$. All selected outputs remain
untouched until the global boundary.
\end{proof}

A local change of basis moves from gate matrices to boundary states
in an amplitude, turning computational-basis endpoints into coherent
product endpoints. The following conjugated-CZ example uses this
relation to exhibit hardness with commuting gates. Let
\[
 V=\frac1{\sqrt5}\begin{pmatrix}2&1\\-1&2\end{pmatrix},
 \qquad W=(V^\dagger\otimes V^\dagger)\CZ(V\otimes V).
\]
Global conjugation by $V^{\otimes n}$ turns all placements of $W$ into
commuting CZ gates. For a circuit $C$ over $\{W\}$, let $C_{\CZ}$
replace each placed $W$ by $\CZ$. Then
\[
 \bra a U_C\ket b
   =\bigl(\bra a(V^\dagger)^{\otimes n}\bigr)
      U_{C_{\CZ}}\bigl(V^{\otimes n}\ket b\bigr).
\]
The endpoints are now product states in the rotated basis.
More explicitly, with
$u=(1,-2,-2,4)^T/5$, one has $W=I-2uu^T$, so
$W^2=I$ and
\[
 W\ket{00}=\frac1{25}
       (23\ket{00}+4\ket{01}+4\ket{10}-8\ket{11}).
\]
The coefficient matrix of this state has determinant $-8/25\ne0$,
so $W$ is entangling. Its unequal nonzero magnitudes give singleton
hardness by Corollary~\ref{app:biased}.
At width $n$, the generated group is isomorphic to
$(\mathbb Z/2\mathbb Z)^{\binom n2}$: the $\CZ_{ij}$ commute and
square to the identity, and evaluating a product on the basis
string supported exactly at $i,j$ detects whether its $(i,j)$
factor occurs. Conjugation preserves this group while rotating its
relation to the computational boundary.
The broader commuting conjugated-CZ family was classified for weak
multiplicative simulation in~\cite[Theorem~5.13]{KFG25}. The present
example gives unconditional $\#\mathrm P$-hardness of exact
amplitudes.

\subsection{Generated groups}
\label{app:generated-groups}

The preceding example gives hardness for a finite commuting group.
The fixed-axis MS pulse $M_\phi(\theta)$, defined with $\sigma_\phi$
in Section~\ref{sec:consequences}, gives a second comparison between
a generated group and its action relative to the boundary basis.
All placements commute, including placements sharing a wire, since
their generators use the same $\sigma_\phi$ on each common wire.
At $\theta=\pi/4$, every pulse has order sixteen. At each fixed width
there are finitely many placements, so they generate a finite abelian
group. Nevertheless, Corollary~\ref{cor:ms-regimes} gives
$\#\mathrm P$-hard exact amplitudes for this singleton library.

For a full diagonalization, let $B=D(e^{i\phi})H$, so
$B^\dagger\sigma_\phi B=Z$ and
\[
 (B^\dagger)^{\otimes2}M_\phi(\theta)B^{\otimes2}
     =e^{-i\theta Z\otimes Z/2}.
\]
Applying this conjugation on every wire gives isomorphic generated
groups at every width. The diagonal-gate model
has polynomial-time computational-basis amplitudes for every $\theta$.
Thus the group action relative to the specified boundaries affects
the complexity of exact evaluation.

This comparison also gives an Ising expression for the MS amplitude.
The eigenvectors of $\sigma_\phi$ have squared overlap $1/2$ with
$\ket0$. Expanding both boundaries in their tensor-product eigenbasis
therefore gives
\[
 \bra{0^n}U_C\ket{0^n}
 =2^{-n}\sum_{s\in\{-1,1\}^n}
       \prod_{(i,j)\in E_C}e^{-i\theta s_i s_j/2},
\]
where $E_C$ lists the pairs of wires acted on by $C$, with multiplicity.
Each edge contributes the eigenvalue of its pulse, so repeated
placements give repeated factors. This is a normalized Ising partition
function with imaginary couplings.

Using only the diagonal part $D=D(e^{i\phi})$ of the preceding
change of basis instead gives
\[
 (D^\dagger)^{\otimes2}M_\phi(\theta)D^{\otimes2}
     =e^{-i\theta X\otimes X/2}.
\]
Since $D^{\otimes n}$ fixes $\ket{0^n}$, this conjugation preserves
the all-zero return amplitude. It identifies the circuit with an
$X$-program at angle $-\theta/2$, whose binary matrix has one row
for each edge of $E_C$, with ones at its two endpoints. This is the
edge--vertex incidence matrix over $\mathbb F_2$, whose row matroid
is the graph's cycle matroid.

Shepherd's Theorem~1 identifies this all-zero amplitude with a
normalized weight enumerator. His Section~2.3 gives the evaluation
classification at algebraic phases, including graphic
matroids~\cite{She10}. For algebraic $e^{-i\theta/2}$, its boundary
in our convention is $\theta\in(\pi/2)\Z$ for tractability, with
hardness at the remaining algebraic phases.
Corollary~\ref{cor:ms-regimes} gives the full
single-pulse classification at arbitrary fixed angles under our
exact-output convention.

Related representations connect IQP output probabilities to Ising
partition functions~\cite{FM17} and exact circuit amplitudes to Tutte
evaluations~\cite{Mann21}. The fixed-point complexity of Tutte
evaluation was classified by Jaeger, Vertigan, and Welsh~\cite{JVW90}.
The gate-library criterion also applies after adjoining gates that
break commutativity, as in Corollary~\ref{cor:cnot}.

An infinite projective group gives a sufficient criterion once
the first two tractable structures have been excluded. If $\G$
contains a nonmonomial gate and an entangling gate, and its placed
gates generate an infinite projective group at some fixed width, then its
exact amplitudes are hard. The first two conditions of
Theorem~\ref{thm:intro-dichotomy} are excluded by hypothesis. The third would
make the group at every fixed width conjugate to a subgroup of the finite
projective Clifford group, a contradiction. The converse fails for
the fixed-axis MS pulse $M_\phi(\pi/4)$ considered in this subsection:
its placements generate a finite abelian group at every width, while
its exact amplitudes are hard. For algebraic libraries, the positive
connection with gate synthesis is the exact-oracle equivalence under
fixed-width projective closure proved in
Theorem~\ref{closure:theorem}.

\section{Parameter families}
\label{app:parameters}

We first allow each gate to vary over its full unitary group, with its
number of qubits fixed. In this ambient space, the three tractable
classes have measure zero whenever some gate acts on at least two qubits.
We then restrict to a specified rational one-parameter family. Such a
curve may lie entirely in the tractable set, and we prove that otherwise
it meets that set at only finitely many parameters. Polynomial equality
tests give a finite or cofinite intersection, and closedness of the
tractable set turns the cofinite case into the whole domain. Complexity
at each parameter value uses its fixed coefficient presentation and the
normal-form output of Appendix~\ref{app:coefficients}.

\subsection{The tractable locus}

Fix the numbers of qubits $k_1,\ldots,k_m\ge1$ on which the $m\ge1$ gates act, and put
\[
 \mathcal U=\prod_{j=1}^mU(2^{k_j}),
 \qquad n_{\mathcal U}=\dim_{\mathbb R}\mathcal U=\sum_{j=1}^m4^{k_j}.
\]
Allow arbitrary complex entries. Let $\mathcal M$ and $\mathcal P$
be the all-monomial and all-product-preserving tuples. Let
$\mathcal D$ consist of tuples admitting a common diagonal Clifford
frame, allowing arbitrary phases, and set
$\mathcal E=\mathcal M\cup\mathcal P\cup\mathcal D$.
Its points are the three structural families of
Theorem~\ref{thm:intro-dichotomy}. We use real semialgebraic dimension
for these sets and normalized Haar measure on the ambient product group
$\mathcal U$. On algebraic tuples, membership in $\mathcal E$ agrees
with the tractable class of Theorem~\ref{thm:algebraic}.
For algebraic gates, a common diagonal phase can be chosen algebraic
by the entry-ratio test of Theorem~\ref{norm:recognition}. This identifies
the algebraic points of $\mathcal D$ with the Clifford-frame case of
that theorem.

\begin{theorem}[The geometry of the three families]
\label{param:geometry}
The set $\mathcal E$ is compact and closed in $\mathcal U$, and
\[
 \dim_{\mathbb R}\mathcal M=\sum_j2^{k_j},\qquad
 \dim_{\mathbb R}\mathcal P=\sum_j(3k_j+1),\qquad
 \dim_{\mathbb R}\mathcal D\le m+1.
\]
If some $k_j\ge2$, then $\mathcal E$ has empty interior and Haar measure
zero. Its complement is open, dense, and of full Haar measure.
Every tuple in the complement has $\#\mathrm P$-hard normal-form
amplitudes. Algebraic tuples in the complement also occur in every
nonempty open subset of $\mathcal U$.
If all $k_j=1$, then $\mathcal P=\mathcal U$.
\end{theorem}
\begin{proof}
For a prescribed basis permutation, each computational-basis vector
carries an independent phase. The monomial unitaries on $k$ qubits
therefore form a torus $(S^1)^{2^k}$ for that permutation. Taking the
finite union over permutations and the product over gates proves
compactness of $\mathcal M$ and gives dimension $\sum_j2^{k_j}$.

The local unitaries on $k$ qubits form the image of the homomorphism
\[
 U(2)^k\longrightarrow U(2^k),\qquad
 (A_1,\ldots,A_k)\longmapsto A_1\otimes\cdots\otimes A_k.
\]
Its kernel consists of tuples $A_j=\mu_jI$ with $\prod_j\mu_j=1$.
To see this, compare eigenvalues of a tensor product equal to the identity,
varying one factor's eigenvalue while fixing the others. Each factor is
scalar. The $k$ local unitaries have $4k$ real parameters, and the
kernel describes $k-1$ independent redistributions of their phases
that leave the tensor product unchanged. The compact image therefore
has dimension $4k-(k-1)=3k+1$. Multiplication by the finitely many wire permutations
and taking products gives the assertions for $\mathcal P$.

For each tuple $(C_1,\ldots,C_m)$ of projective Clifford representatives,
there is one shared frame phase $z$ and an independent global phase
$\lambda_j$ for each gate. The resulting map
\[
 (z,\lambda_1,\ldots,\lambda_m)\longmapsto
 \bigl(\lambda_jD(z)^{\otimes k_j}C_j
                  (D(z)^\dagger)^{\otimes k_j}\bigr)_{j=1}^m,
 \qquad D(z)=\diag(1,z),
\]
is smooth on the compact torus $(S^1)^{m+1}$. The finite union of its
images is $\mathcal D$, proving compactness and the dimension upper
bound. All three sets are semialgebraic, since their parametrizations
are polynomial in real and imaginary coordinates and have semialgebraic
domains. Their union $\mathcal E$ is therefore compact and closed.
Closedness will make the hard locus open and, in
Theorem~\ref{param:finite}, extend tractability from all but finitely
many admissible parameters to the whole domain.

For positive $k$, $2^k<4^k$ and $3k+1\le4^k$, with equality in the
second inequality only at $k=1$. Also $m+1<4m\le n_{\mathcal U}$.
If some $k_j\ge2$, each constituent is a finite union of smooth images
of compact manifolds of dimension below $n_{\mathcal U}$. In finitely
many coordinate charts these maps are Lipschitz. Lower-dimensional
Lipschitz images have zero ambient volume. Haar measure is normalized
smooth volume, so $\mathcal E$ is null and has empty interior.
Closedness makes its complement open and dense.

It remains to find an algebraic hard tuple in every nonempty open
subset of $\mathcal U$. For this, algebraic unitaries are dense:
Hermitian matrices with rational real and imaginary parts are dense
among Hermitian matrices, and their Cayley
transforms $(I+iA)(I-iA)^{-1}$ have algebraic entries and are dense in
the unitaries without eigenvalue $-1$. This latter set is itself dense.
Taking products gives density in $\mathcal U$. Each nonempty open set
intersects the open complement of $\mathcal E$, which therefore contains
an algebraic point. Theorem~\ref{thm:intro-dichotomy}, or equivalently
Theorem~\ref{fg:extension}, proves hardness at every tuple in that
complement. When all gates are one-qubit gates, every gate preserves product states.
\end{proof}

When some prescribed gate size satisfies $k_j\ge2$, every library outside $\mathcal E$
has a neighbourhood of hard libraries, while every tractable library
has hard libraries arbitrarily close to it. These are pointwise
exact-evaluation conclusions, with the presentation fixed separately
at each library.

\subsection{One-parameter rational families}

To study a rational curve of gate libraries, we need membership tests
that become univariate polynomial equalities along the curve.
Proposition~\ref{hol:constructible} supplies a Boolean combination of
polynomial equalities in the complex matrix entries. For fixed gate
dimensions, it uses the predicate
\[
 \Psi(\mathbf U)=\Mono(\mathbf U)\ \lor\
   \exists S\in\mathrm{GL}_2(\mathbb C)\,
                 \ReTract(\mathcal H(\mathbf U,S))
\]
on all complex matrix tuples. Its effective Boolean formula uses
finitely many polynomial equalities
\begin{equation}
 p_1(\mathbf U)=0,\ldots,p_t(\mathbf U)=0
 \label{param:tests}
\end{equation}
with algebraic coefficients. The proposition identifies this formula
with the three quantum structures on algebraic unitary tuples.
The following lemma extends that identification to arbitrary complex
entries, so the same tests apply at every admissible parameter.

\begin{lemma}[Structural comparison over the complex numbers]
\label{param:complex-bridge}
For every unitary tuple $\mathbf U$ of the prescribed matrix dimensions, with arbitrary
complex entries,
\[
 \Psi(\mathbf U)\quad\Longleftrightarrow\quad
 \mathbf U\in\mathcal E.
\]
\end{lemma}
\begin{proof}
Both sides can be expressed by first-order formulas over real closed
fields with algebraic coefficients. For $\Psi$, split every complex
entry and every complex witness into real and imaginary coordinates
in the formula from Proposition~\ref{hol:constructible}. Equations,
non-equations, matrix inverses with nonzero determinant, and the
quantifier over $S$ all have first-order polynomial descriptions.
For $\mathcal E$, monomiality is a finite zero-pattern condition,
and product preservation is a finite disjunction of tensor
factorizations into unitary $2\times2$ matrices. A common diagonal
Clifford frame is described by a finite choice of projective Clifford
representatives, one unit phase for the frame, and one overall phase
for each gate. The corresponding matrix and unit-modulus equations
are polynomial in real and imaginary parts. The representatives can
be chosen algebraic, so all constants used in these formulas are
algebraic. Unitarity is also a polynomial condition.

Over the real algebraic numbers, the sentence asserting this
equivalence for every unitary tuple is
Theorem~\ref{norm:reduction}. The observation at the start of this
appendix identifies its algebraic-phase convention with the
interpretation of $\mathcal E$ over that field. Apply the real
closed field transfer principle stated in
Section~\ref{fg:specialization}, with the common ordered field
generated by the finitely many algebraic constants. The sentence
therefore holds over $\mathbb R$, which proves the asserted
equivalence for arbitrary complex entries. This argument transfers
the structural identity $\Psi(\mathbf U)\iff\mathbf U\in\mathcal E$.
The complexity conclusion at transcendental coefficients comes from
Theorem~\ref{fg:extension}.
\end{proof}

For a rational family $\G(z)$, its phase domain is the unit circle
with all poles of its matrix entries removed. For the real-parameter
version below, the domain is the real axis with those poles removed.
The next theorem uses the polynomial tests to locate a finite set of
candidates, then uses closedness of $\mathcal E$ to settle the remaining
case.

\begin{theorem}[Finite exceptions in rational families]
\label{param:finite}
Let $\G(z)$ be a fixed finite list of matrix-valued rational functions
with algebraic coefficients, unitary whenever $|z|=1$ and no entry has
a pole. Exactly one of the following holds: $\G(z)\in\mathcal E$
at every phase in the domain, or it belongs to $\mathcal E$ at only
finitely many phases. In the second case all exceptional phases are
algebraic. The alternative and the finite exceptional set are
effectively computable from the rational functions. One phase in the
domain with $\G(z)\notin\mathcal E$ therefore implies $\#\mathrm P$-hard normal-form
amplitudes at all but finitely many fixed phases in the domain.
The same conclusions hold for a rational family unitary at every
real point of its domain, with algebraic exceptional real parameters.
\end{theorem}
\begin{proof}
Substitute the rational entries into the tests~\eqref{param:tests}
and clear the entry denominators to write
\[
 p_i(\G(z))=\frac{a_i(z)}{b_i(z)},\qquad 1\le i\le t,
\]
where $a_i,b_i$ have algebraic coefficients and each $b_i$ is nonzero
on the domain. There a test holds exactly when $a_i(z)=0$.
An identically zero numerator gives a constant true test. A nonzero
numerator gives a false test away from its finitely many roots.
Let $B$ be the union of all entry poles and all roots of the
nonzero numerator polynomials. This is a computable finite set of
algebraic points. Outside $B$, each atomic test, and hence the whole
Boolean formula, has a fixed truth value. We call this the generic
value. Testing which numerators vanish identically computes it.

By Lemma~\ref{param:complex-bridge}, the formula describes membership
in $\mathcal E$ at every unitary point, including transcendental ones.
Thus its generic value first gives a finite or cofinite tractable set
within the domain. If the value is false, all tractable phases lie
among the roots used to form $B$. Compute these roots and retain only
those on the unit circle and away from the entry poles. At each
remaining candidate the gate entries are algebraic, so
Theorem~\ref{norm:recognition} decides membership in $\mathcal E$.
The accepted candidates form exactly the finite exceptional set.

If the generic value is true, closedness upgrades the cofinite set
to the entire domain. Fix a phase $z_0$ in the domain and approach
it through phases outside the finite set $B$. The rational entries
are continuous at $z_0$, and the approaching libraries belong to the
closed set $\mathcal E$ by Theorem~\ref{param:geometry}. Hence
$\G(z_0)\in\mathcal E$. This proves tractability throughout the
domain.

The computed generic value therefore decides the alternative, and
the finite root tests compute the exceptional set when it exists.
For a real parameter, retain real candidates away from the poles and
approach any domain point through real points outside $B$. The same
closedness argument applies. The general dichotomy gives the stated
pointwise hardness conclusions.
\end{proof}

In the finite-exception case, the number of exceptions is bounded
by $Cd$ for fixed gate dimensions, where $d\ge1$ bounds the numerator
and denominator degrees of the matrix entries and $C$ is a computable
constant depending only on those dimensions. The tests
in~\eqref{param:tests} have fixed number and degree. Substitution and
clearing denominators gives univariate numerators of degree $O(d)$,
and every exception is a root of a nonzero numerator.

Laurent-polynomial parametrizations in $z=e^{i\theta}$ are included,
with one shared phase fixed before the circuit is given. The algebraic
exceptional points are the values of $z$ on the unit circle.
Algebraicity of the rational-function coefficients supplies the
effective polynomial tests and finite root computations. The evaluated
phase itself may be transcendental.

\section{Exact evaluation from fixed-width gate closures}
\label{app:closure}

We show how projective closure on a fixed local block gives an exact
amplitude reduction for circuits of growing size. The block width is
fixed, while the total number of wires and the number of target-gate
occurrences may grow. Each occurrence may act on a different ordered
tuple of wires. The reduction expresses the amplitude as a polynomial
in the target matrix entries and recovers its value at the target from
values on finitely many available words. The key is that the monomial
vectors of those words span a finite-dimensional complex space, whose closedness
turns a limiting relation into an exact linear identity.

Fix a finite algebraic gate library $\G$ and a positive integer $q$.
All amplitudes use the number-field encoding of
Appendix~\ref{app:prelim}. Put $D=2^q$, and let
$\Omega\subseteq U(D)$ be a finite alphabet of algebraic matrices.
An available word is a product of letters from $\Omega$, including the
empty word $I_D$. Only letters in $\Omega$ are used in these products.
For example, $\Omega$ may consist of all placements of gates from $\G$
on these $q$ wires. This alphabet is finite because $q$ and $\G$ are
fixed. Let $\pi:U(D)\to PU(D)$ be the quotient by scalar phases.
For $V\in U(D)$, membership of $\pi(V)$ in the projective closure means
that there are available words $w_j$ and scalars $|\zeta_j|=1$ such that
$\zeta_jw_j\to V$. The reduction below recovers amplitudes with the
actual phase of $V$.

\begin{theorem}[Fixed-width projective closure]
\label{closure:theorem}
Suppose every letter of $\Omega$ has a fixed implementation over $\G$.
If $V\in U(D)$ has algebraic entries and $\pi(V)$ is in the topological
closure of the projective classes of available words, then
\[
 \Amp(\G\cup\{V\})\equiv_T^{\poly}\Amp(\G).
\]
The conclusion holds for any fixed finite collection of such targets.
It also holds for a $k$-qubit target, $k\le q$, if
$V\otimes I_{2^{q-k}}$ satisfies the closure condition. The polynomial
running time may depend on $q$.
\end{theorem}
\begin{proof}
We first write the target amplitude as a homogeneous polynomial.
Consider one $q$-qubit target $V$. An instance with no target occurrences
is already a circuit over $\G$, so suppose it has $r\ge1$ occurrences.
Replace every occurrence by the same indeterminate $D\times D$ matrix
$X$, retaining its ordered wire tuple. These positions all originally
use $V$, so the desired amplitude is the value $h(V)$ of the resulting
polynomial. Their different wire placements affect the coefficients
and index contractions. The polynomial $h(X)$ is homogeneous of degree
$r$ in the $d=D^2$ entries of $X$, because each summand in the contraction
contains one entry from each designated occurrence. All background
gates retain their positions.

Let $\nu_r(X)$ list all monomials of degree $r$ in these $d$ entries.
Its length is $N=\binom{r+d-1}{d-1}$. There is a linear functional
$\ell_h$ such that $h(X)=\ell_h(\nu_r(X))$. We will express
$\nu_r(V)$ as a linear combination of available-word vectors and then
apply $\ell_h$ through amplitude queries. This computes $h(V)$ directly.
Choose a fixed number field $F$ containing the entries of $\Omega$, and set
\[
 W_r=\operatorname{span}_F
       \{\nu_r(g):g\text{ is an available word}\}\subseteq F^N.
\]
We next construct a basis of $W_r$ represented by short available words.
Start with $\nu_r(I_D)$, labelled by the empty word, and maintain a queue
of accepted basis vectors. For each accepted vector $\nu_r(g)$ and each
$A\in\Omega$, compute $Ag$ and $\nu_r(Ag)$. Accept this vector and
its word precisely when it is independent of the accepted vectors.
Process each accepted vector once.

Left multiplication $X\mapsto AX$ induces a linear substitution
$T_r(A):F^N\to F^N$ with
$\nu_r(AX)=T_r(A)\nu_r(X)$. If a candidate is a linear combination
of accepted vectors, its image under $T_r(A)$ is the same combination
of their images. Thus extending accepted vectors suffices to account
for successors of rejected candidates as well. The search computes
each required image by evaluating $\nu_r(Ag)$.

When the queue is empty, the accepted span contains $\nu_r(I_D)$ and
is invariant under every $T_r(A)$. Induction on word length shows that
it contains every $\nu_r(g)$, so it equals $W_r$. At most $N$ vectors
are accepted and at most $|\Omega|N$ candidates are tested. Every
accepted word has length at most $N-1$. Indeed, the search-parent chain
of a length-$t$ word contains $t+1$ accepted nodes, starting with the
empty word and adding one letter at each step. Each node contributes
a distinct independent vector. Write the resulting basis as
$\nu_r(g_1),\ldots,\nu_r(g_s)$, $s\le N$.

The projective closure condition now places the target vector in the
complex span of this basis. It gives available words $w_j$ and phases
$\zeta_j$ such that $\zeta_jw_j\to V$. Homogeneity absorbs these phases
into scalar coefficients:
\[
 \nu_r(\zeta_jw_j)=\zeta_j^r\nu_r(w_j)\in W_r\otimes_F\mathbb C.
\]
Here $W_r\otimes_F\mathbb C$ is the space spanned by the same basis
vectors over $\mathbb C$. It is finite-dimensional and therefore closed,
so $\nu_r(V)$ belongs to it. Let $E$ be a fixed number field containing $F$,
the entries of $V$, and all background gate entries. The linear system
\begin{equation}
 \nu_r(V)=\sum_{j=1}^s c_j\nu_r(g_j)
 \label{closure:linear-system}
\end{equation}
has a complex solution and hence an $E$-valued solution, since field
extension preserves the ranks of a matrix and its augmentation.
The limiting sequence establishes consistency of this finite system.
The algorithm solves the system exactly over $E$, using the computed
basis and the entries of $V$. Applying $\ell_h$ gives the query identity
\begin{equation}
 h(V)=\sum_{j=1}^s c_jh(g_j).
 \label{closure:oracle-sum}
\end{equation}
We realize each value $h(g_j)$ as a legal amplitude query by replacing
every designated occurrence with the same word $g_j$ on its own ordered
wires, using the fixed implementations of its letters. This local
replacement preserves acyclicity. If the original circuit has
$m_{\mathrm{bg}}$ background gates and each letter implementation uses
at most $L_\Omega$ gates, a query has at most
\[
 m_{\mathrm{bg}}+rL_\Omega(N-1)
\]
gates. There are $s\le N$ queries. Including wire labels and boundary
strings, their total encoding length is polynomial in the original
input size. The oracle values supply the required evaluations of
$\ell_h$, so its coefficients need not be computed.

It remains to control the exact arithmetic. In a fixed rational basis
of $E$, multiplication uses fixed rational structure constants, as in
Appendix~\ref{app:prelim}. For products of $O(N)$ fixed matrices, a fixed
denominator raised to $O(N)$ clears all coordinate denominators.
Repeated multiplication bounds the absolute values of the resulting
integer coordinates exponentially in $N$, giving polynomial bit length.
Degree-$r$ monomials in these entries therefore have bit length polynomial
in $rN$, as do the coordinates of $\nu_r(V)$. Independence tests and
equation~\eqref{closure:linear-system} involve systems of dimension at
most $N$ over a fixed field. Clearing denominators and bounding
determinants gives polynomial coordinate lengths for their exact
solutions. The oracle outputs have bit length polynomial in the query
size by Appendix~\ref{app:prelim}. Since $d$ is fixed, $N$ is polynomial
in $r$, and equation~\eqref{closure:oracle-sum} is computable in time
polynomial in the original input size.
Its result is converted to the prescribed output field by fixed linear
algebra.

For finitely many target types, eliminate them successively, treating
the remaining fixed targets as background gates at each stage. The
number of stages is fixed, so composing their polynomial-time reductions
still gives a polynomial-time reduction.

For a $k$-qubit target with $k<q$, first allocate $q-k$ fresh private
wires at each occurrence, initialized and finally pinned to zero.
Regard that occurrence as $V\otimes I_{2^{q-k}}$, which preserves these
zeros and leaves the target amplitude unchanged. Apply the preceding
polynomial argument to this padded circuit. Each substituted word may
act on its private wires, whose final zero pins are included in $h$.
Thus every query is a legal global amplitude, and
equation~\eqref{closure:oracle-sum} recovers the padded target value.
The reverse reduction simply regards a circuit over $\G$ as one over
the larger library.
\end{proof}

Thus two fixed finite algebraic libraries have equivalent exact-amplitude
problems whenever every generator of each, padded if necessary, belongs
to the other library's projective closure on a fixed block. In particular,
equality of the two fixed-width closures suffices, at a width containing
all generators, with both closures taken in the same computational
coordinates. This common coordinate system keeps the specified
computational-basis boundaries aligned in the two reductions.


\begin{thebibliography}{MWXZ25}
\raggedright
\bibitem[Aar05]{Aar05}
Scott Aaronson.
Ten semi-grand challenges for quantum computing theory.
July 2005.
\url{https://www.scottaaronson.com/writings/qchallenge.html}.

\bibitem[ADH97]{ADH97}
Leonard M. Adleman, Jonathan DeMarrais, and Ming-Deh A. Huang.
Quantum computability.
\emph{SIAM Journal on Computing}, 26(5):1524--1540, 1997.
\href{https://doi.org/10.1137/S0097539795293639}{doi:10.1137/S0097539795293639}.

\bibitem[AG04]{AG04}
Scott Aaronson and Daniel Gottesman.
Improved simulation of stabilizer circuits.
\emph{Physical Review A}, 70:052328, 2004.
\href{https://doi.org/10.1103/PhysRevA.70.052328}{doi:10.1103/PhysRevA.70.052328}.

\bibitem[Aru20]{Arute20}
F. Arute et al.\ (Google AI Quantum and Collaborators).
Hartree--Fock on a superconducting qubit quantum computer.
\emph{Science}, 369(6507):1084--1089, 2020.
\href{https://doi.org/10.1126/science.abb9811}{doi:10.1126/science.abb9811}.

\bibitem[Bac17]{Bac17}
Miriam Backens.
A new Holant dichotomy inspired by quantum computation.
In \emph{ICALP 2017}, LIPIcs 80, 16:1--16:14, 2017.
\href{https://doi.org/10.4230/LIPIcs.ICALP.2017.16}{doi:10.4230/LIPIcs.ICALP.2017.16}.

\bibitem[Bac21]{Bac21}
Miriam Backens.
A full dichotomy for $\operatorname{Holant}^{c}$, inspired by quantum computation.
\emph{SIAM Journal on Computing}, 50(6):1739--1799, 2021.
\href{https://doi.org/10.1137/20M1311557}{doi:10.1137/20M1311557}.

\bibitem[Bac25]{Bac25}
Miriam Backens.
Erratum: A full dichotomy for $\operatorname{Holant}^{c}$, inspired by quantum computation.
\emph{SIAM Journal on Computing}, 54(3):814--818, 2025.
\href{https://doi.org/10.1137/24M167723X}{doi:10.1137/24M167723X}.

\bibitem[BFK18]{BFK18}
Adam Bouland, Joseph F. Fitzsimons, and Dax Enshan Koh.
Complexity classification of conjugated Clifford circuits.
In \emph{CCC 2018}, LIPIcs 102, 21:1--21:25, 2018.
\href{https://doi.org/10.4230/LIPIcs.CCC.2018.21}{doi:10.4230/LIPIcs.CCC.2018.21}.

\bibitem[BMZ16]{BMZ16}
Adam Bouland, Laura Man\v{c}inska, and Xue Zhang.
Complexity classification of two-qubit commuting Hamiltonians.
In \emph{CCC 2016}, LIPIcs 50, 28:1--28:33, 2016.
\href{https://doi.org/10.4230/LIPIcs.CCC.2016.28}{doi:10.4230/LIPIcs.CCC.2016.28}.

\bibitem[BSS26]{BSS26}
J\k{e}drzej Burkat, Sergii Strelchuk, and Micha\l{} Studzi\'nski.
The power of power-of-SWAP: Postselected quantum computation with the
exchange interaction, 2026.
\href{https://arxiv.org/abs/2603.28527v2}{arXiv:2603.28527v2}.

\bibitem[Buc65]{Buc65}
Bruno Buchberger.
\emph{Ein Algorithmus zum Auffinden der Basiselemente des
Restklassenringes nach einem nulldimensionalen Polynomideal}.
PhD thesis, University of Innsbruck, 1965.
English translation in \emph{Journal of Symbolic Computation},
41(3--4):475--511, 2006.
\href{https://doi.org/10.1016/j.jsc.2005.09.007}{doi:10.1016/j.jsc.2005.09.007}.

\bibitem[CFS20]{CFS20}
Jin-Yi Cai, Zhiguo Fu, and Shuai Shao.
From Holant to quantum entanglement and back.
In \emph{ICALP 2020}, LIPIcs 168, 22:1--22:16, 2020.
\href{https://doi.org/10.4230/LIPIcs.ICALP.2020.22}{doi:10.4230/LIPIcs.ICALP.2020.22}.

\bibitem[CGW16]{CGW16}
Jin-Yi Cai, Heng Guo, and Tyson Williams.
A complete dichotomy rises from the capture of vanishing signatures.
\emph{SIAM Journal on Computing}, 45(5):1671--1728, 2016.
\href{https://doi.org/10.1137/15M1049798}{doi:10.1137/15M1049798}.

\bibitem[CGW18]{CGW18}
Jin-Yi Cai, Heng Guo, and Tyson Williams.
Clifford gates in the Holant framework.
\emph{Theoretical Computer Science}, 745:163--171, 2018.
\href{https://doi.org/10.1016/j.tcs.2018.06.010}{doi:10.1016/j.tcs.2018.06.010}.

\bibitem[Choi75]{Choi75}
Man-Duen Choi.
Completely positive linear maps on complex matrices.
\emph{Linear Algebra and its Applications}, 10(3):285--290, 1975.
\href{https://doi.org/10.1016/0024-3795(75)90075-0}{doi:10.1016/0024-3795(75)90075-0}.

\bibitem[CH96]{CH96}
Nadia Creignou and Miki Hermann.
Complexity of generalized satisfiability counting problems.
\emph{Information and Computation}, 125(1):1--12, 1996.
\href{https://doi.org/10.1006/inco.1996.0016}{doi:10.1006/inco.1996.0016}.

\bibitem[CLO15]{CLO15}
David A. Cox, John Little, and Donal O'Shea.
\emph{Ideals, Varieties, and Algorithms: An Introduction to Computational
Algebraic Geometry and Commutative Algebra}.
Fourth edition. Undergraduate Texts in Mathematics, Springer, 2015.
\href{https://doi.org/10.1007/978-3-319-16721-3}{doi:10.1007/978-3-319-16721-3}.

\bibitem[CLX09]{CLX09}
Jin-Yi Cai, Pinyan Lu, and Mingji Xia.
Holant problems and counting CSP.
In \emph{STOC 2009}, pages 715--724, 2009.
\href{https://doi.org/10.1145/1536414.1536511}{doi:10.1145/1536414.1536511}.

\bibitem[CLX11a]{CLX11}
Jin-Yi Cai, Pinyan Lu, and Mingji Xia.
Computational complexity of Holant problems.
\emph{SIAM Journal on Computing}, 40(4):1101--1132, 2011.
\href{https://doi.org/10.1137/100814585}{doi:10.1137/100814585}.

\bibitem[CLX11b]{CLX11star}
Jin-Yi Cai, Pinyan Lu, and Mingji Xia.
Dichotomy for $\operatorname{Holant}^{*}$ problems of Boolean domain.
In \emph{SODA 2011}, pages 1714--1728, 2011.
\href{https://doi.org/10.1137/1.9781611973082.132}{doi:10.1137/1.9781611973082.132}.

\bibitem[DN06]{DN06}
Christopher M. Dawson and Michael A. Nielsen.
The Solovay--Kitaev algorithm.
\emph{Quantum Information and Computation}, 6(1):81--95, 2006.
\href{https://arxiv.org/abs/quant-ph/0505030}{arXiv:quant-ph/0505030}.

\bibitem[dBH22]{DH22}
Niel de Beaudrap and Steven Herbert.
Fast stabiliser simulation with quadratic form expansions.
\emph{Quantum}, 6:803, 2022.
\href{https://doi.org/10.22331/q-2022-09-15-803}{doi:10.22331/q-2022-09-15-803}.

\bibitem[Deb16]{Debnath16}
S. Debnath, N. M. Linke, C. Figgatt, K. A. Landsman, K. Wright, and C. Monroe.
Demonstration of a small programmable quantum computer with atomic qubits.
\emph{Nature}, 536:63--66, 2016.
\href{https://doi.org/10.1038/nature18648}{doi:10.1038/nature18648}.

\bibitem[DVC00]{DVC00}
W. D\"ur, G. Vidal, and J. I. Cirac.
Three qubits can be entangled in two inequivalent ways.
\emph{Physical Review A}, 62:062314, 2000.
\href{https://doi.org/10.1103/PhysRevA.62.062314}{doi:10.1103/PhysRevA.62.062314}.

\bibitem[FGHP99]{FGHP99}
Stephen Fenner, Frederic Green, Steven Homer, and Randall Pruim.
Determining acceptance possibility for a quantum computation is hard
for the polynomial hierarchy.
\emph{Proceedings of the Royal Society of London A}, 455:3953--3966, 1999.
\href{https://doi.org/10.1098/rspa.1999.0485}{doi:10.1098/rspa.1999.0485}.

\bibitem[Fey82]{Fey82}
Richard P. Feynman.
Simulating physics with computers.
\emph{International Journal of Theoretical Physics}, 21:467--488, 1982.
\href{https://doi.org/10.1007/BF02650179}{doi:10.1007/BF02650179}.

\bibitem[Fox20]{Foxen20}
B. Foxen et al.\ (Google AI Quantum).
Demonstrating a continuous set of two-qubit gates for near-term quantum algorithms.
\emph{Physical Review Letters}, 125:120504, 2020.
\href{https://doi.org/10.1103/PhysRevLett.125.120504}{doi:10.1103/PhysRevLett.125.120504}.

\bibitem[FM17]{FM17}
Keisuke Fujii and Tomoyuki Morimae.
Commuting quantum circuits and complexity of Ising partition functions.
\emph{New Journal of Physics}, 19:033003, 2017.
\href{https://doi.org/10.1088/1367-2630/aa5fdb}{doi:10.1088/1367-2630/aa5fdb}.

\bibitem[Got99]{Got99}
Daniel Gottesman.
The Heisenberg representation of quantum computers.
In \emph{Group22: Proceedings of the XXII International Colloquium
on Group Theoretical Methods in Physics}, pages 32--43.
International Press, 1999.
\href{https://arxiv.org/abs/quant-ph/9807006}{arXiv:quant-ph/9807006}.

\bibitem[GS22]{GS22}
Daniel Grier and Luke Schaeffer.
The classification of Clifford gates over qubits.
\emph{Quantum}, 6:734, 2022.
\href{https://doi.org/10.22331/q-2022-06-13-734}{doi:10.22331/q-2022-06-13-734}.

\bibitem[GST26]{GST26}
Jincheng Guan, Shuai Shao, and Zhuxiao Tang.
A dichotomy for Boolean complex Holant problems with conjugate-closed
signature sets, 2026.
\href{https://arxiv.org/abs/2609.13132v1}{arXiv:2609.13132v1}.

\bibitem[JVW90]{JVW90}
F. Jaeger, D. L. Vertigan, and D. J. A. Welsh.
On the computational complexity of the Jones and Tutte polynomials.
\emph{Mathematical Proceedings of the Cambridge Philosophical Society},
108(1):35--53, 1990.
\href{https://doi.org/10.1017/S0305004100068936}{doi:10.1017/S0305004100068936}.

\bibitem[Jam72]{Jam72}
Andrzej Jamio{\l}kowski.
Linear transformations which preserve trace and positive
semidefiniteness of operators.
\emph{Reports on Mathematical Physics}, 3(4):275--278, 1972.
\href{https://doi.org/10.1016/0034-4877(72)90011-0}{doi:10.1016/0034-4877(72)90011-0}.

\bibitem[Joh11]{Johnston11}
Nathaniel Johnston.
Characterizing operations preserving separability measures via linear
preserver problems.
\emph{Linear and Multilinear Algebra}, 59(10):1171--1187, 2011.
\href{https://doi.org/10.1080/03081087.2011.596540}{doi:10.1080/03081087.2011.596540}.

\bibitem[JM08]{JM08}
Richard Jozsa and Akimasa Miyake.
Matchgates and classical simulation of quantum circuits.
\emph{Proceedings of the Royal Society A}, 464:3089--3106, 2008.
\href{https://doi.org/10.1098/rspa.2008.0189}{doi:10.1098/rspa.2008.0189}.

\bibitem[JvN14]{JvN14}
Richard Jozsa and Maarten Van den Nest.
Classical simulation complexity of extended Clifford circuits.
\emph{Quantum Information and Computation}, 14(7--8):633--648, 2014.
\href{https://arxiv.org/abs/1305.6190}{arXiv:1305.6190}.

\bibitem[KFG25]{KFG25}
Chaitanya Karamchedu, Matthew Fox, and Daniel Gottesman.
A criterion for post-selected quantum advantage, 2025.
\href{https://arxiv.org/abs/2411.02369v2}{arXiv:2411.02369v2}.

\bibitem[Koh17]{Koh17}
Dax Enshan Koh.
Further extensions of Clifford circuits and their classical simulation
complexities.
\emph{Quantum Information and Computation}, 17(3--4):262--282, 2017.
\href{https://arxiv.org/abs/1512.07892}{arXiv:1512.07892}.

\bibitem[LMW26]{LMW26}
Chenghua Liu, Boning Meng, and Juqiu Wang.
A full complexity dichotomy for complex-valued Boolean Holant problems, 2026.
\href{https://arxiv.org/abs/2610.00081v1}{arXiv:2610.00081v1}.
\bibitem[LW18]{LW18}
Jiabao Lin and Hanpin Wang.
The complexity of Boolean Holant problems with nonnegative weights.
\emph{SIAM Journal on Computing}, 47(3):798--828, 2018.
\href{https://doi.org/10.1137/17M113304X}{doi:10.1137/17M113304X}.

\bibitem[Man21]{Mann21}
Ryan L. Mann.
Simulating quantum computations with Tutte polynomials.
\emph{npj Quantum Information}, 7:141, 2021.
\href{https://doi.org/10.1038/s41534-021-00477-0}{doi:10.1038/s41534-021-00477-0}.

\bibitem[MS08]{MarkovShi08}
Igor L. Markov and Yaoyun Shi.
Simulating quantum computation by contracting tensor networks.
\emph{SIAM Journal on Computing}, 38(3):963--981, 2008.
\href{https://doi.org/10.1137/050644756}{doi:10.1137/050644756}.

\bibitem[MWXZ25]{MWXZ25}
Boning Meng, Juqiu Wang, Mingji Xia, and Jiayi Zheng.
From an odd arity signature to a Holant dichotomy.
In \emph{CCC 2025}, LIPIcs 339, 23:1--23:20, 2025.
\href{https://doi.org/10.4230/LIPIcs.CCC.2025.23}{doi:10.4230/LIPIcs.CCC.2025.23}.

\bibitem[OCo03]{OConnor03}
John J. O'Connor.
Full finite symmetry groups in 3 dimensions.
\emph{Topics in Geometry}, Lecture 12, 2003.
\url{https://jjoc45.github.io/geometry/Lectures/L12.html}.

\bibitem[Pla23]{Pla23}
Daniel Plaumann.
\emph{Real Algebraic Geometry}.
Tutorial lecture notes, Institut Henri Poincar\'e, October 2023.
\url{https://wwwold.mathematik.tu-dortmund.de/~dplauman/notes/rag-ihp.pdf}.

\bibitem[SC20]{SC20}
Shuai Shao and Jin-Yi Cai.
A dichotomy for real Boolean Holant problems.
In \emph{FOCS 2020}, pages 1091--1102, 2020.
\href{https://doi.org/10.1109/FOCS46700.2020.00105}{doi:10.1109/FOCS46700.2020.00105}.

\bibitem[Sch78]{Sch78}
Thomas J. Schaefer.
The complexity of satisfiability problems.
In \emph{STOC 1978}, pages 216--226, 1978.
\href{https://doi.org/10.1145/800133.804350}{doi:10.1145/800133.804350}.

\bibitem[She10]{She10}
Dan Shepherd.
Binary matroids and quantum probability distributions, 2010.
\href{https://arxiv.org/abs/1005.1744}{arXiv:1005.1744}.

\bibitem[TD02]{TD02}
Barbara M. Terhal and David P. DiVincenzo.
Classical simulation of noninteracting-fermion quantum circuits.
\emph{Physical Review A}, 65:032325, 2002.
\href{https://doi.org/10.1103/PhysRevA.65.032325}{doi:10.1103/PhysRevA.65.032325}.

\bibitem[Val02]{Val02}
Leslie G. Valiant.
Quantum circuits that can be simulated classically in polynomial time.
\emph{SIAM Journal on Computing}, 31(4):1229--1254, 2002.
\href{https://doi.org/10.1137/S0097539700377025}{doi:10.1137/S0097539700377025}.

\bibitem[Val08]{Val08}
Leslie G. Valiant.
Holographic algorithms.
\emph{SIAM Journal on Computing}, 37(5):1565--1594, 2008.
\href{https://doi.org/10.1137/070682575}{doi:10.1137/070682575}.

\bibitem[Vid03]{Vid03}
Guifr\'e Vidal.
Efficient classical simulation of slightly entangled quantum computations.
\emph{Physical Review Letters}, 91:147902, 2003.
\href{https://doi.org/10.1103/PhysRevLett.91.147902}{doi:10.1103/PhysRevLett.91.147902}.

\bibitem[Xia26]{Xia26}
Mingji Xia.
The framework to unify all complexity dichotomy theorems for Boolean
tensor networks, 2026.
\href{https://arxiv.org/abs/2603.09417v1}{arXiv:2603.09417v1}.
\bibitem[Yal25]{Yale25}
Christopher G. Yale, Ashlyn D. Burch, Matthew N. H. Chow, Brandon P. Ruzic,
Daniel S. Lobser, Brian K. McFarland, Melissa C. Revelle, and Susan M. Clark.
Realization and calibration of continuously parameterized two-qubit gates on
a trapped-ion quantum processor.
\emph{IEEE Transactions on Quantum Engineering}, 6:3102117, 2025.
\href{https://doi.org/10.1109/TQE.2025.3600216}{doi:10.1109/TQE.2025.3600216}.

\bibitem[YY99]{YY99}
Tomoyuki Yamakami and Andrew C. Yao.
$\mathrm{NQP}_{\mathbb C}=\mathrm{co}\text{-}\mathrm{C}_{=}\mathrm{P}$.
\emph{Information Processing Letters}, 71(2):63--69, 1999.
\href{https://doi.org/10.1016/S0020-0190(99)00084-8}{doi:10.1016/S0020-0190(99)00084-8}.
\end{thebibliography}
\end{document}